\documentclass[useAMS,usenatbib,usegraphicx]{mn2e}
\usepackage{amsmath,amssymb}
\usepackage{epsfig,subfigure,color}
\usepackage[T1]{fontenc}
\usepackage{ae,aecompl}

\def\apjl{ApJ}
\def\apjs{ApJS}
\def\ao{Appl.~Opt.}

\def\aap{A\&A}

\newcommand{\vecx}{\ensuremath{{\bmath{x}}}}
\newcommand{\veck}{\ensuremath{{\bmath{k}}}}

\newcommand{\vecg}{\ensuremath{{\bmath{g}}}}
\newcommand{\vecU}{\ensuremath{{\bmath{U}}}}
\newcommand{\vecf}{\ensuremath{{\bmath{f}}}}
\newcommand{\vecnab}{\ensuremath{{\bmath{\nabla}}}}

\newcommand{\Fb}{\ensuremath{{\bmath{\mathcal{F}}}}}
\newcommand{\Qb}{\ensuremath{{\bmath{\mathcal{Q}}}}}

\newcommand{\dd}{\ensuremath{\mathrm{d}}}

\def\gb {\vecg}
\def\Ub {\vecU}

\newcommand{\nyx}{\textsc{Nyx}}
\newcommand{\oiv}{\textsc{Oiv}}
\newcommand{\ovi}{\textsc{Ovi}}

\title[Cosmological Fluid Mechanics with Adaptively Refined LES]{Cosmological Fluid Mechanics with Adaptively Refined Large Eddy Simulations}
\author[W. Schmidt et al.]{\parbox{18cm}{
        W.~Schmidt$^{1}$\thanks{E-mail:schmidt@astro.physik.uni-goettingen.de}, 
        A.~S.~Almgren$^{2}$,
        H.~Braun$^{1}$,
        J.~F.~Engels$^{1}$,
        J.~C.~Niemeyer$^{1}$,
        J.~Schulz$^{3}$,
        R.~R.~Mekuria$^{1,4}$,
		A.~J.~Aspden$^{2,5}$,
        J.~B.~Bell$^{2}$}\vspace{0.3cm}\\
$^{1}$Institut f\"ur Astrophysik, Universit\"at G\"ottingen, Friedrich-Hund-Platz 1, D-37077 G\"ottingen, Germany\\
$^{2}$Center for Computational Sciences and Engineering, Lawrence Berkeley National Laboratory, Berkeley, CA 94720, USA\\
$^{3}$Institut f\"ur Numerische und Angewandte Mathematik, Universit\"at G\"ottingen, Lotzestra{\ss}e 16-18, D-37083 G\"ottingen, Germany\\
$^{4}$University of the Witwatersrand, 1 Jan Smuts Avenue Braamfontein 2000, Johannesburg, South Africa\\
$^{5}$School of Engineering, Cranfield University, College Road, Cranfield, Bedfordshire, MK43 0AL, United Kingdom}

\begin{document}

%\date{Accepted ... Received ...; in original form ...}
\date{Revised version, March 2014}

\pagerange{\pageref{firstpage}--\pageref{lastpage}} \pubyear{2013}

\maketitle

\label{firstpage}

\begin{abstract}

We investigate turbulence generated by cosmological structure formation by means of large eddy simulations using adaptive mesh refinement. 
In contrast to the widely used implicit large eddy simulations, which resolve a limited range of length scales and treat the effect of turbulent velocity fluctuations below the grid scale solely by numerical dissipation, we apply a subgrid-scale model for the numerically unresolved fraction of the turbulence energy. For simulations with adaptive mesh refinement, we utilize a new methodology that allows us to adjust the scale-dependent energy variables in such a way that the sum of resolved and unresolved energies is globally conserved. 
We test our approach in simulations of randomly forced turbulence, a gravitationally bound cloud in a wind, and the Santa Barbara cluster. To treat inhomogeneous turbulence, we introduce an adaptive Kalman filtering technique that separates turbulent velocity fluctuations on resolved length scales from the non-turbulent bulk flow. From the magnitude of the fluctuating component and the subgrid-scale turbulence energy, a total turbulent velocity dispersion of several 100 km/s is obtained for the Santa Barbara cluster, while the low-density gas outside the accretion shocks is nearly devoid of turbulence. The energy flux through the turbulent cascade and the dissipation rate predicted by the subgrid-scale model correspond to dynamical time scales around 5 Gyr, independent of numerical resolution.
\end{abstract}

\begin{keywords}
methods: numerical -- galaxies: clusters: intracluster medium -- intergalactic medium -- hydrodynamics -- turbulence -- magnetic fields.
\end{keywords}

\section{Introduction}

Throughout cosmological structure formation, turbulence is generated in the baryonic gas component as a result
of gravitational accretion into the potential wells of dark matter halos, merger events, and feedback processes from
galaxies and AGNs. This was mainly deduced from numerical studies, 
\citep[e.~g.][]{IapNie08,XuLi09,BrueggScann09,ZhuFeng10,PaulIap11,BorKravt11,VazzaBrun11,IapSchm11,VazzaBruegg13}. 
Direct observations of turbulence on cosmological 
scales, however, are difficult. There is no clear evidence for turbulence in cluster cores yet, because current
observational facilities cannot unambiguously detect turbulent line broadening of X-ray emissions
\citep{SunyNor03,RebusChur08,SandFab11}. Improved measurements are expected from the upcoming ASTRO-H mission 
\citep{BiffiDol12}. On the other hand, turbulence can be inferred from measurements
of random magnetic fields with a magnitude of a few $\mu$G in the ICM \citep{VogtEns05,FerrGov08,BonaFer10,BonaGov11}.
\citet{GasChur13} have recently demonstrated that a comparison of 
the power spectrum of gas density perturbations in simulations with observations of the Coma cluster implies subsonic turbulence.
There are also some indications of non-thermal line broadening in the IGM. \citet{RauchSar01} detected turbulence on 
kpc scales in \oiv\ absorbing clouds, which is likely to be caused by stellar feedback. 
\citet{OppDave09} showed that properties of \ovi\ absorbers are well reproduced by numerical simulations only
if the turbulent velocity dispersion is increased by a fudge factor, which can be interpreted as enhancement due to numerically
unresolved turbulence. Large scale motions of the IGM probed by Ly$\alpha$
forest absorption in QSO sight lines, however, appear to be dominated by coherent accretion flows, 
without clear evidence for turbulent fluctuations \citep{RauchBeck05}. 
This reflects a vorticity-free evolution in the nearly linear regime.

Indications of the presence of turbulence in the ICM warrant a thorough theoretical investigation. 
On a fundamental level, the production of turbulence is expected as a result of dynamical instabilities
when gas is accreted by galaxy clusters and surrounding filaments and during mergers. 
The dissipation of kinetic energy counteracts large-scale injection. The nature and the typical length scale of energy
dissipation is still highly uncertain \citep{NaraMed01,ReyKer05,Lazar06,RoedKraft13}. 
For the typical density and temperature of the ICM,
a strongly anisotropic viscosity associated 
with non-collisional magneto-turbulent dissipation processes
follows from theoretical considerations \citep{ParrCourt12,SantosGouv13}.
Nevertheless, a much simpler hydrodynamical description of baryonic gas dynamics is applied in many cosmological
simulations, which amounts to the following basic assumptions:
\begin{enumerate}
\item Non-collisional dissipative processes occur below the grid scale and anisotropies decay
on sufficiently small time scales such that the baryonic gas effectively behaves as a fluid on
numerically resolved length scales. 
\item Magnetic fields are subdominant for the numerically resolved turbulent cascade, although
a sizable fraction of the kinetic energy may be converted into magnetic energy by turbulent dynamo
action in MHD simulations.
\end{enumerate}
Large eddy simulations (LES) then
allow for the numerical treatment of non-linear fluid dynamics,
independent of the details of the dissipation mechanism. The only question is how
numerically resolved modes couple to the unresolved modes just below the grid scale and vice versa. 
In the simplest case,
this is modeled by the truncation error terms of the finite-volume discretization, a method that is
known as implicit large eddy simulation (ILES). In this article, we devise and 
test a subgrid-scale (SGS) model that explicitly accounts for the non-linear energy transfer
across the grid scale. This gives rise to both turbulent pressure and diffusion on resolved scales
\citep[see][]{SchmFeder11}. 
To begin with, we consider only hydrodynamical turbulence. However, 
LES have the potential to be generalized to magnetohydrodynamics.
Eventually, the first assumption stated above could also be relaxed by absorbing non-collisional
and anisotropic effects into SGS closures. 

While ILES dissipate kinetic energy directly into heat through the action of numerical viscosity, 
the additional dynamical variable
for the kinetic energy of numerically unresolved turbulence in LES provides an intermediate energy reservoir. 
Apart from energy dissipation, this is relevant for adaptive mesh refinement (AMR), which is indispensable
for grid-based cosmological simulations. As a consequence of momentum- and energy-conserving averaging and
interpolation, the kinetic energies in a particular spatial region differ between the refinement levels in AMR simulations. 
The standard method to keep consistent energy variables is to compensate these differences by adjusting the
internal energy if the grid is refined or values from higher refinement levels are averaged down to coarser grids. 
In contrast, energy can be transferred between the resolved
and unresolved kinetic energy variables in adaptively refined LES. 
This is naturally related to the scale-dependence of turbulent velocity fluctuations 
\citep[see also][]{MaierIap09}. 
A further difficulty when running LES of cosmological structure formation is that
turbulence is far from being homogeneous and stationary. 
This necessitates modifications to the SGS model. Here, we adopt the shear-improved method that
separates the resolved velocity field into a bulk flow and turbulent fluctuations by utilizing
an adaptive temporal low-pass filter \citep{LevTosch07,CahuBou10}. As we will show, this allows us to avoid
spurious production of SGS turbulence by the shear associated with accretion flows. Moreover, we
are able to compute the total (i.~e., resolved plus unresolved) turbulent velocity dispersion 
as opposed to the non-turbulent bulk flow.

For the implementation, we chose the cosmological $N$-body and fluid dynamics
code \textsc{Nyx} \citep{AlmBell12}. The code features the \textsc{BoxLib} framework
for block-structured AMR, which is particularly suitable for LES. A further advantage
is the high fidelity of the available hydro solvers, which is important for the numerical
treatment of dynamical instabilities and turbulent flows \citep{AlmBeck10}. 
In this article, a detailed exposition of the methodology of adaptively refined
LES is given in Section~\ref{sc:consrv}, and in Appendix \ref{sc:algo}.
In Sect~\ref{sc:num_tests}, we consider test cases. Firstly, statistically stationary
and homogeneous forced turbulence is computed in a periodic box with nested grids. In
particular, we investigate the effects of exponential expansion as an idealized model
of cosmic turbulence. 
Secondly, we revisit a simple model for a minor merger, i.~e., the infall of 
a subcluster into the ICM of big cluster, by computing the evolution of a gravitationally bound cloud 
in a wind \citep{IapiAda08}. Thirdly, we perform simulations of the Santa Barbara cluster, which
is a well known test problem for cosmological codes.
Apart from the matter-dominated cosmology, the gas dynamics is adiabatic and 
feedback processes are neglected. We thus keep the physics as simple as possible
in these simulations. This allows us to focus on fluid-dynamical aspects. 
In Section~\ref{sc:sb_sgs}, we elaborate on differences
introduced by the shear-improved model and the dependence on numerical resolution.
In particular, we analyze the statistics of the turbulent velocity dispersion
in the Santa Barbara cluster.
This also enables us to estimate the magnetic field 
on the basis of a simple energy equipartition argument.
We summarize our results and discuss perspectives for future applications
in Section~\ref{sc:concl}. 

\section{Energy Conservation on Adaptive Meshes and Dynamical Equations}
\label{sc:consrv}

On adaptive meshes, the energy variables must be corrected when new regions are refined or ghost cells of
finer grids are interpolated from coarser grids in order to maintain global conservation.
This is simply a consequence of the kinetic energy difference between refinement levels, which is
quadratic in the momentum, being incommensurate with the conservative interpolation 
of the momenta. The same problem occurs if data from finer grids are projected to coarser
grid representations by conservative averaging. Usually, the discrepancies are compensated
via the internal energy, i.~e., numerical cooling or heating is introduced. By using an SGS model, on the other hand, 
we are able exchange energy between the resolved and unresolved components of the turbulence energy, 
without changing the internal energy. We can equate coarse and fine grid total kinetic energies,

\begin{equation}
	(\rho K_{\rm tot})_{\rm crs} = \frac{1}{N}\sum_{n}(\rho K_{\rm tot})_n,
\end{equation}
by writing:
\begin{equation}
	\label{eq:energy_balance}
	\frac{1}{2}\frac{(\rho U)_{\rm crs}^2}{\rho_{\rm crs}} + (\rho K)_{\rm crs} = 
	\frac{1}{N}\sum_{n}\left(\frac{1}{2}\rho_{n} U_{n}^2 + \rho_n K_{n}\right),
\end{equation}
where 
\begin{equation}
	\label{eq:crs}	
	\rho_{\rm crs} := \overline{\rho} = \frac{1}{N}\sum_n\rho_{n},\quad
	(\rho\vecU)_{\rm crs} := \overline{\rho\vecU} = \frac{1}{N}\sum_n(\rho\vecU)_{n}
\end{equation}
are, respectively, the baryonic mass density and momentum of a coarse grid cell, say, at refinement level $l$.
The index $n$ runs over $N$ finer grid cells (usually chosen to be eight) at the next higher refinement level $l+1$.\footnote{The cell index $n$ has to be distinguished from the indices $i$, $j$, and $k$, which are used for vector or tensor components in the following section. Summation over $n$ is explicitly written, while the Einstein summation convention
applies to vector or tensor indices}.
The cells sizes are related by $\Delta_{l+1}=\Delta_{l}/r$, where the refinement ratio $r=2$ or $4$. 
The coarse-grid SGS turbulence energy, on the other hand, is determined by
the balance equation~(\ref{eq:energy_balance}), which can be written as
\begin{equation}
	\label{eq:rho_K_balance}
	(\rho K)_{\rm crs} = \overline{\rho K} + \Delta(\overline{\rho K}),\quad
%\end{equation}
\mbox{where}\quad
%\begin{equation}
	\overline{\rho K} = \frac{1}{N}\sum_{n}(\rho K)_n
\end{equation}
and
\begin{equation}
	\label{eq:delta_energy}
	\Delta(\overline{\rho K}) =
	\frac{1}{N}\sum_{n}\frac{1}{2}\rho_{n} U_{n}^2 -
	\frac{1}{2}\frac{(\rho U)_{\rm crs}^2}{\rho_{\rm crs}}\,.
\end{equation}
If the flow is fully turbulent, the exchange of energy between the resolved and unresolved kinetic energy 
fractions is a natural consequence of the scaling properties of turbulent
velocity fluctuations. Since the unresolved kinetic energy fraction statistically decreases from coarser to finer levels
(assuming a power law for the velocity fluctuations on a given scale), typically $\Delta(\overline{\rho K})>0$
and $\overline{\rho K}<(\rho K)_{\rm crs}$. As a consequence, we must numerically ensure that the SGS turbulence
energy does not become negative if grid cells are refined. 
This is achieved by the algorithm described in Appendix~\ref{sc:ftec}, which also determines
how the energy difference $\Delta(\overline{\rho K})$ is distributed among the finer cells. 
The projection from finer to coarser grids, on the other hand, is completely specified by the above equations. 

Averaging the data from finer to coarser cells is an example for a filter operation, which
plays a central role in the theory of LES (see \citealt{Sagaut}). For any coarse cell, 
the Leonard tensor
\begin{equation}
\begin{split}
	\boldsymbol{\tau}_{\rm L} 
	&= -\frac{1}{N}\sum_{n}\rho_n \Ub_{n}\otimes\Ub_{n}+
	\frac{(\rho \Ub)_{\rm crs}\otimes(\rho \Ub)_{\rm crs}}{\rho_{\rm crs}}\\
	&\equiv -\overline{\rho \Ub\otimes\Ub}+
	\frac{\overline{\rho \Ub}\otimes\overline{\rho \Ub}}{\overline{\rho}}
\end{split}
\end{equation}
specifies the stresses associated with turbulent velocity fluctuations between length scales $\Delta_{l+1}$ and $\Delta_{l}$.
The energy correction defined by equation~(\ref{eq:delta_energy}) is given by the trace of this tensor:
\begin{equation}
	\Delta(\overline{\rho K}) = -2\,\mathrm{tr}\,\boldsymbol{\tau}_{\rm L}\,.
\end{equation}
There is an important correspondence between the above relation and the definition of the SGS turbulence energy 
in terms of the trace of the SGS turbulence stress tensor in the framework of the \cite{Germano92} consistent decomposition,
on which our SGS model is based (see \citealt{SchmNie06}). If we formally consider the limit 
$\Delta_{l+1}/\Delta_{l}\rightarrow 0$, then the Leonard tensor $\boldsymbol{\tau}_{\rm L}$ becomes the turbulence stress tensor $\boldsymbol{\tau}$ and 
$\Delta(\overline{\rho K})$ becomes $(\rho K)_{\rm crs}$. In contrast to $\Delta(\overline{\rho K})$ and 
$\boldsymbol{\tau}_{\rm L}$, which are associated with the narrow window of length scales between refinement levels
$l$ and $l+1$, $(\rho K)_{\rm crs}$ and $\boldsymbol{\tau}$ formally encompass all length scales from the physical
dissipation range up to $\Delta_{l+1}$.
The conservation of the total kinetic energy, as expressed
by equation~(\ref{eq:energy_balance}), can thus be seen as consequence of the consistent scale separation of the
fluid dynamics equations, 
which is a necessary condition for satisfying their fundamental conservation properties.

In simulations of self-gravitating systems, a large component of the velocity $\Ub$ can be coherent accretion flow toward dense
structures. This particularly applies to cosmological structure formation. Since the non-turbulent component can have strong
gradients, it generally contributes to the kinetic energy differences between levels. As a consequence, the identification with
turbulence energy differences, as expressed by equation~(\ref{eq:delta_energy}) does not apply in this form. In this case, an estimate
of the genuinely turbulent energy fraction is required. As detailed in Appendix~\ref{sc:ctec}, combining the
rationale outlined above with the assumption of power-law scaling of the turbulent energy fraction turns out to work very well.
The non-turbulent fraction is then compensated through internal energy, corresponding to the standard method
in AMR simulations. A correction of the SGS turbulence energy based on Kolmogorov scaling has already been utilized by \citet{MaierIap09}. 
However, they do not employ an inter- but rather an \emph{intra}-level correction, meaning that energy is simply exchanged between the 
resolved kinetic and SGS energies in single cells after interpolation (grid refinement) or averaging (projection to coarser grid) 
so that the SGS turbulence energy is adjusted to the shifted grid scale according to the Kolmogorov two-thirds law. But the energy balance 
between the levels, which is expressed by equation~(\ref{eq:energy_balance}), is not observed at all. This has the drawback that neither 
the sum of resolved and SGS turbulence energies nor the momentum can be globally conserved. The method we propose here, on the other hand, 
is fully conservative, except for rare exceptions if energies become negative and cannot be corrected in a conservative fashion.  

We model the behaviour of the baryonic gas using the Euler equations with SGS terms and a partial differential equation for
turbulent energy density $\rho K$ below the grid scale \citep{SchmNie06,SchmFeder11}, which follow from
a compressible generalization of the the filtering formalism of \citet{Germano92}. To solve these equations in an
expanding space with scale factor $a(t)$, we transform them to comoving coordinates, as specified in \citet{AlmBell12}:
\begin{align}
	\label{eq:mass_les}
	\frac{\partial \rho}{\partial t} =& - \frac{1}{a} \nabla \cdot (\rho \Ub) \, , \\
	\label{eq:momt_les}
	\frac{\partial (a \rho \Ub)}{\partial t} =& 
	-             \nabla \cdot (\rho \Ub\otimes\Ub) 
	-             \nabla p
	+             \nabla \cdot \boldsymbol{\tau}
	+             \rho \gb  \, , \\
	\label{eq:energy_int_les}
	\begin{split}
	\frac{\partial (a^2 \rho e)}{\partial t} =&
	 - a \nabla \cdot (\rho \Ub e) - a p \nabla \cdot \Ub\\
	&+ a \dot{a} [2 - 3 (\gamma - 1) ] \rho e  + a\rho \varepsilon\, ,
	\end{split}\\
	\label{eq:energy_les}
	\begin{split}
	\frac{\partial (a^2 \rho E)}{\partial t} =& - a \nabla \cdot (\rho \Ub E + p \Ub)
	 + a \rho \Ub \cdot \gb \\
	&+ a \dot{a} [2 - 3 (\gamma - 1)] \rho e 
	 + a \nabla \cdot (\Ub\cdot\boldsymbol{\tau})\\
	&- a(\Sigma - \rho \varepsilon) \, , 	
	\end{split}
\end{align}
and
\begin{equation}
	\label{eq:k_les}
	\begin{split}
	\frac{\partial (a^2\rho K)}{\partial t} =&
	 - a\nabla \cdot \left(\rho \Ub K\right) 
	 + a\nabla \cdot \left(\rho \kappa_{\rm sgs}\nabla K\right) \\
	&+ a(\Sigma - \rho \varepsilon)\, ,
	\end{split}
\end{equation}
where $\rho$ is the comoving baryonic density, related to the proper density by $\rho = a^3 \rho_{\rm proper}$, 
$\Ub$ is the proper peculiar velocity (i.~e., the physical velocity minus the Hubble flow), 
the specific internal energy $e$ is related to the pressure $p$ by 
$\rho e = p/(\gamma - 1)$, where $\gamma$ is the adiabatic exponent, and $E = e + \Ub \cdot \Ub/2$ is the total 
specific energy on resolved scales.  The gravitational acceleration vector,
$\gb = - \nabla \phi$, where $\phi$ is the peculiar potential,
acts as forcing term in simulations of cosmological structure formation. 
The additional terms introduced by the SGS model are explained below. For details about the
numerical solution of the gas dynamics equations, see Appendix~\ref{sc:time_integr} and \citet{AlmBell12}.

The non-linear interaction between resolved and unresolved turbulent eddies is described by
the SGS turbulence stress tensor $\boldsymbol{\tau}$. Since the right-hand side of the
momentum equation~(\ref{eq:momt_les}) has exactly the same form as for non-expanding fluids, 
we conjecture that $\boldsymbol{\tau}$ can be adopted from turbulence studies in
static space. Thereby, we also ensure that there is no contribution from the completely smooth
Hubble flow to the energy transfer across  the cutoff scale $\Delta$
that separates resolved from unresolved scales. This scale is only shifted in time by the scale factor $a(t)$. 
An explicit validation of this model assumption is not computationally feasible
at present, but we will show that reasonable results are obtained in various test
cases. For compressible turbulence, \citet{SchmFeder11} propose the following closure:
\begin{equation}
 \label{eq:tau_nonlin}
 \begin{split}
  \tau_{ij} =&\, 2C_{1}\Delta\rho(2 K_{\mathrm{sgs}})^{1/2}S_{\! ij}^{\ast}
   -4C_{2}\rho K\frac{U_{i,k}U_{j,k}}{|\nabla\otimes\Ub|^{2}}\\
  &-\frac{2}{3}(1-C_{2})\rho K\delta_{ij}\,.
  \end{split}
\end{equation}
where $|\nabla\otimes\Ub|:=(2U_{i,k}U_{i,k})^{1/2}$ is the norm of the resolved velocity derivative,
\begin{equation}
	\label{eq:strain}
	S_{ij}^{\ast} = S_{ij} - \frac{1}{3}\delta_{ij}d =
	\frac{1}{2} (U_{i,j} + U_{j,i}) - \frac{1}{3}\delta_{ij}U_{k,k}
\end{equation}
is the trace-free rate-of-strain tensor, and $\Delta$ is the grid scale in comoving coordinates. For highly
compressible turbulence in the supersonic regime, \citet{SchmFeder11} find $C_1 = 0.02$ and $C_2 = 0.7$, largely independent
of the Mach number, the forcing type, and the equation of state. The commonly used
eddy-viscosity closure is obtained for $C_1=0.095$ and $C_2=0$. This is a good approximation for 
weakly or moderately compressible turbulence \citep{SchmNie06}.
The SGS turbulence production and dissipation terms in equation~(\ref{eq:k_les}) are defined as
\begin{equation}
	\label{eq:prod}
	\Sigma = \tau_{ij} S_{ij} 
\end{equation}
and
\begin{equation}
	\label{eq:diss}
	\varepsilon = \frac{C_\varepsilon K^{3/2}}{\Delta}\,.
\end{equation}
We use a dissipation coefficient $C_{\epsilon}\approx 1.58$, which follows from the assumption of
an approximate balance between the mean production and dissipation rates in compressible turbulence
\citep{SchmFeder11}. A constant dissipation coefficient around unity 
is a very robust parameter of developed turbulence \citep{Frisch,Sagaut}.
Only during the transition from laminar to turbulent flow, the growth of $K$ will be slightly inhibited because of the
overestimate of the dissipation rate resulting from a constant value of $C_{\epsilon}$.
A time-dependent in-situ estimation of $C_{\epsilon}$ in LES was proposed by \citet{SchmNie06},
but it is questionable whether this method is applicable to the strongly inhomogeneous turbulence in cluster simulations.
The coefficient $\kappa_{\rm sgs} = C_{\kappa}\Delta K^{1/2}$ in equation~(\ref{eq:k_les}) is the SGS diffusivity with $C_{\kappa}=0.4$ \citep{SchmNie06}.
The production rate $\Sigma$ approximates the kinetic energy flux through the turbulent cascade across the
grid scale $\Delta$.
Here we assume that the Reynolds number of turbulence is sufficiently high that the damping of turbulent eddies by the microscopic viscosity of the fluid occurs entirely on the subgrid scales. Particularly for the ICM, this assumption requires a careful investigation, which is left for future work. Because of the numerical viscosity of finite-volume methods,
part of the numerically resolved kinetic energy is also dissipated directly into internal energy.

The formulation of the SGS model with constant coefficients presumes that turbulence is statistically
stationary and homogeneous. While this is the case for the simulations of forced turbulence in periodic boxes
presented in Section~\ref{sc:turb_box}, neither of these assumptions are satisfied for turbulence produced in the 
course of cosmological structure formation, where the forcing results from gas accretion by dark matter halos and mergers.
These processes occur episodically and are not space filling. 
To address this problem, we adopt the shear-improved method for LES of inhomogeneous 
terrestrial turbulent flows \citep{LevTosch07}.\footnote{
	While the original shear-improved model was formulated as a variant of the Smagorinsky model
	for incompressible LES, it can easily be carried over to other structural closures 
	that depend on the velocity derivative.
} The basic assumption of this method is that the production of turbulence is caused by
the shear associated with turbulent velocity fluctuations relative to an ensemble-averaged bulk flow. 
The accretion flows from voids towards
clusters and filaments can posses significant gradients, but can be separated from genuine turbulent velocity
fluctuations by a suitable filter. As pointed out in
Section~\ref{sc:consrv}, non-turbulent bulk flow is caused by the gravitational attraction of matter concentrations in 
cosmological simulations. Since we cannot compute ensemble averages in
LES, the bulk flow is locally approximated by Eulerian time averaging over the past history of the flow (corresponding to 
a temporal low-pass filter). This is computationally very efficient because the time averages can be iteratively computed. 
To implement the time averaging, \cite{CahuBou10} propose a Kalman filtering procedure (see Appendix~\ref{sc:kalman}
for a brief description of the algorithm). With this technique,  
the local velocity $\Ub$ can be separated into the mean flow $[\Ub]$ and a random component $\Ub^\prime=\Ub-[\Ub]$, 
which corresponds to the turbulent fluctuations, even if the mean flow is not constant in time. 
The turbulent stresses are then computed from the rate-of-strain tensor of the fluctuating component.
In the adiabatic cluster simulations considered in this work, turbulence is mostly
subsonic. For this reason, we use the simple eddy-viscosity closure with $C_1=0.095$ and $C_2=0$.
The shear-improved closure for the turbulent stresses is thus defined by\footnote{
	\citet{LevTosch07} define the turbulent viscosity of the Smagorinsky model in terms of the
	of the fluctuating component. To obtain a linear dependence on $U_{i,j}^{\prime}$ for the
	$K$-equation model, we define the SGS stress tensor rather than the turbulent viscosity analogous to 
	the shear-improved Smagorinsky model.}
\begin{equation}
  \label{eq:tau_si}
  \begin{split}
	\tau_{ij}^{\rm (SI)}=&\, 2C_{1}\Delta\rho(2 K_{\mathrm{sgs}})^{1/2}
	\left[\frac{1}{2} (U_{i,j}^\prime + U_{j,i}^\prime)
	      -\frac{1}{3}\delta_{ij}U_{k,k}^\prime\right] \\
	&-\frac{2}{3}\rho K\delta_{ij}\,.
  \end{split}
\end{equation}
To calculate $[\Ub]$ with the Kalman filter, two free parameters have to be specified: 
First, the characteristic velocity the bulk flow would assume if it were statistically stationary and, 
second, a time scale over which the flow evolves. In Sect~\ref{sc:sb_calibr}, we will show how these parameters 
can be chosen for cluster simulations. 

%------------------------------------------------------------------------
% Results from numerical simulations
%------------------------------------------------------------------------

\begin{figure*}
\centering
  \includegraphics[width=0.49\linewidth]{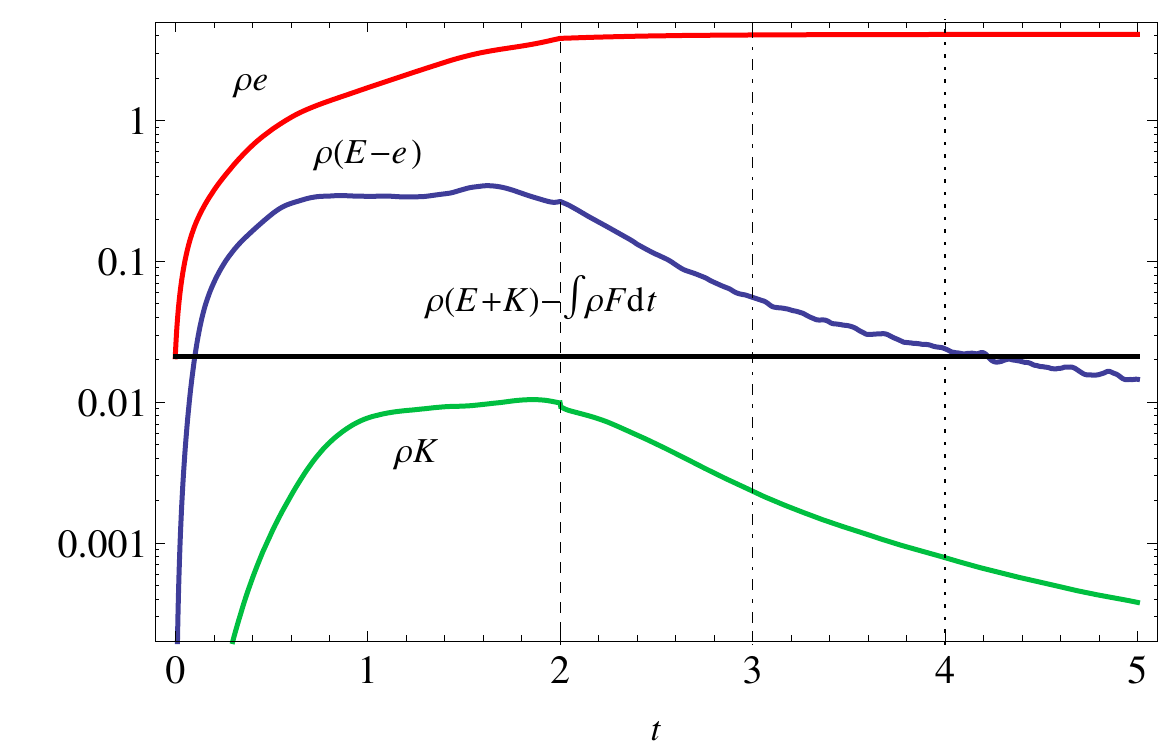}
  \includegraphics[width=0.502\linewidth]{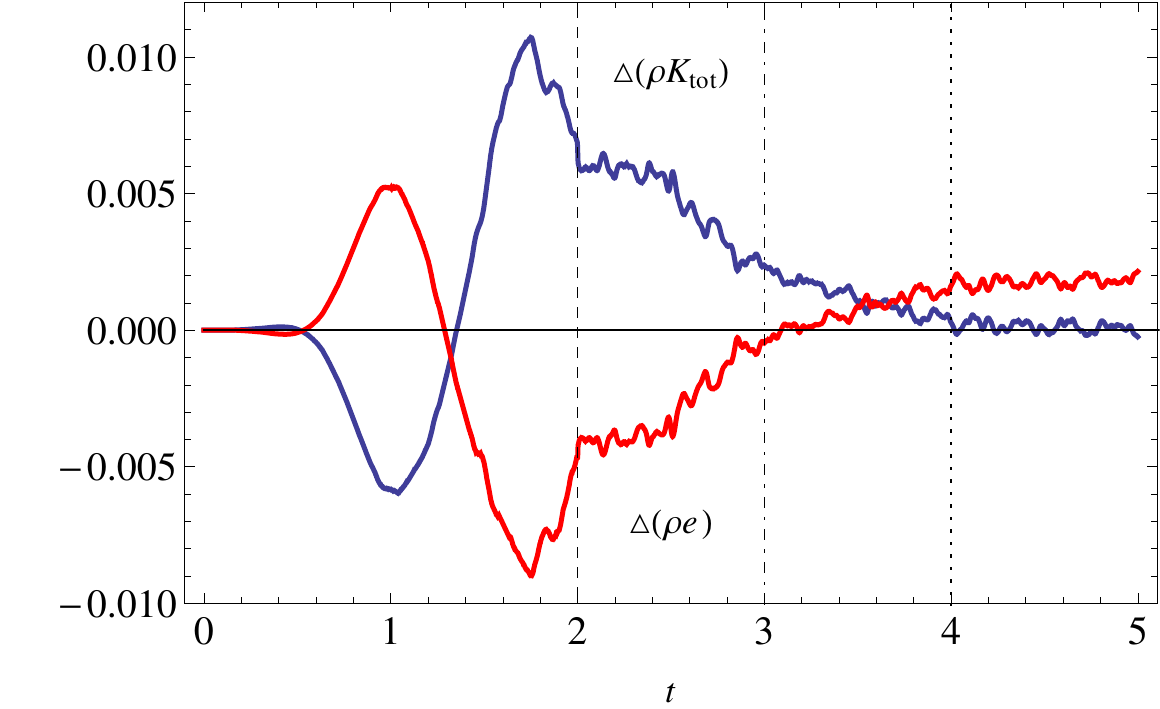}
\caption{Left plot: time evolution of the the mean internal, kinetic, conserved, and SGS turbulence energies 
  in an LES of forced adiabatic turbulence. Right plot: differences of the 
  mean kinetic and internal energies between LES and ILES with the same forcing (see equation~\ref{eq:K_tot}
  for the definition of $\rho K_{\rm tot}$). Time is normalized by the forcing time scale.
  The vertical lines indicate the insertion of nested refined grids with $1/2$,
  $1/4$ and $1/8$ of the root-grid cell size at $t=2.0$, $3.0$, and $4.0$, respectively 
  (also see figure~\ref{fig:adiab_slices}).}
\label{fig:adiab_evol}
\end{figure*}

\begin{figure*}
\centering
  \includegraphics[height=0.38\linewidth]{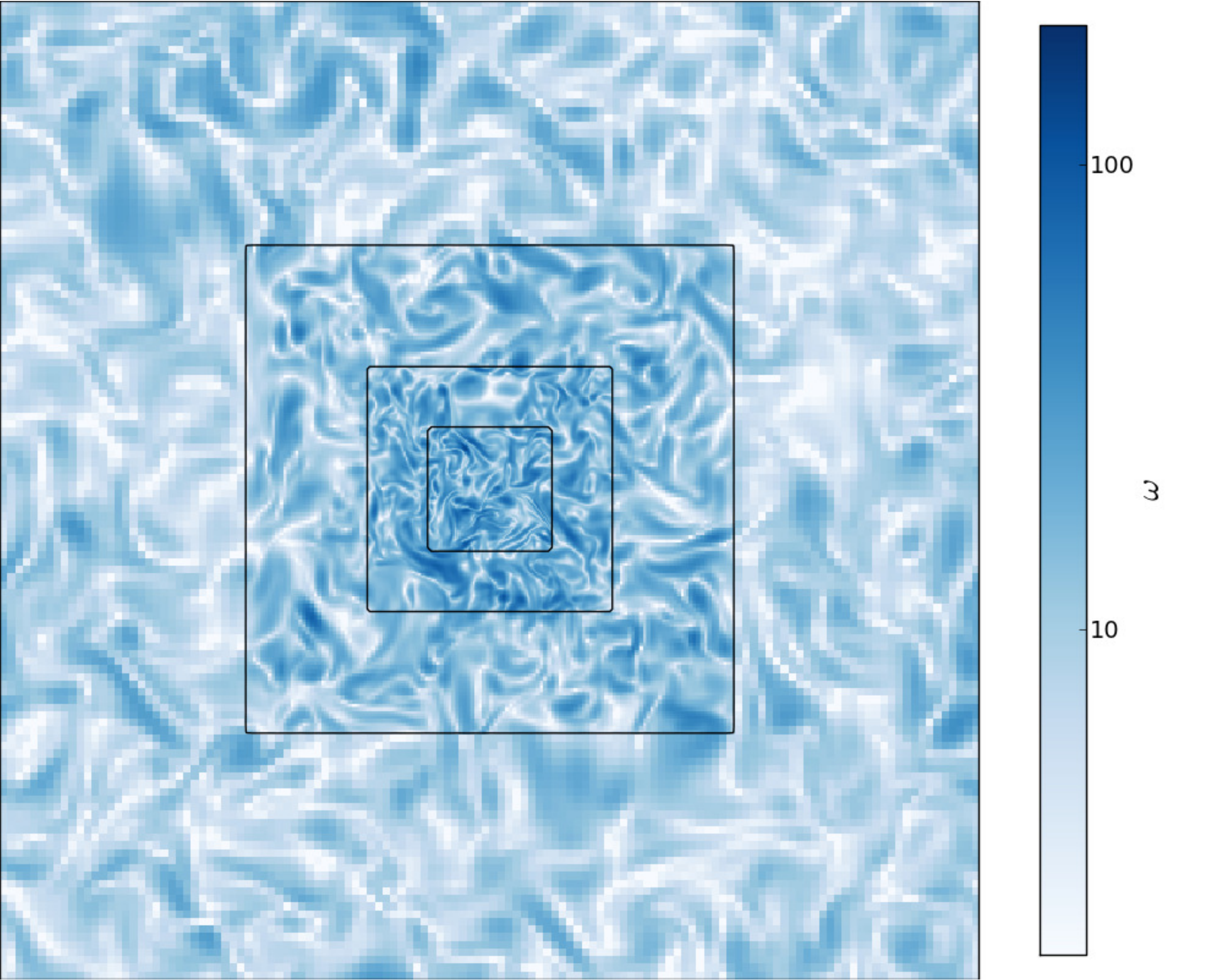}\quad
  \includegraphics[height=0.38\linewidth]{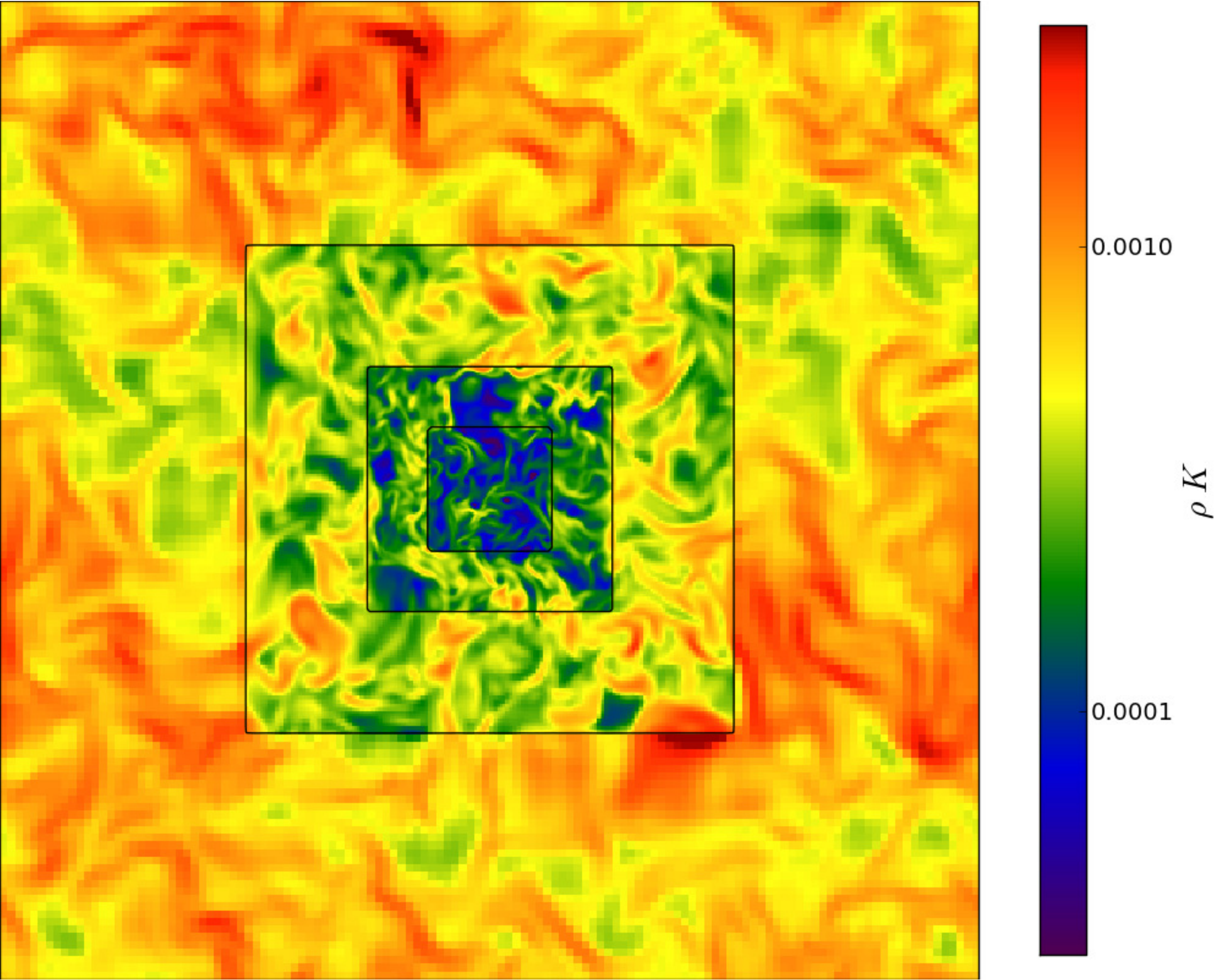}
\caption{Slices of the vorticity magnitude $\omega$ (left) and SGS turbulence energy $\rho K$ (right) for the
  forced adiabatic turbulence simulation at $t=4.025$. The boundaries of refined regions are indicated by
  black lines. 
  The root grid resolution is $128^3$, the effective resolution at the highest refinement level
  is $1024^3$.}
\label{fig:adiab_slices}
\end{figure*}

\section{Numerical tests}
\label{sc:num_tests}

\subsection{Forced turbulence}
\label{sc:turb_box}

Numerical simulations of statistically homogeneous turbulence produced by a random force field in a box with periodic boundary conditions
can be utilized to infer basic statistical properties of the SGS turbulence energy $\rho K$ \citep[see also][]{SchmFeder11}. 
Here, we insert nested grids and make use of the energy balance between finer and coarser grids outlined in
Section~\ref{sc:consrv} (fully turbulent energy compensation; see Section~\ref{sc:ftec} for details). 
To produce turbulence, 
we apply the large-scale forcing of \citet{AspNiki08}, which is composed from random plane waves with low wavenumbers
and time-dependent phase shifts.
The resulting force field $\vecf$, which replaces the gravitational acceleration $\gb$ in 
equations~(\ref{eq:momt_les}) and~(\ref{eq:energy_les}), 
is solenoidal (divergence-free). Obviously, $\vecf$ is not intended to mimic the rotation-free gravitational field $\gb$.
It merely provides an idealized mechanism of energy injection that excites eddy-like motions on length scales comparable
to the size of the box. For highly compressible flows, arguably, a mix of solenoidal and compressive modes should be used \citep{SchmFeder09a,KritUsty10}. 
Since we focus on turbulence at small or moderate Mach numbers, however, solenoidal forcing is suitable. For the same
reason, we use the eddy-viscosity closure for the SGS turbulence stress tensor. It turns out, that reasonable results
are obtained even for mildly supersonic turbulence.\footnote{For highly compressible, supersonic turbulence,
\citet{SchmFeder11} show that the prediction of the turbulence energy flux $\Sigma$ suffers from systematic biases.
In this case, the generalized closure~(\ref{eq:tau_nonlin}) with $C_2>0$ is preferable.} In the following, 
we normalize all quantities to unit mean density and unit box size, which is possible because hydrodynamic turbulence without 
gravity is scale-free. The forcing time scale $T$ is also chosen to be unity if $a=1$. 

First, we verify the conservation of the total energy. For adiabatic gas, this can be done by computing the global
average $\langle\rho(E+K)\rangle(t)$ at time $t$ and by subtracting the time-integrated work done by the forcing. This must equal the
initial energy density, which is $\langle\rho e\rangle(0)=\rho_0 e_0$, where $e_0$ is given by the initial temperature $T_0$:
\[
	\langle\rho(E+K)\rangle(t) + \int_0^t\langle\rho\vecU\cdot\vecf\rangle \dd t' = \rho_0 e_0
\]
Indeed, we find this to be the case for an LES with a root-grid resolution of $128^3$ and several
nested grids with higher resolution. In Fig.~\ref{fig:adiab_evol} (left panel), 
the evolution of the mean internal, kinetic, and SGS energies are plotted
together with the above expression for the total energy minus injected energy. 
In the initial phase, the forcing raises the
resolved kinetic energy $\langle\rho(E-e)\rangle$ and the SGS turbulence energy $\langle\rho K\rangle$ until
energy dissipation balances the injection of energy. This leads to the plateau of the mean kinetic energy, for $t$ greater than
about $1.0$. At $t=2.0$, the grid is refined from $\Delta_0=1/128$ to $\Delta_1=1/256$ in one eighth of the volume (i.~e.,
the linear size of the refined region is half the domain size). At the same time, the forcing is switched off and the
turbulence begins to decay. This prevents further heating of the gas due the 
dissipation of kinetic energy into internal energy, which would otherwise continue to heat the gas, resulting in a rapid decrease in Mach number.\footnote{A further reason is that the code infrastructure does not allow us to calculate the time-integrated work
done by the forcing on non-uniform grids.} 
In our simulation, the RMS Mach number $\mathcal{M}_{\rm rms}=\langle \rho U^2/(\gamma P)\rangle^{1/2}$ 
reaches a peak value of about $0.7$ at time $t\approx 0.5$ and decreases to less than $0.1$ at $t=5.0$.
For the grid refinement and the subsequent computation of numerical fluxes at the
boundaries between the refined region and the coarser root grid as well as the projection of the fine-grid solution
to the root grid, we apply the algorithms explained in Section~\ref{sc:consrv} and in Appendix~\ref{sc:refine}. 
As demonstrated by the statistics in Fig.~\ref{fig:adiab_evol}, energy conservation holds to very high precision 
also for the non-uniform grid structure. 
At time $t=3.0$ and $t=4.0$, further refined grids with resolutions $\Delta_2=1/512$ and $\Delta_3=1/1024$ are inserted, 
covering $1/4$ and $1/8$ of the linear domain size in each spatial direction. 
The resulting grid structure is illustrated together with 
slices of the SGS turbulence energy shortly after the insertion of the third refinement level
in Fig.~\ref{fig:adiab_slices}. This plot also shows that $\rho K$ systematically decreases from coarser to finer grids,
as is expected from the scaling behavior of the turbulence energy. 
By defining the total kinetic energy as the sum of the resolved and subgrid-scale energies, i.~e.,
\begin{equation}
	\label{eq:K_tot}
	\rho K_{\rm tot} = \frac{1}{2}\rho U^2 + \rho K,
\end{equation}
we can compare the difference of the mean kinetic energies for LES and ILES (see right plot in Fig.~\ref{fig:adiab_evol}).
In the case of the ILES, $K$ vanishes identically. The initial growth of $\rho K$ causes stronger dissipation in the LES,
which results in a steeper increase of the internal energy compared to ILEs around $t\approx 1$.
Once turbulence is developed, however, $\rho K_{\rm tot}$ tends to be larger in the LES because of the contribution from numerically unresolved scales. Energy conservation then implies a lower average of $\rho e$ in the LES.
This can be interpreted as delayed dissipation in LES, as resolved kinetic energy is first converted into SGS turbulence energy, which in turn is dissipated into heat. The fluctuations in the decay phase are caused by pressure waves. The asymmetry between the energy differences, which entails a positive internal energy difference in the late decay phase, is simply a consequence of the not exactly identical flow realizations in the two simulations (as a result, the mean rate of energy injection, $\langle\rho\vecU\cdot\vecf\rangle$, is slightly different).

\begin{figure*}
\centering
  \includegraphics[width=0.48\linewidth]{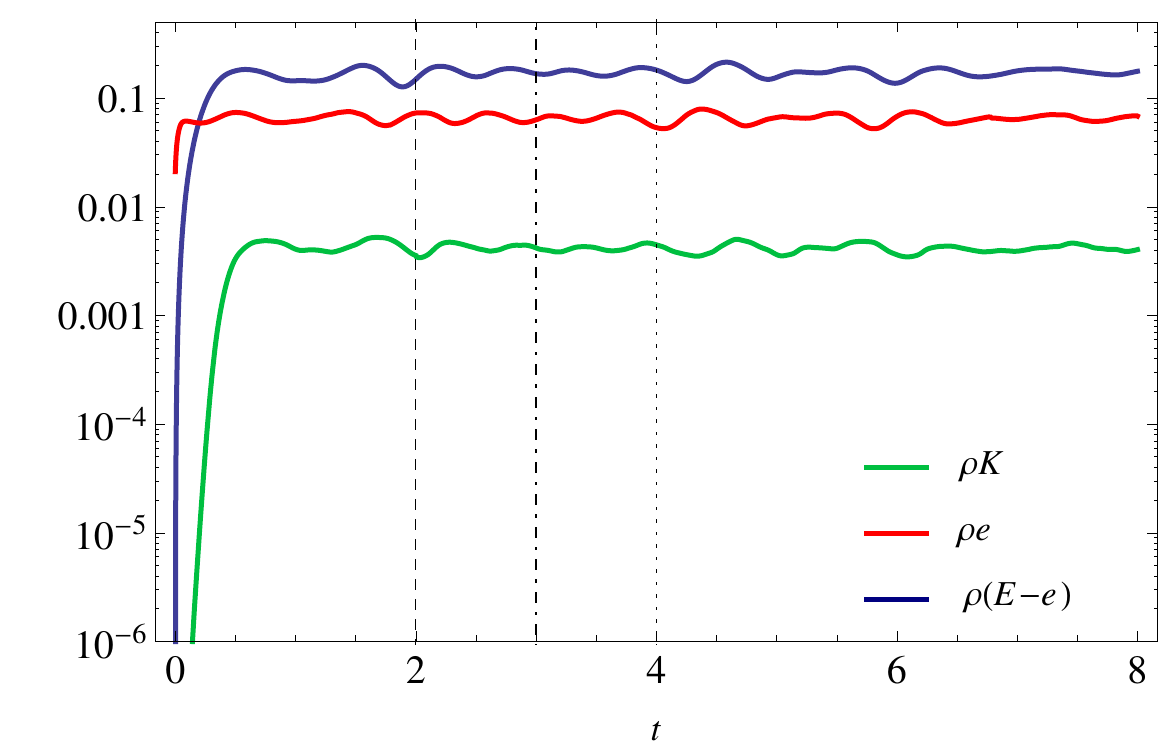}
  \includegraphics[width=0.48\linewidth]{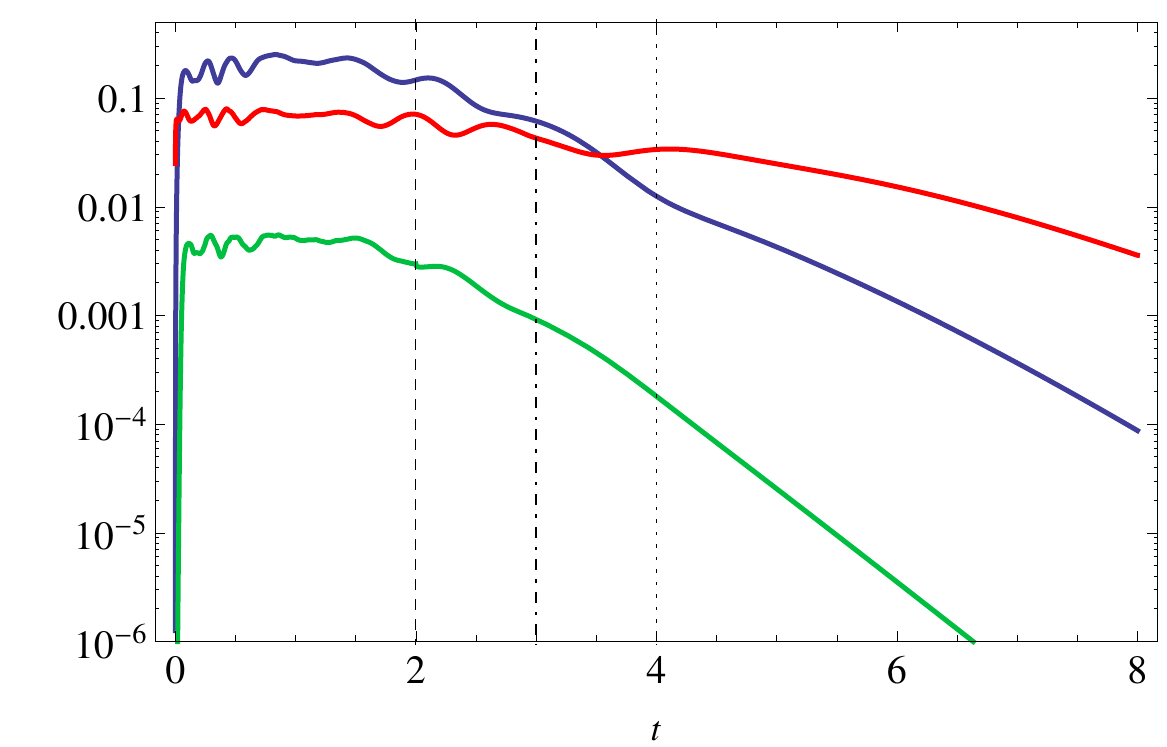}
 \caption{Time evolution of the the mean internal, kinetic, and SGS turbulence energies in a static-box LES of 
  isothermal turbulence with  $\mathcal{M}_{\rm rms}\approx 2.3$ (left panel) and for an expanding box with normalized 
  Hubble constant $H=1$ (right panel).
  The vertical lines indicate the insertion of nested refined grids with $1/2$,
  $1/4$ and $1/8$ of the root-grid cell size at $t=2.0$, $3.0$, and $4.0$, respectively.}
\label{fig:isoth_evol}
\end{figure*}

\begin{figure*}
\centering
 \includegraphics[width=0.48\linewidth]{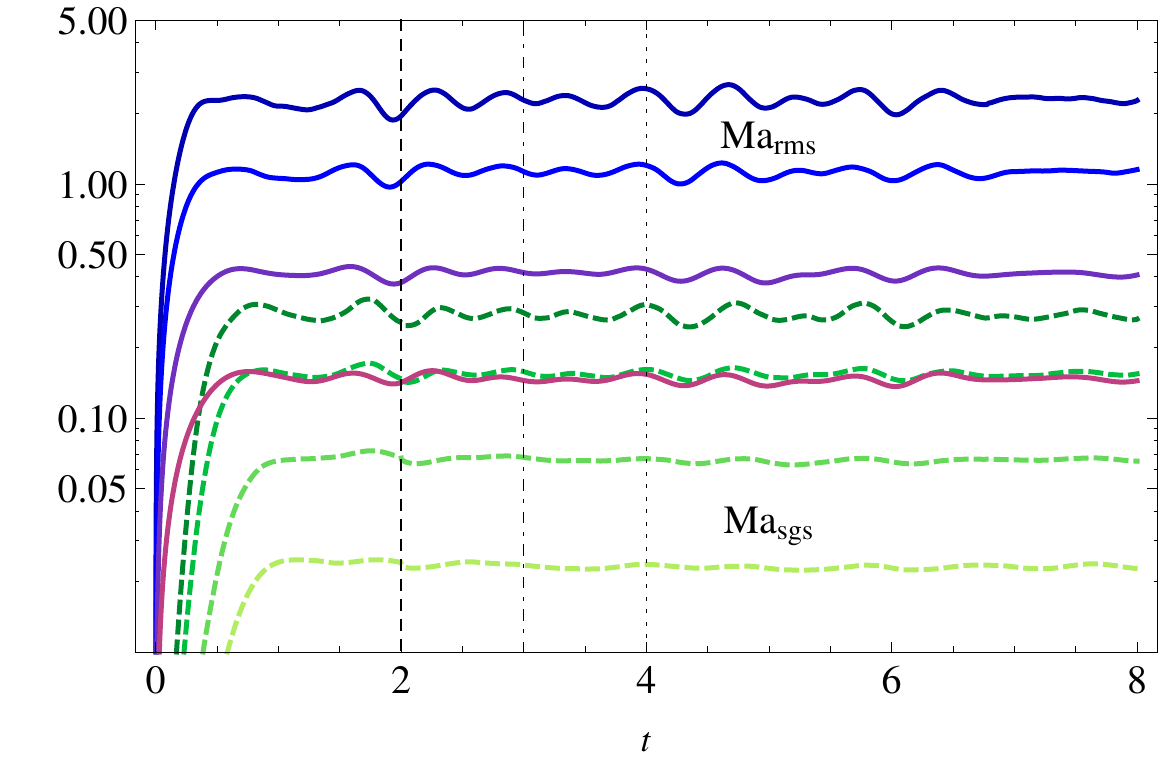}
  \includegraphics[width=0.48\linewidth]{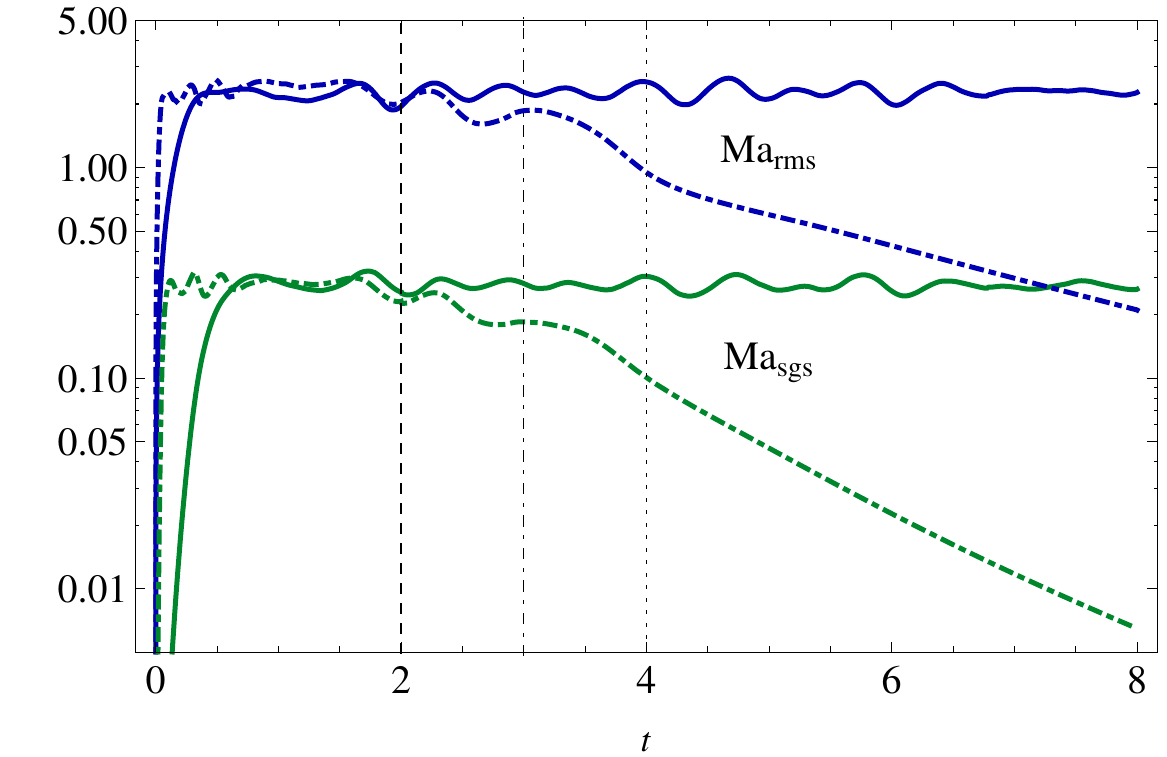}
\caption{In the left panel, the RMS Mach numbers of resolved (solid lines) and 
  subgrid-scale (dashed lines) velocity fluctuations
  are plotted as functions of time for a series of LES with different mean temperatures of the gas (top lines correspond
  to the lowest temperature, bottom lines to the highest temperature).
  The right panel compares LES with $H=0$ (solid lines) and $H=1$ (dot-dashed lines).}
\label{fig:isoth_mach_evol}
\end{figure*}

The next test problem we consider is compressible turbulence in a steady state, 
using the same nested grid structures as for the adiabatic turbulence simulation.
To maintain statistically stationary turbulence at an approximately constant Mach number, heat produced by the dissipation of 
kinetic energy has to be removed from the system. Rather than solving a strictly isothermal system, we add a simple source term,
\begin{equation}
    a\Lambda = a\alpha\rho(e-e_{0}),
\end{equation}
to the energy equations~(\ref{eq:energy_int_les}) and (\ref{eq:energy_les}), where $e_0$ corresponds to the initial temperature $T_0$.
The coefficient $\alpha<0$ regulates cooling ($e>e_{0}$) or heating ($e<e_{0}$) of the gas. For $T_0=1$ and $\alpha=-50$, the gas reaches a nearly isothermal state after a short transient phase and the kinetic energy saturates after about one dynamical time (see left plot in Fig.~\ref{fig:isoth_evol}). 
In this case, moderately supersonic turbulence with a time-averaged RMS Mach number 
$\mathcal{M}_{\rm rms}\approx 2.3$ is produced. 
By varying the initial temperatures $T_0$, different values of 
$\mathcal{M}_{\rm rms}$ are obtained. In Fig.~\ref{fig:isoth_mach_evol}, we show the evolution of $\mathcal{M}_{\rm rms}$ 
together with SGS turbulent Mach number $\mathcal{M}_{\rm sgs}=\langle 2\rho K/(\gamma P)\rangle^{1/2}$
for four different runs, ranging from weakly compressible to transonic and supersonic turbulence. 
One can also see that the ratio of $\mathcal{M}_{\rm rms}$ to $\mathcal{M}_{\rm sgs}$ is roughly the same for all runs, corresponding to the fixed numerical resolution. 

\begin{figure*}
\centering
 \includegraphics[width=0.48\linewidth]{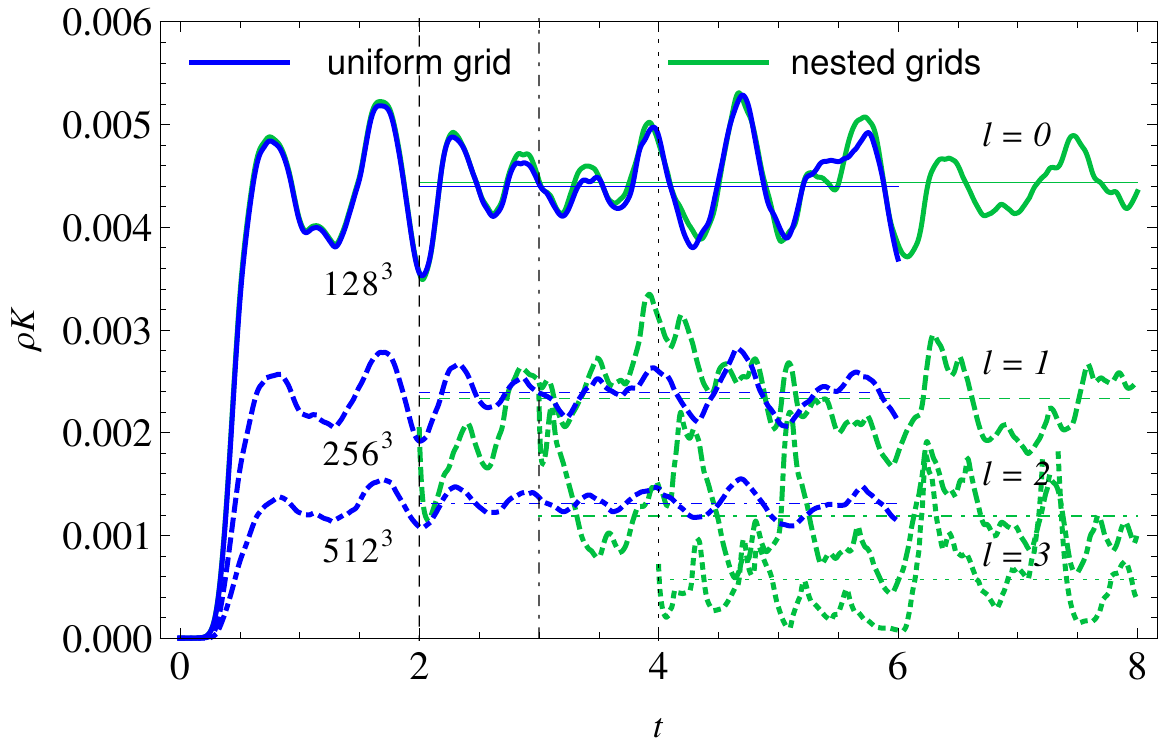}
 \includegraphics[width=0.48\linewidth]{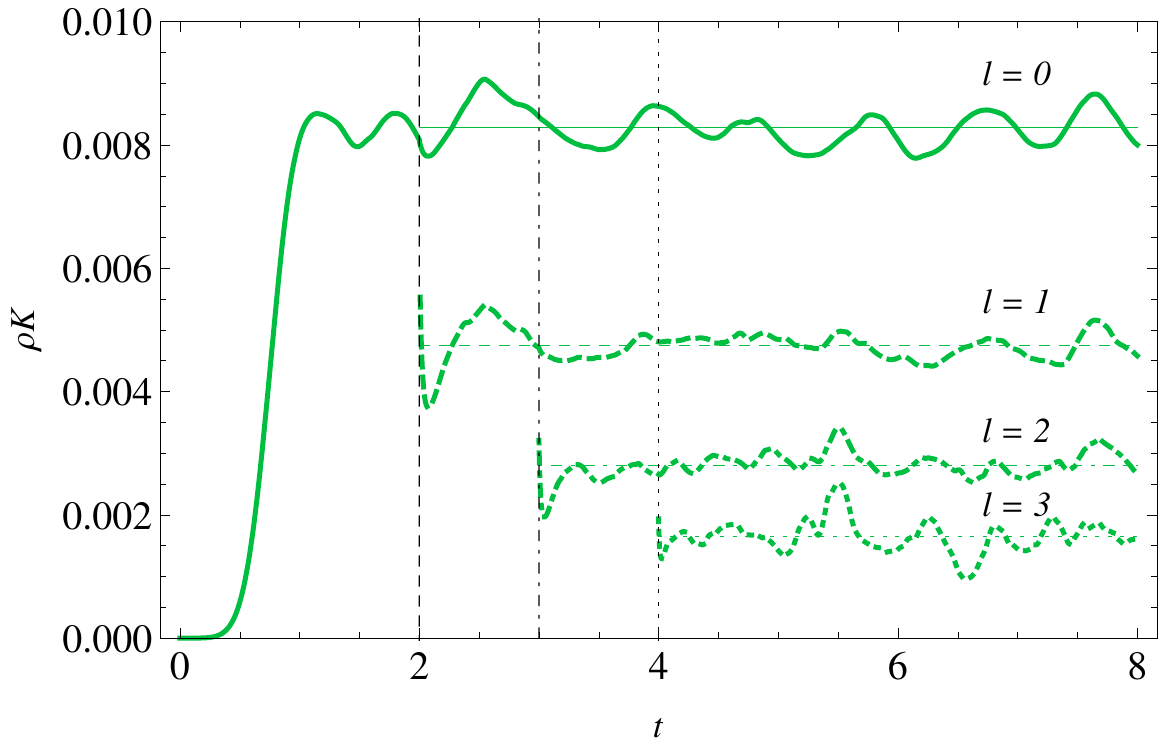}
\caption{Level-wise mean values of the SGS turbulence energy $\rho K$ as functions of time for 
  $\mathcal{M}_{\rm rms}\approx 2.3$ (left panel) and $\mathcal{M}_{\rm rms}\approx 0.15$ (right panel).
  Time averages are indicated by the thin horizontal lines.
  For the supersonic case, a comparison is made to a series of uniform-grid LES with the resolutions indicated
  in the plot, corresponding to the levels $l=0$, $1$, and $2$ of the runs with nested grids.}
\label{fig:isoth_levels_evol}
\end{figure*}

To investigate the effect of a time-dependent scale factor $a(t)$ in the system of equations~(\ref{eq:mass_les})--(\ref{eq:k_les}), we compute the evolution of forced turbulence in a box with 
$H=\dot{a}/a=\mathrm{const}$. 
Thus, the scale factor is given by $a(t)\propto\exp(Ht)$, corresponding to a de Sitter universe.
Regarding the forcing, one has to keep in mind that the
gas dynamics equations are invariant under the transformation $a=1\rightarrow a=a'$ for any constant $a'$ if the time
is rescaled as $t\rightarrow t'=t/a'$. To satisfy this invariance property for forced turbulence, we define the random forcing modes as functions of $t'=t/a(t)$ for any expansion law with time-dependent scale factor $a(t)$. Here, we choose $H=1$ and $a(2)=1$. Then $a(t)<1$
for $t<2$, and a time interval shorter than unity is required so that the force $\vecf(\vecx,t/a)$ goes through one period. 
Once $a(t)$ grows beyond unity, the forcing is evolving increasingly slow. This results in the graph of the mean kinetic energy shown in the right plot in Fig.~\ref{fig:isoth_evol}. It can be seen that the oscillations gradually slow down
as time progresses and the scale factor increases. Another important effect is that the energy of turbulence decreases
rapidly for $a(t)>1$, although the magnitude of the forcing is approximately constant. This can be understood as follows.
By adding equations~(\ref{eq:energy_les}) and~(\ref{eq:k_les}) and subtracting equation~(\ref{eq:energy_int_les}), an equation for the total 
kinetic energy $\frac{1}{2}\rho U^2 + \rho K$ is obtained. Since spatial averaging over the whole domain cancels all flux terms if the
boundary conditions are periodic, the mean energy is given by
\begin{equation}
	\frac{\dd}{\dd t}  \left(a^2\left\langle\frac{1}{2}\rho U^2 + \rho K\right\rangle\right) 
		= a\left\langle \rho\left[\vecU\cdot(\vecf-\nabla P)-\epsilon\right]\right\rangle
\end{equation}
For $a=\mathrm{const.}$ this simply expresses the equilibrium between energy injection and dissipation in the
statistically stationary state, for which the left-hand side vanishes. For a time-dependent scale factor,
on the other hand, the time derivative can be split by applying the product rule:
\begin{equation}
  \begin{split}
	\frac{\dd}{\dd t}  \left\langle\frac{1}{2}\rho U^2 + \rho K\right\rangle =&\,
		\frac{1}{a}\left\langle \rho\left[\vecU\cdot(\vecf-\nabla P)-\epsilon\right]\right\rangle \\
		&-H\langle \rho (U^2 + 2K)\rangle
  \end{split}
\end{equation}
From this form of the equation, it becomes immediately clear that the Hubble damping term on the very right
dominates over the forcing term once the expansion time scale becomes short compared to the effective dynamical time scale of the system, i.~e., if $\dot{a}=H a\gg 1$ or $a\gg 1$. The mean turbulence energy is then exponentially decaying.
Consequently, the expansion causes Hubble damping in combination with a freeze-out of turbulence. 
The scaling of turbulence prior to the freeze-out is imprinted as fixed ratios between the
exponentially decaying mean values of $\rho K$ at the different refinement levels.
The opposite effect for contracting systems with $H<0$ was investigated by \cite{RobGold12}.
The internal energy also decreases, but at a slower rate. The corresponding evolution of the RMS Mach numbers 
(see Fig.~\ref{fig:isoth_mach_evol}) shows that the compressibility of turbulence is reduced by expansion.

\begin{figure}
\centering
 \includegraphics[width=\linewidth]{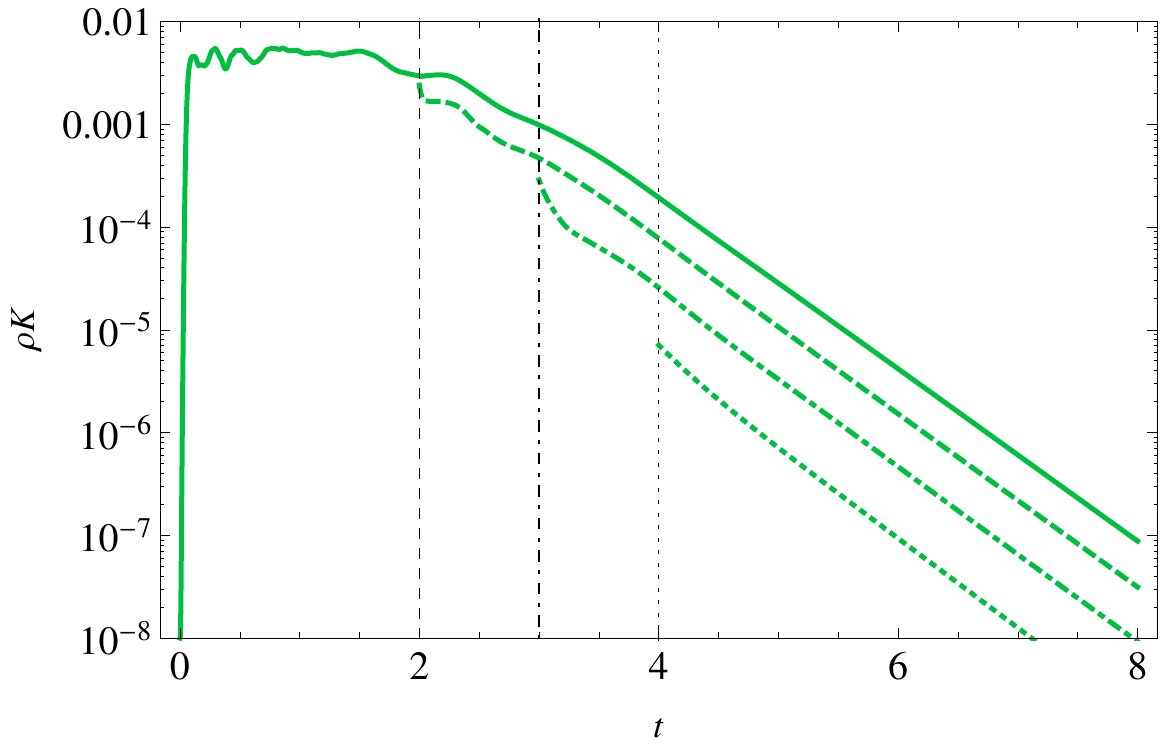}
\caption{Level-wise mean values of the SGS turbulence energy in the LES with expanding box ($H=1$).}
\label{fig:isoth_levels_exp_evol}
\end{figure}

To infer the scale-dependence of the SGS turbulence energy, mean values of $\rho K$ are plotted for each
refinement level in Figs~\ref{fig:isoth_levels_evol} and~\ref{fig:isoth_levels_exp_evol}. For
LES with a non-expanding box, one can clearly see that $\langle\rho K\rangle$ decreases at higher
refinement levels. This trend reflects the smaller fraction of unresolved
turbulence energy, as the grid scale decreases with the refinement level.
A comparison between the LES with the highest and lowest Mach numbers shows a steeper decrease
and stronger fluctuations of $\langle\rho K\rangle$ for higher Mach number. This
is a consequence of the stronger compression effects in supersonic flow. For the levels
$l=0$, $1$, and $2$, we compare the time-averaged mean values to LES on uniform grids with the 
corresponding grid scales $\Delta_0=1/128$,
$\Delta_1=1/256$, and $\Delta_2=1/512$. Although the refined grids fill only small fractions
of the total volume, the right plot in Fig.~\ref{fig:isoth_levels_evol} shows that the
spatio-temporal averages of $\rho K$ agree very closely. This is an important consistency
check for the application of the SGS model to non-uniform grids. In the case of
an expanding box, the scaling properties of $\rho K$ are 
preserved even if Hubble damping dominates (see Fig.~\ref{fig:isoth_levels_exp_evol}). This
is a consequence of the freeze-out of the turbulent cascade. Although non-linear interactions
cease due to the rapid expansion, the eddy hierarchy remains imprinted (this can be interpreted
as a form of ``fossil turbulence''). 

\begin{figure*}
\centering
\includegraphics[width=\linewidth]{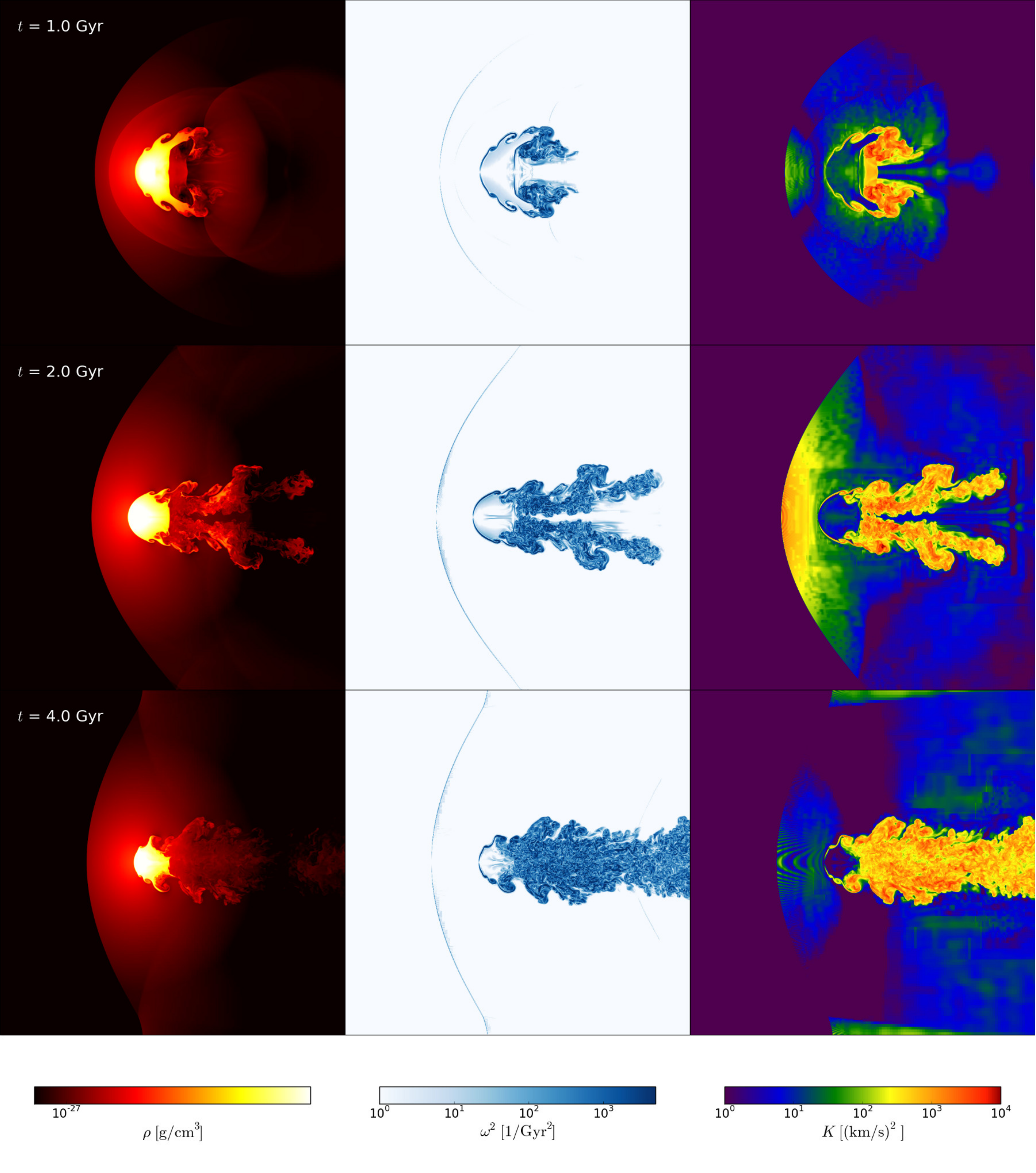}
\caption{Slices of the gas density $\rho$, the vorticity $\omega$, and the specific SGS turbulence energy $K$ 
	for the minor-merger model of \citet{IapiAda08} after $1$ (top), $2$ (middle), and $4\;\mathrm{Gyr}$ (bottom).}
\label{fig:msc_slices}
\end{figure*}

\subsection{A simple model for minor mergers}
\label{sc:minor_merger}

\citet{IapiAda08} consider a simple model for the infall of a low-mass sub cluster into the ICM of a big cluster.
The initial condition for the subcluster is a spherically symmetric isothermal gas cloud in hydrostatic equilibrium. 
The gas density profile of the cloud is defined by a so-called beta profile with core radius $r_{\rm core}=250\;\mathrm{kpc}$.
The subcluster is bound by a static gravitational field, corresponding to a dark matter halo with a King profile with a cutoff at $r=6r_{\rm core}$ (in this work; \citealt{IapiAda08} set $r=5r_{\rm c}$). To mimic the surrounding ICM in the frame of reference of the subcluster, the cloud is embedded into a uniform wind with an inflow boundary condition $U=U_{\rm in}$
imposed at one face of the computational domain, while the boundary conditions at the other faces are outflowing.
The domain size is $4\;\mathrm{Mpc}$, corresponding to $16$ core radii.
The wind velocity $U_{\rm in}$ and the chosen gas densities and temperatures are typical for minor mergers
\citep[for details, see][]{IapiAda08}. Based on this model, we performed AMR simulations with the shear-improved
SGS model, constrained turbulent energy compensation (see Appendix~\ref{sc:ctec}) and the refinement method explained in Appendix~\ref{sc:refinement}. Instead of refinement by overdensity, however,
refinement by the rate of compression is applied, with a statistical threshold analogous to equation~(\ref{eq:omega_thresh}).
The rate of compression is the substantial derivative of the divergence $\vecnab\cdot\vecU$ and traces, for example, shocks 
\citep[see][]{IapiAda08,Schmidt}. With a root-grid resolution of $128^3$ cells and three levels of refinement, 
we achieve a maximal resolution of $4\;\mathrm{kpc}$ in turbulent regions and in the vicinity of shocks.

\begin{figure}
\centering
 \includegraphics[width=\linewidth]{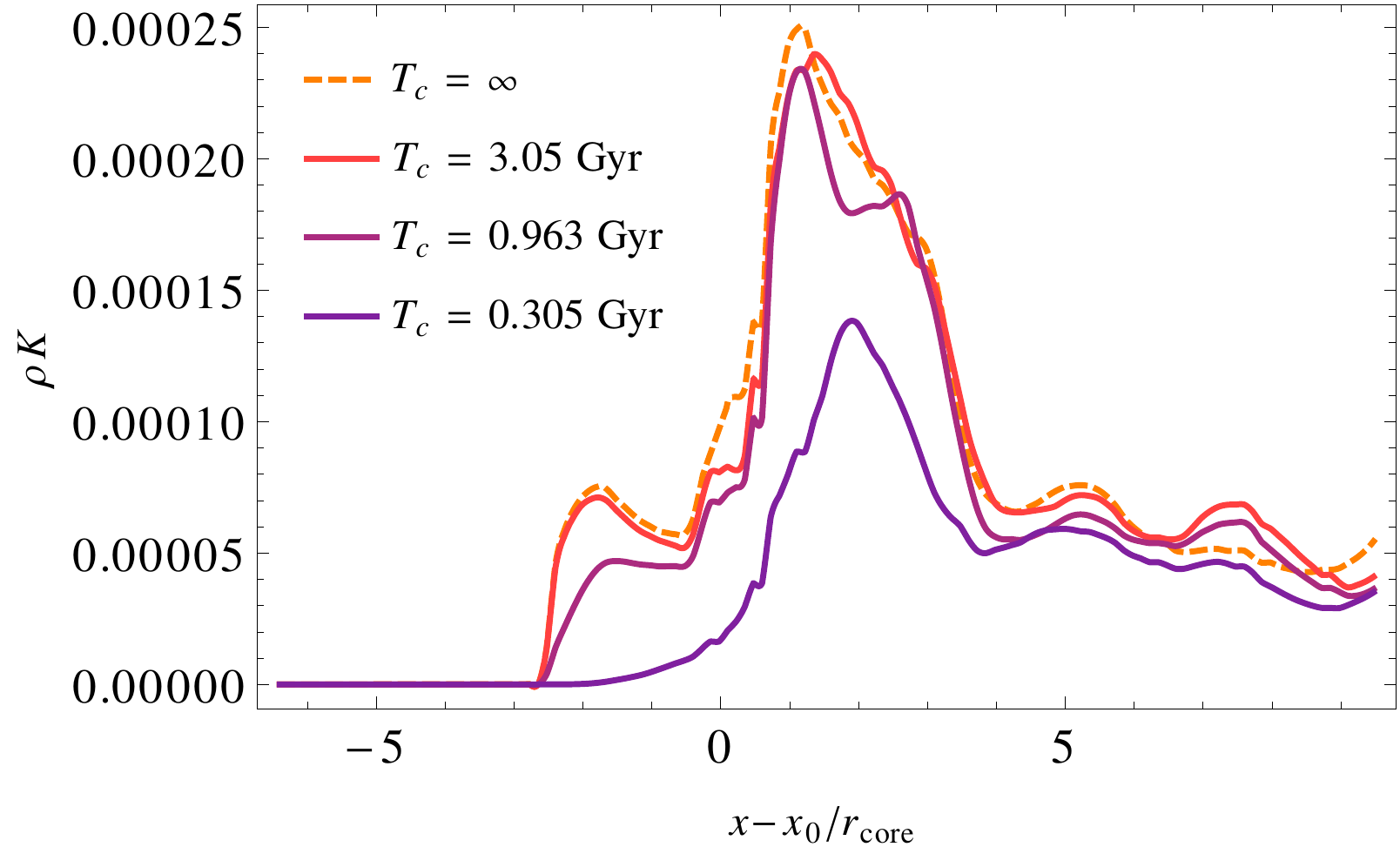}
\caption{SGS turbulence energy averaged over transversal slices at normal distance $x-x_0$ from the initial center of 
	mass of the subcluster in a wind. The distance is normalized to the core radius $r_{\rm core}$ of the
	subcluster.}
\label{fig:msc_profiles_energy_sgs}
\end{figure}

\begin{figure}
\centering
 \includegraphics[width=\linewidth]{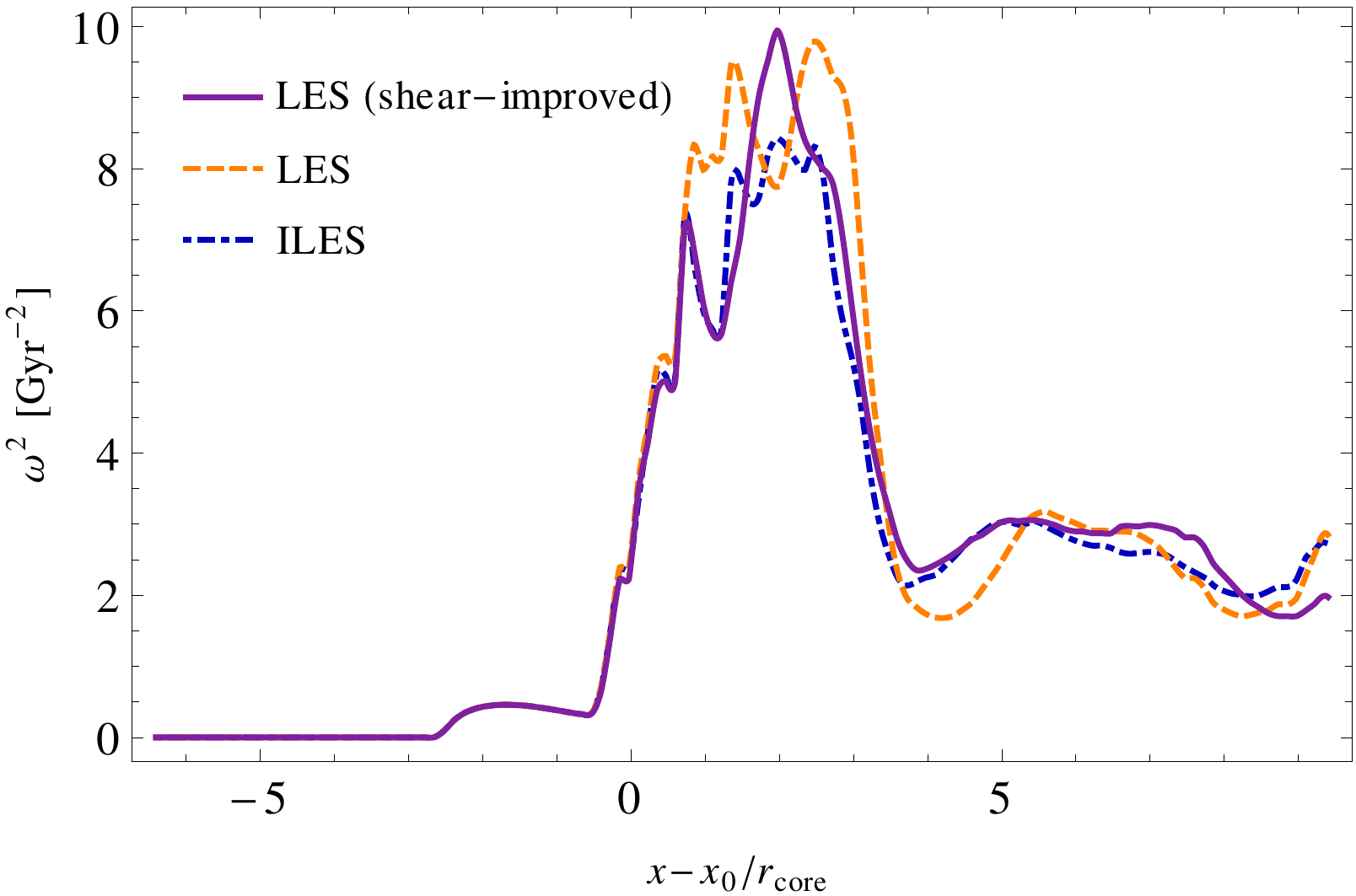}
\caption{Averaged squared vorticity plotted as in Fig.~\ref{fig:msc_profiles_energy_sgs}.}
\label{fig:msc_profiles_vorticity}
\end{figure}

As illustrated by the sequence of snapshots in Fig~\ref{fig:msc_slices}, the shear exerted onto the subcluster strips off material from the bound cloud (left column). The ensuing Kelvin-Helmholtz instability sheds vortices and produces a turbulent wake in the downstream direction of the wind (middle column). In front of the subcluster, a bow shock forms. Owing to the highly inhomogeneous flow structure, this is a particularly challenging test problem for SGS models. With the standard model for homogeneous flow, the rate of strain associated with the bow shock produces a large amount of spurious SGS turbulence energy that is advected by the wind. This can be seen by averaging $\rho K$ over slices perpendicular to the wind direction.
In Fig.~\ref{fig:msc_profiles_energy_sgs}, averaged values of $\rho K$ after about $3\;\mathrm{Gyr}$ of evolution
are plotted as function of the normalized coordinate $\tilde{x}=(x-x_{0})/r_{\rm core}$, where $x_{0}$ is the position of the center of mass of the cloud at time $t=0$. The dashed line shows the result for a run with the standard SGS model for homogeneous turbulence. While $\rho K$ is zero in the wind entering through the left face of the domain, the bow shock induces a steep increase around $\tilde{x}\approx -2.5$. Since the shock does not develop hydrodynamical instabilities (see Fig~\ref{fig:msc_slices}), the production of $\rho K$ at the shock front is an artifact of the simple eddy-viscosity closure, where the turbulence stress tensor is computed from the rate-of-strain tensor defined by equation~(\ref{eq:strain}). 
By considering the vorticity profile shown in Fig.~\ref{fig:msc_profiles_vorticity}, one can see that 
the large peak of $\rho K$ between $\tilde{x}\approx 0$ and $4$ is associated with the vortex shedding from the cloud, followed by the turbulent wake. Although the SGS turbulence energy contributes less than one percent to the kinetic energy of the flow
(and an even smaller fraction of the total energy), the non-linear evolution results in noticeable differences
of the vorticity compared to an an ILES without explicit SGS model.

\begin{figure}
\centering
 \includegraphics[width=\linewidth]{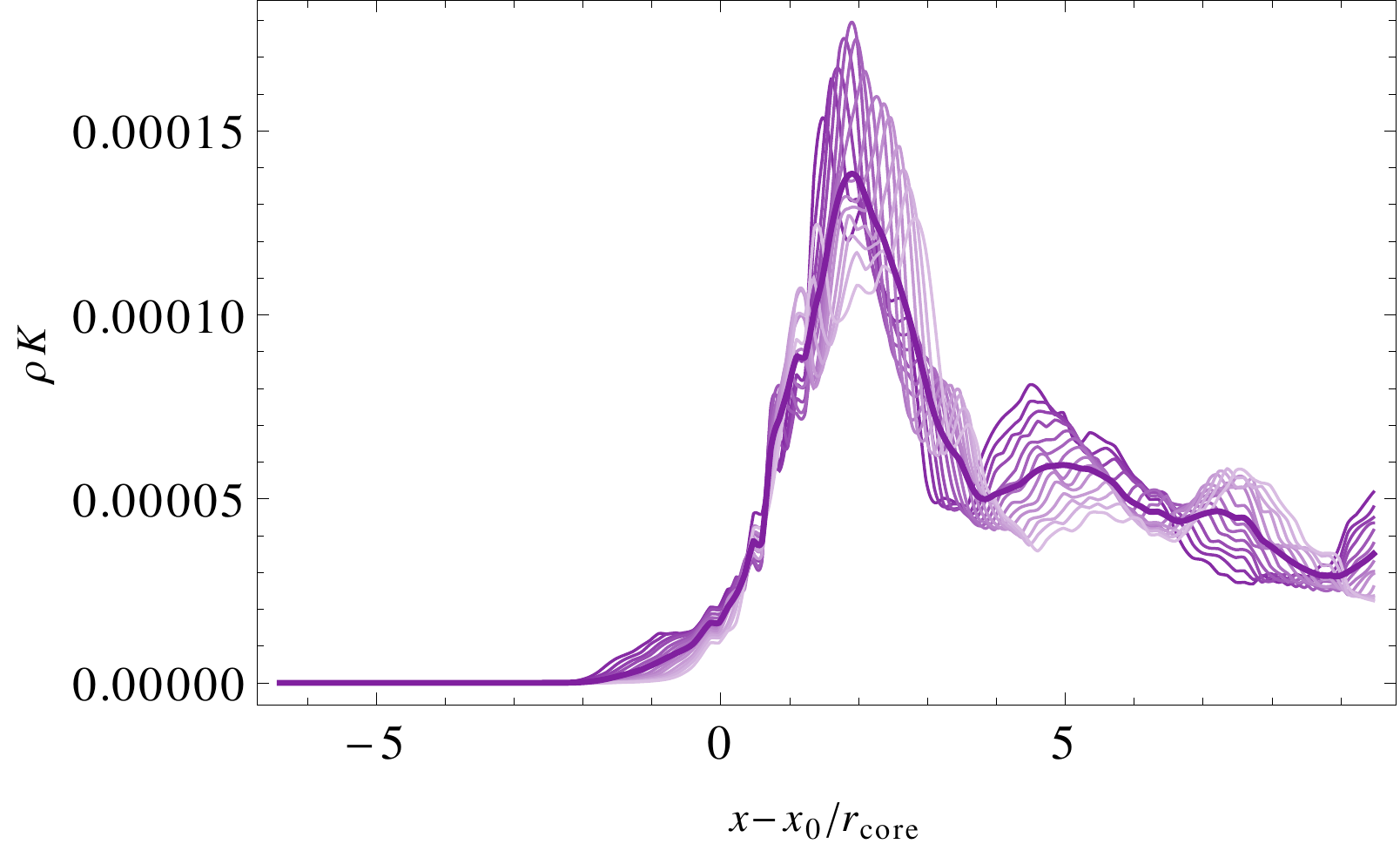}
\caption{Time variation of the SGS turbulence energy profiles (thin lines) for the snapshots contributing to the 
	time-averaged profile (thick line) in the case $T_{\rm c}\approx 0.3\;\mathrm{Gyr}$.}
\label{fig:msc_profiles_energy_sgs_evol}
\end{figure}

The profile of $\rho K$ changes markedly if the shear-improved model based on
equation~(\ref{eq:tau_si}) is applied with the Kalman filter parameters $U_{c} = U_{\rm in}$ and $T_{\rm c}=2r_{\rm core}/U_{\rm in}$, where $U_{\rm in}$ is assumed to be the characteristic turbulent velocity and $2r_{\rm core}=500\;\mathrm{kpc}$ the associated integral length scale (see Appendix~\ref{sc:kalman}). The resulting smoothing time is $T_{\rm c}\approx 0.3\;\mathrm{Gyr}$. As can be seen in Fig.~\ref{fig:msc_profiles_energy_sgs}, the spurious jump of $\rho K$ at the bow shock is largely removed in this case. Furthermore, the pronounced inhomogeneity of the flow around the cloud results in a reduction of the peak, while the average SGS turbulence energy changes only little in the turbulent wake. For better comparability, all profiles shown in Fig.~\ref{fig:msc_profiles_energy_sgs} were averaged over the time interval between 
$t\approx 3\;\mathrm{Gyr}$ and $t+T_{\rm c}\approx 3.3\;\mathrm{Gyr}$. The modulation of the instantaneous profiles in this interval shows the imprint of the vortex-street-like structure of the turbulent wake on large scales 
(see Fig.~\ref{fig:msc_profiles_energy_sgs_evol}). Even in this case, there is some spurious production caused by the 
bow shock, although it diminishes as the shock becomes stationary (see also the lateral slices 
of $\rho K$ in Fig.~\ref{fig:msc_slices}).
Since the integral length scale is more or less fixed, the relation $T_{\rm c}=2r_{\rm core}/U_{\rm c}$
leaves only one degree of freedom for the Kalman filter. Figure~\ref{fig:msc_profiles_energy_sgs} shows 
the effect of a longer smoothing
time scale by setting $U_{c}$ to some fraction of $U_{\rm in}$ (it cannot reasonably be assumed that $U_{\rm c}>U_{\rm in}$). 
One can clearly see the transition to the standard SGS model ($T_{\rm c}=\infty$) as $T_{\rm c}$ increases.

The results from the minor-merger simulations illustrate both the utility and the limitations of the Kalman filter. Although the filter is adaptive, it does not perfectly separate the bow shock form the random component while the shock front is still changing its shape. Nevertheless, the shear-improved SGS model greatly improves the prediction of the SGS turbulence energy compared to the standard model. One also has to bear in mind that the perfect symmetry of the bow shock is rather artificial. As we shall see, shocks in clusters are subject to hydrodynamical instabilities and are sites of actual turbulence production.

\subsection{The Santa Barbara cluster}
\label{sc:sb_calibr}

As a test case for cosmological simulations, we perform adaptively refined LES of the Santa Barbara cluster with the constrained turbulent energy compensation (CTEC) method. As motivated in Section~\ref{sc:consrv} and detailed in Appendix~\ref{sc:ctec}, this method is applicable if the flow has a significant non-turbulent component. For the same reason, the shear-improved SGS model is applied. The initial conditions are specified in \citet{HeitRick05}. Dark matter and adiabatic gas dynamics are evolved in a box of $64\,\mathrm{Mpc}$ comoving size, as detailed in \citet{AlmBell12}. As we will show below, 
the eddy-viscosity closure defined by equation~(\ref{eq:tau_si}) is applicable to this problem because
the turbulent velocity fluctuations are mostly subsonic. To track down not only dense gas, 
but also strongly turbulent regions in clusters and filaments, refinement is based on overdensity and
vorticity modulus in our simulations (see Appendix~\ref{sc:refinement}). 
In this case, the filtered velocity component $[\Ub]$ calculated with the Kalman filter can be interpreted
as the velocity of the bulk flow driven by gas accretion into the gravitational wells of the dark-matter halos, i.~e., 
$[\Ub]\simeq\Ub_{\rm bulk}$. 
As we shall see in the following, the choice of the filter parameters $U_{\rm c}$ and $T_{\rm c}$ can be narrowed down
to physically reasonable values by statistical results form the test runs.

\begin{figure*}
\centering
 \includegraphics[width=\linewidth]{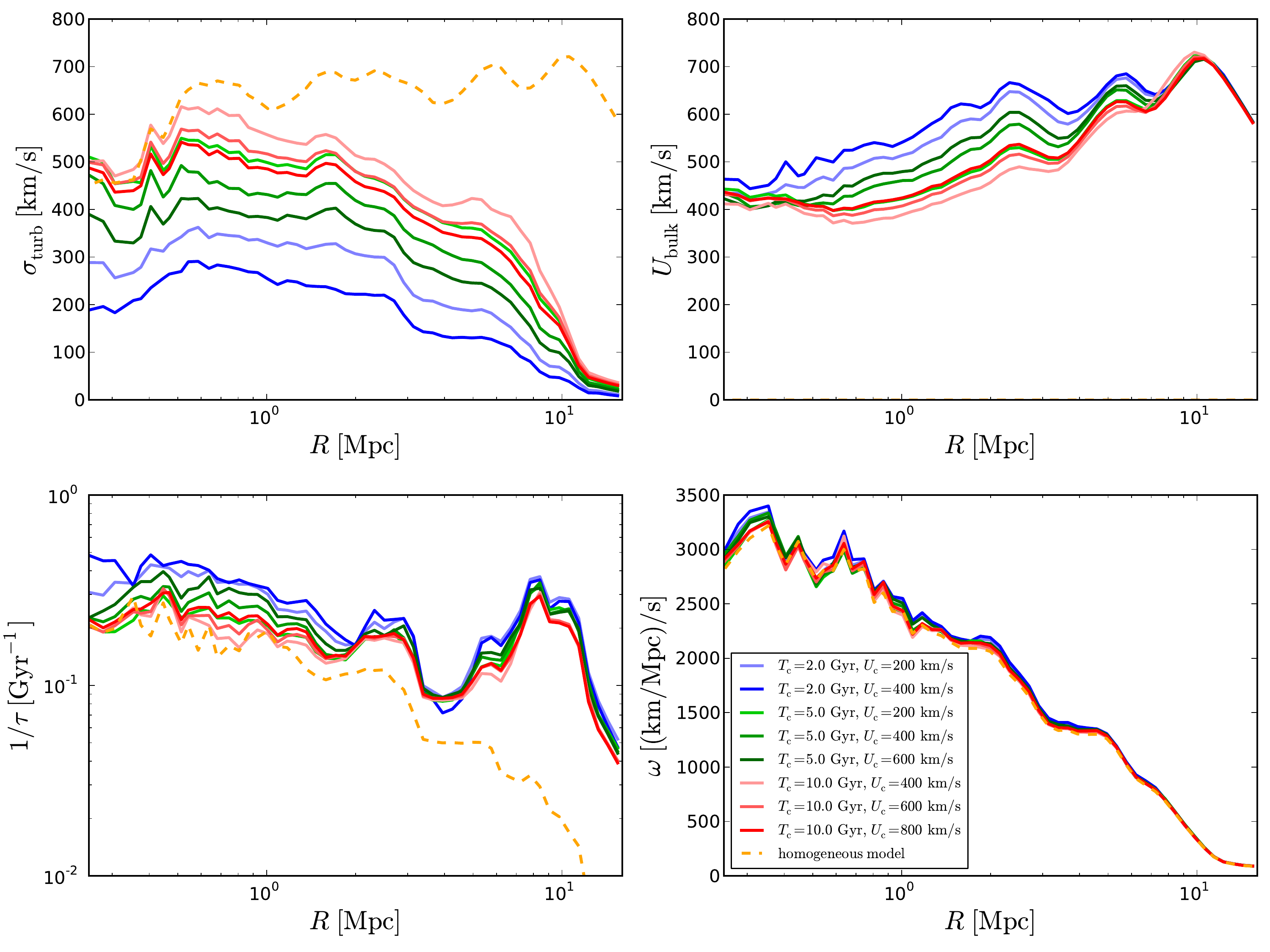}
\caption{Radial profiles of the turbulent velocity dispersion $\sigma_{\rm turb}$, the bulk velocity $U_{\rm bulk}$, 
	the inverse turbulence production time scale $1/\tau$ (see equation~\ref{eq:tau_dyn}), and the vorticity modulus 
	$\omega$ in adaptively refined LES of the Santa Barbara cluster.
	Solid lines correspond to the shear-improved SGS model
	with different choices of the Kalman filter parameters $U_{\rm c}$ and $T_{\rm c}$ and the dashed line
	to the standard SGS model for homogeneous turbulence.}
\label{fig:profiles_sgs}
\end{figure*}

Figure~\ref{fig:profiles_sgs} shows statistics for a suite of test runs, 
where $T_{\rm c}$ was varied between $2$ to $10$ Gyr and $U_{\rm c}$ between $200$ to $800\;\mathrm{km/s}$.
For each test run, a root-grid resolution of $128^3$ and two levels of refinement by a factor $2$ were used.\footnote{
	Owing to the development history of \nyx, a simpler hydrodynamical scheme with piece-wise linear reconstruction
	was used for these test simulations. For all other simulations, the more advanced scheme with piecewise parabolic 
	reconstruction was applied. The differences between the two hydro solvers does not have a significant impact
	on the calibration of the Kalman filter parameter.
} One of the key quantities that follows from the Kalman filter in conjunction with the SGS 
model is the local turbulent velocity dispersion, encompassing both resolved and unresolved velocity
fluctuations:
\begin{equation}
  \label{eq:sigma_turb}
   \sigma_{\rm turb}^2 = U^{\prime\,2} + 2K.
\end{equation}
Although $\sigma_{\rm turb}$ is defined as a local quantity, we will mainly consider statistics of $\sigma_{\rm turb}$,
which can be properly interpreted as velocity dispersions.
Radial profiles of $\sigma_{\rm turb}$ and $U_{\rm bulk}$ centered around the density maximum at redshift $z=0$ are plotted in the
top panels of Fig.~\ref{fig:profiles_sgs}. One can see that the turbulent velocity dispersion is maximal at radii of a few
Mpc, corresponding to the central region of the cluster, and declines outwards. The bulk velocity, on the other hand, is larger outside, but also contributes to the flow in the cluster interior.

For a shorter smoothing time scale $T_{\rm c}$, the instantaneous velocity has a larger weight in the iterative update of the mean flow, which results in a reduction of the fluctuating component. This suppresses the production of $K$ (equation~\ref{eq:tau_si}). 
The maximal SGS turbulence energy is obtained with the standard model, which corresponds to the limiting case of a vanishing mean flow 
and $\Ub'=\Ub$. This is consistent only for isotropic homogenous turbulence. For comparison, the profiles obtained with the standard model are plotted as dashed lines in Fig.~\ref{fig:profiles_sgs}. In this case, we formally set $\sigma_{\rm turb}^2 = U^2 + 2K$ and $U_{\rm bulk}=0$, although one must keep in mind that $\sigma_{\rm turb}$ does \emph{not} correspond to the turbulent velocity dispersion, but the total velocity dispersion. For non-stationary turbulence, $T_{\rm c}$ should
correspond to the time scale over which turbulence develops. As an estimate of this time scale, we use the
turbulence production time scale,
\begin{equation}
  \label{eq:tau_dyn}
  \tau = \frac{\rho\sigma_{\rm turb}^2}{2\Sigma},
\end{equation}
where $\Sigma$ is the turbulence energy flux across the grid scale (equation~\ref{eq:prod}).
In the inertial subrange, both the flux $\Sigma$ and the energy $\rho\sigma_{\rm turb}^2$ are, in principle, scale-invariant. 
Consequently, $\tau$ is at least approximately independent of the grid scale. Since $\Sigma$ can become zero or negative, while
$\sigma_{\rm turb}$ is positive, we computed profiles of $\tau^{-1}$ (bottom left plot in Fig.~\ref{fig:profiles_sgs}).
For the central part of the cluster, a quite robust estimate is $\tau\approx 5\;\mathrm{Gyr}$.
This value is well matched by the profiles for $T_{\rm c}=5\;\mathrm{Gyr}$, while the choices $T_{\rm c}=2$ and $10\;\mathrm{Gyr}$ 
do not agree with the typical values of $\tau$. The pronounced peak in the profiles of $1/\tau$ at $R\approx 10\;\mathrm{Mpc}$
corresponds to a steep outward decrease of $\sigma_{\rm turb}$. This peak, which is not seen for the homogeneous model,
is a signature of turbulence production by the accretion shocks.

The effect of $U_{\rm c}$ is inverse to $T_{\rm c}$, i.~e., a higher velocity scale $U_{\rm c}$ results in a \emph{lower} turbulent velocity dispersion
(see top left plot in Fig.~\ref{fig:profiles_sgs}).\footnote{For large $U_{\rm c}$, the variance $\sigma^{2\,(n)}_{\delta[U_i]}$ is
	large and the Kalman gain tends toward unity (equation~\ref{eq:kalman_gain}). In this case, the mean flow closely follows the
	local velocity and, consequently, the fluctuation is small.}
Since the typical value of $\sigma_{\rm turb}$ should be close to $U_{\rm c}$ in 
the statistically stationary regime, which we can reasonably assume for the cluster core at redshift zero, it is easy 
to single out a consistent choice of $U_{\rm c}$. We only obtain a turbulent velocity dispersion around $400\;\mathrm{km/s}$
for $U_{\rm c}=400\;\mathrm{km/s}$ and $T_{\rm c}=5\;\mathrm{Gyr}$. For the other tested parameters, we find that
$\sigma_{\rm turb}$ is too high or too low in comparison to $U_{\rm c}$. 

The profile of the vorticity modulus $\omega$ shows only little variation among the different runs (bottom right plot in Fig.~\ref{fig:profiles_sgs}). 
This is reasonable because the decomposition into fluctuating and bulk-flow components affect the
hydrodynamical equations only through the SGS stresses $\tau_{ij}$, which amount to 
small corrections to the Euler equations. The vorticity tends to be higher for lower values of $T_{\rm c}$ and larger $U_{\rm c}$, corresponding to the behavior of the SGS turbulence energy. If $\rho K$ is higher, the turbulent stresses become larger, which reduces the vorticity. This effect is strongest for the homogeneous model and weakest for the parameter pair $T_{\rm c}=2\;\mathrm{Gyr}$ and $U_{\rm c}=400\;\mathrm{km/s}$.

\begin{figure*}
\centering
\includegraphics[width=\linewidth]{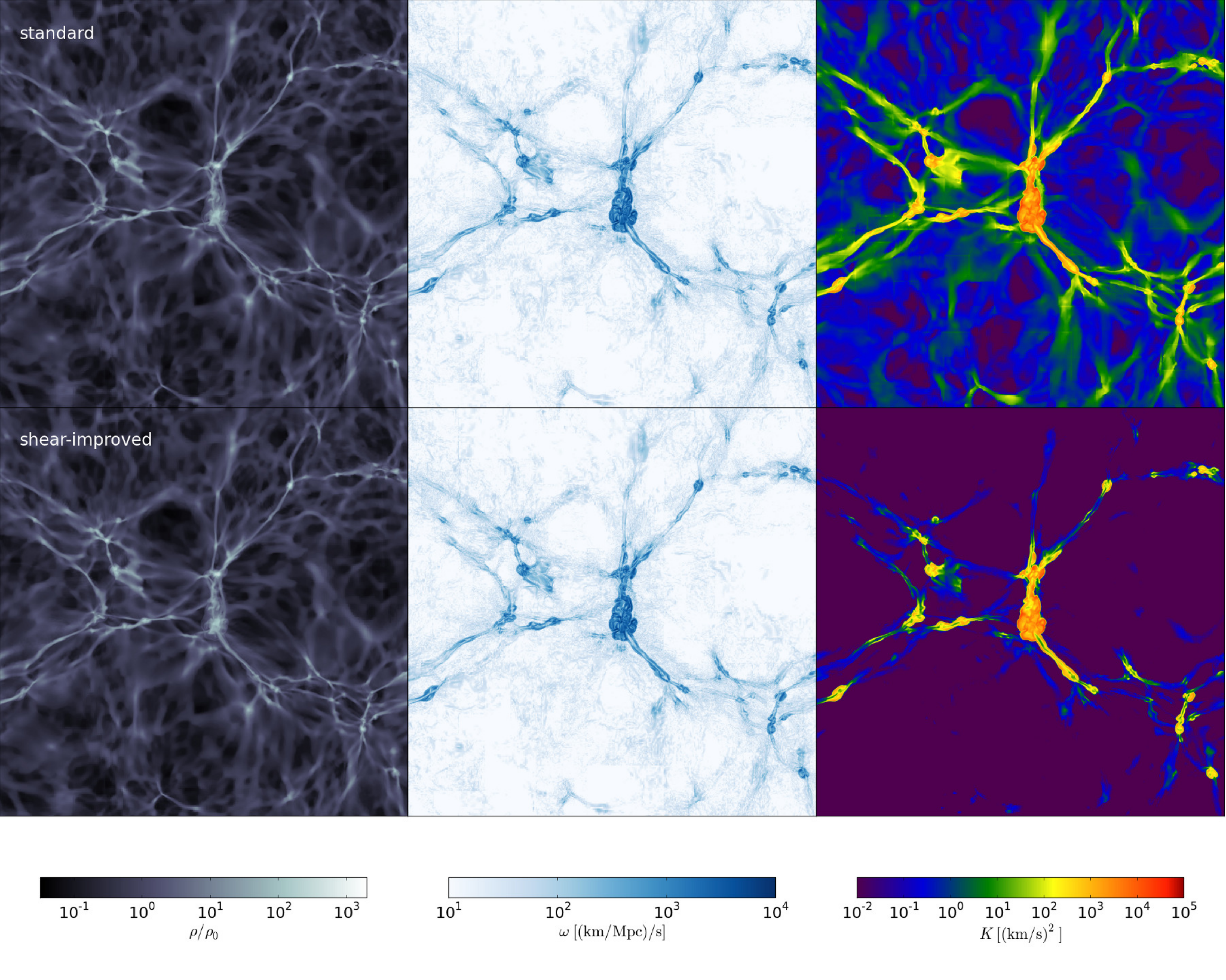}
\caption{Slices of the gas density $\rho$, the vorticity $\omega$, and the specific SGS turbulence energy $K$ 
	for the Santa Barbara cluster at redshift $z=2.0$ with the standard (top) and the shear-improved (bottom) SGS models.}
\label{fig:slices_z2}
\end{figure*}

\begin{figure*}
\centering
\includegraphics[width=\linewidth]{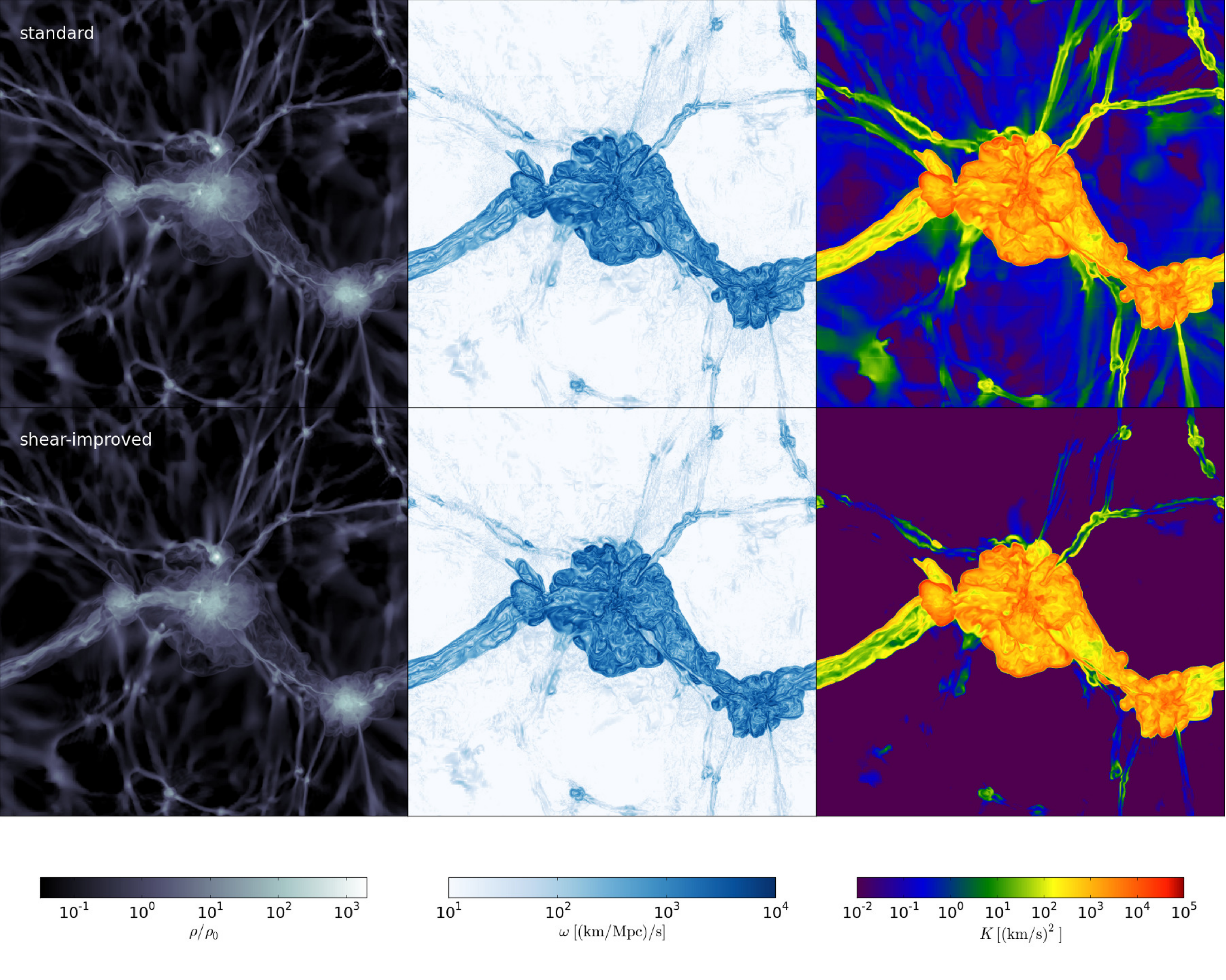}
\caption{Slices as in Fig.~\ref{fig:slices_z0} for the Santa Barbara cluster at redshift $z=0.0$.}
\label{fig:slices_z0}
\end{figure*}

\section{Properties of Turbulence in the Santa Barbara Cluster}
\label{sc:sb_sgs}

The structure of the Santa Barbara Cluster is illustrated by
slices of the baryonic gas density, vorticity modulus, and specific SGS turbulence energy from
simulations performed with a $256^3$ root grid and two levels of refinement in Figs~\ref{fig:slices_z2} 
($z=2$) and~\ref{fig:slices_z0} ($z=0$). 
The gas density is normalized by the mean total matter density $\rho_0$. 
In both cases, one can see the typical web-like structure, with clusters forming at the vertices of filaments.
The filaments have thin threads of strongly compressed gas, surrounded by low-density material. As indicated by the vorticity, 
turbulence fills the ICM and the thick filaments between large clusters. The outer accretion shocks appear as sharp borders of 
the turbulent gas.
In contrast to the standard SGS model, for which $K$ spreads out far into the IGM, $K$ is confined to the high-vorticity
regions if the shear-improved model is applied with the fiducial parameters $T_{\rm c}=5\;\mathrm{Gyr}$ and 
$U_{\rm c}=400\;\mathrm{km/s}$. In the following, we analyze and compare the properties of turbulence
for the two variants of the SGS model. 

\subsection{Numerically resolved flow}

\begin{figure*}
\centering
  \includegraphics[width=\linewidth]{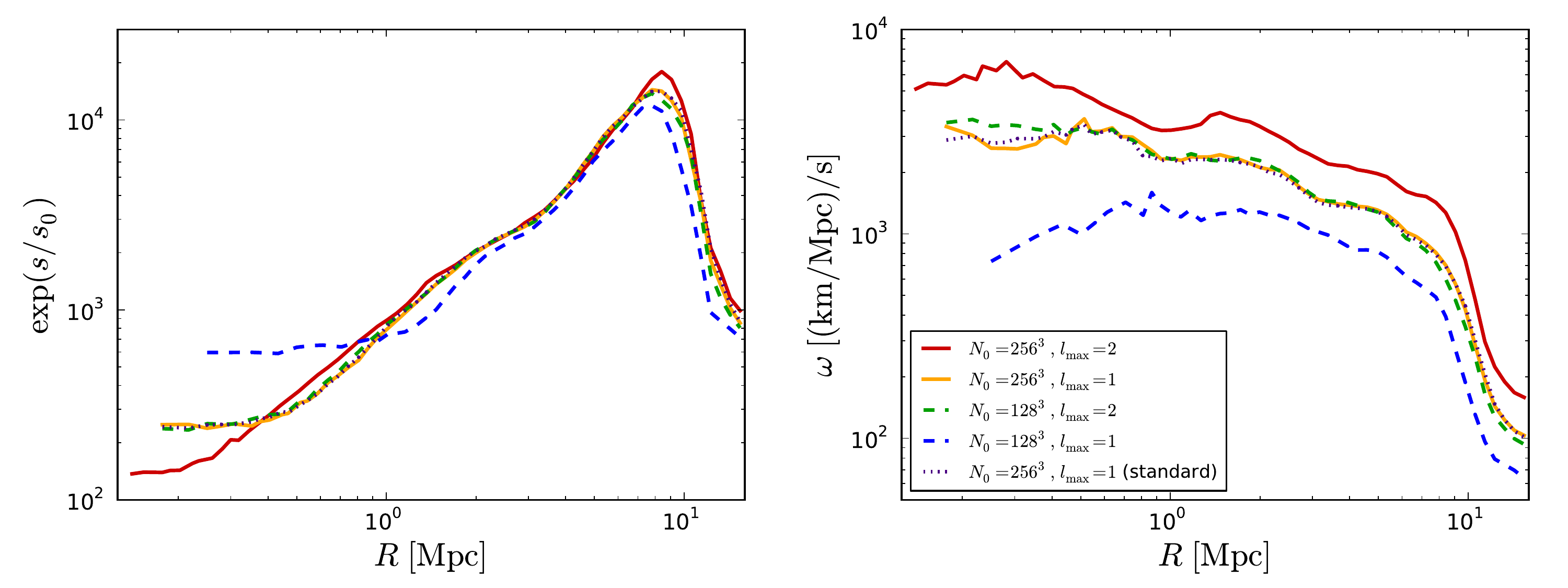}
\caption{Radial profiles of the entropy (left) and the vorticity (right) at $z=0$ for the shear-improved SGS 		model at different numerical resolutions. A comparison run with the standard SGS model for homogenous turbulence
	is shown as dotted line.}
\label{fig:SB_profiles_res_si}
\end{figure*}

Figure\ref{fig:SB_profiles_res_si} shows the dependence of the entropy and vorticity modulus
on numerical resolution. We follow the usual convention and specify the entropy of the adiabatic gas by
\begin{equation}
  \exp\left(\frac{s}{s_0}\right) = \left(\frac{T}{T_0}\right)\left(\frac{\rho_0}{\rho}\right)^{\gamma-1},
\end{equation}
where $\rho_0$ is the mean baryonic mass density, and $T_0=1000\;\mathrm{K}$ the initial gas temperature.
For comparison, also the profiles for an intermediate-resolution run with the standard SGS model are plotted.
Regardless of the applied SGS model and the numerical resolution, the radial entropy profiles agree very
well for radii greater than about $1\;\mathrm{Mpc}$. We only observe shallower profiles, which are artifacts of lower
resolution, in the cluster core.  
The vorticity, which is given by the velocity derivative, becomes systematically higher as the resolution increases, but the
profiles have similar shapes with a very steep decline around $R\approx 10\;\mathrm{Mpc}$. 
As can be seen in Fig.~\ref{fig:slices_z0}, this feature is due to the confinement of vorticity to the regions 
interior to the accretion shocks. Another important result is that the profiles for the runs performed with a $256^3$ root grid and one refinement level
almost perfectly reproduce the profiles from the test run with a $128^3$ 
root grid and two levels of refinement. Both runs have the same effective resolution.
This impressively demonstrates the reliability of our refinement method.
Compared to the standard SGS model, there are small differences between the vorticity profiles 
in the cluster core. The entropy profile, however, is not significantly affected by the SGS model.
This makes sense because the mass density is dominated by gravity and SGS effects induce only
minute changes in the thermal energy.
 
\begin{figure}
\centering
 \includegraphics[width=\linewidth]{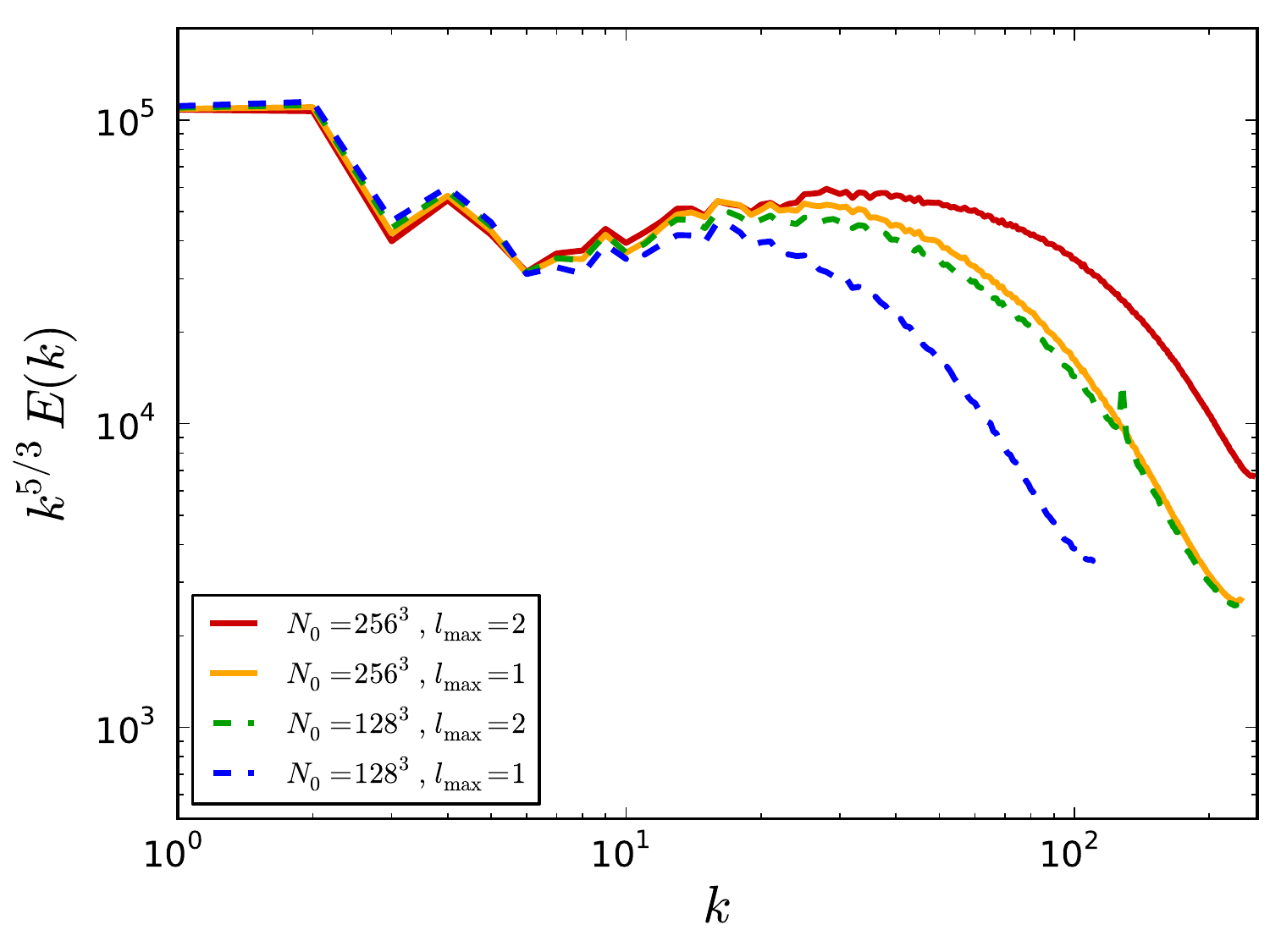}\quad
\caption{Compensated energy spectrum functions for the same simulations as in Fig.~\ref{fig:SB_profiles_res_si}
	(only shear-improved SGS model).}
\label{fig:sb_spect_si}
\end{figure}

\begin{figure}
\centering
 \includegraphics[width=\linewidth]{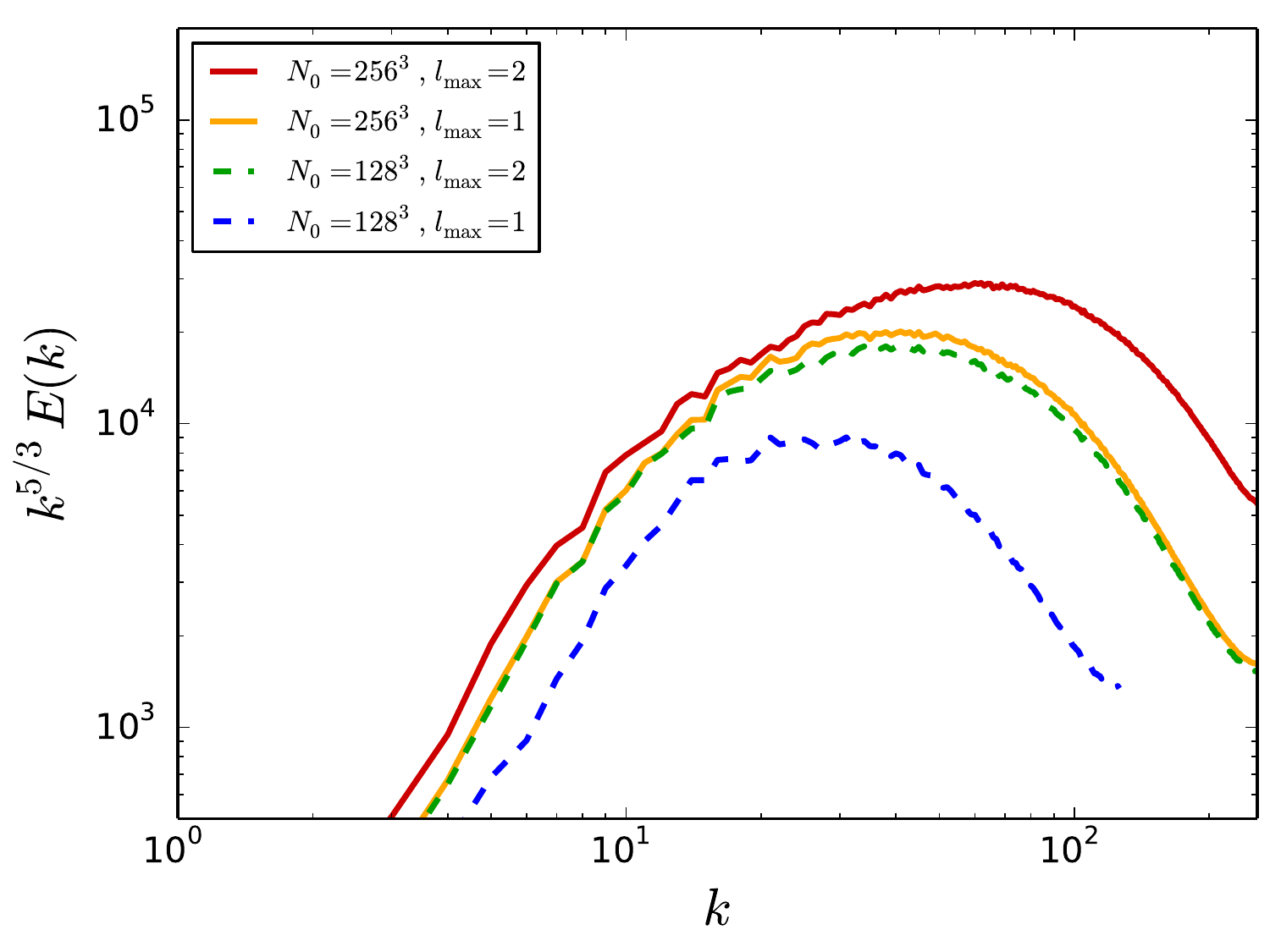}
\caption{Compensated energy spectrum functions for the fluctuating velocity component.}
\label{fig:sb_spect_si_fluc}
\end{figure}

An important question regards the scaling of turbulence. The energy spectrum function $E(k)$, for which the Kolmogorov theory
of incompressible isotropic turbulence predicts $E(k)\propto k^{-5/3}$, is obtained by
integration of $\frac{1}{2}|\hat{\Ub}(\veck)|^2$, where $\hat{\Ub}(\veck)$ is the Fourier spectrum of the velocity $\Ub$,
over spherical shells of radius $k$ in Fourier space. For a comparison with the Kolmogorov spectrum, it is favorable
to plot compensated spectrum functions, $k^{5/3}E(k)$. We normalize the wave numbers such that $k=1$ corresponds to the
domain size of $64\;$Mpc.
The results are shown in Fig.~\ref{fig:sb_spect_si}. 
For the calculation of the Fourier transforms, the AMR data were interpolated to uniform $512^3$ grids, with
the exception of the lowest resolution case, for which a $256^3$ grid was used. For the run with the highest resolution, 
only the data up to the first refinement level were taken.\footnote{This turned out to be necessary because of memory 
restrictions for postprocessing.} In the maximally refined regions, however, these data are averaged down from the
second level. 
Since the bulk flow contributes to the total velocity spectrum, Fig.~\ref{fig:sb_spect_si_fluc} shows
spectra of the fluctuating component $\Ub^{\prime}$ of the velocity. 
As expected, the subtraction of the bulk flow removes a large fraction of the power at small wave numbers, 
but some power is unavoidably removed at higher wave numbers because of the adaptive nature of the filter.
 
An upper limit for the scale of turbulence energy injection by gravity is the size of the region 
inside the accretion shocks, which is about $20\;$Mpc in diameter, corresponding to a wavenumber around $3$. 
The slope in the intermediate range of wavenumbers, however, is not easy to pinpoint. There are several effects that
influence the compensated spectra in Fig.~\ref{fig:sb_spect_si}. At least a decade of wavenumbers below the numerical
cutoff wavenumber (512 at the highest refinement level) are significantly affected by numerical dissipation. 
The resulting bottleneck effect, which appears as a bump-like feature in the 
spectrum (see Fig.~\ref{fig:sb_spect_si}), is observed in many simulations of forced isotropic turbulence
\citep[e.~g.,][]{SchmHille06,KritNor07,AspNiki08}.
If the dynamical range between the driving scale and the cutoff scale is too narrow, 
the bottleneck effect distortes the inertial range and obscurs Kolmogorov scaling.  
Moreover, gravity inside and mergers could very well drive 
turbulence on length scales smaller than the radius of the outer accretion shock. 
This would result in an overlap between the range
of driving scales and the bottleneck. This is indeed suggested by the maxima of the compensated spectra of the 
fluctuating component of the velocity field at wave numbers above 10 (see Fig.~\ref{fig:sb_spect_si_fluc}), which imply
driving scales of turbulence roughly between $1$ and $3\;$Mpc.
Going to significantly higher resolution with our aggressive refinement method turned out to be infeasible. 
Consequently, we cannot reliably infer the scaling
of turbulence in our simulations. However, this might be achieved by combining dynamical refinement on turbulence
with the zoom-in technique.

\subsection{Turbulent velocity dispersion, production, and dissipation}

\begin{figure}
\centering
 \includegraphics[width=\linewidth]{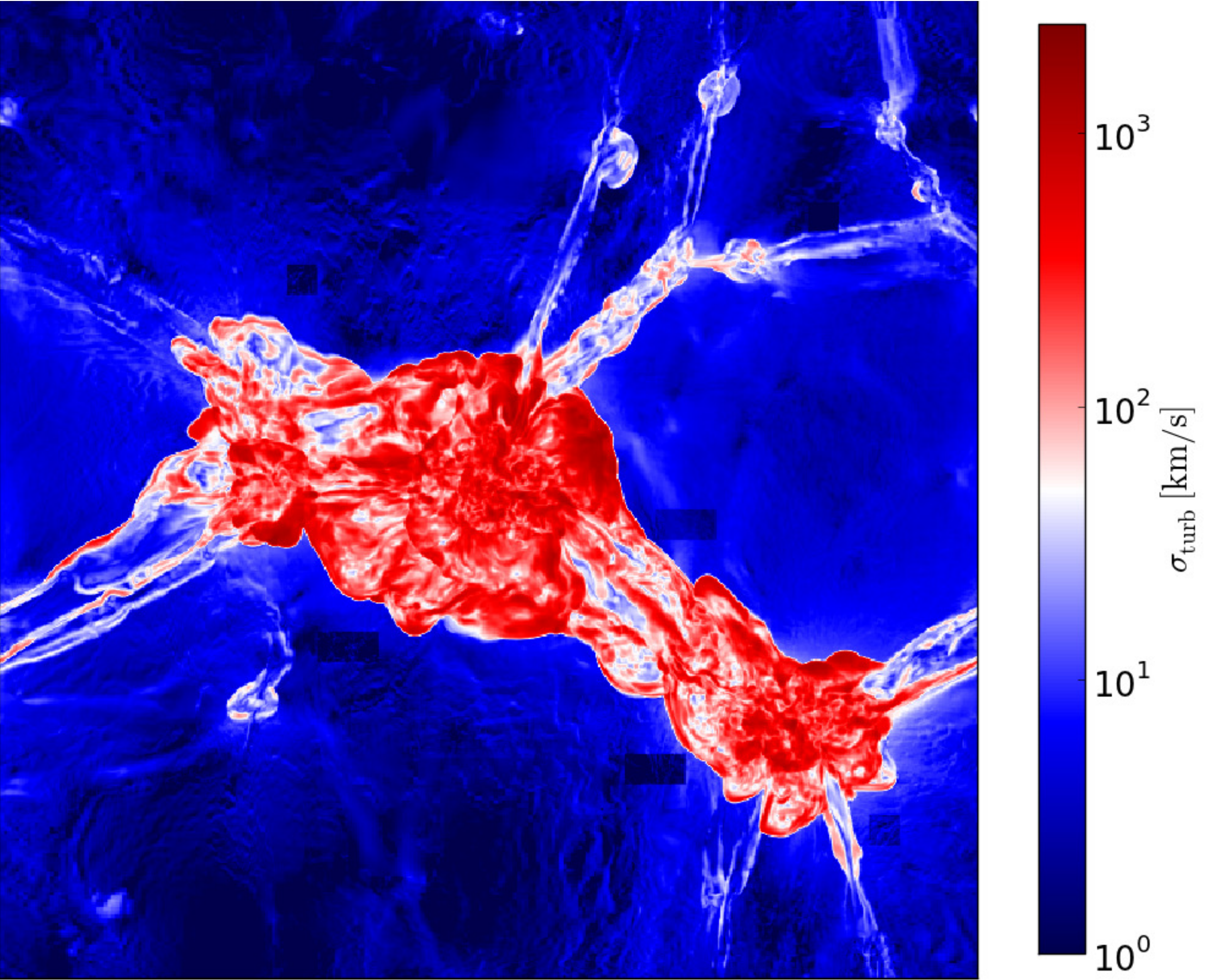}
\caption{Slice of the turbulent velocity dispersion at redshift $z=0.0$.}
\label{fig:sb_slice_sigma_turb}
\end{figure}

\begin{figure}
\centering
 \includegraphics[width=\linewidth]{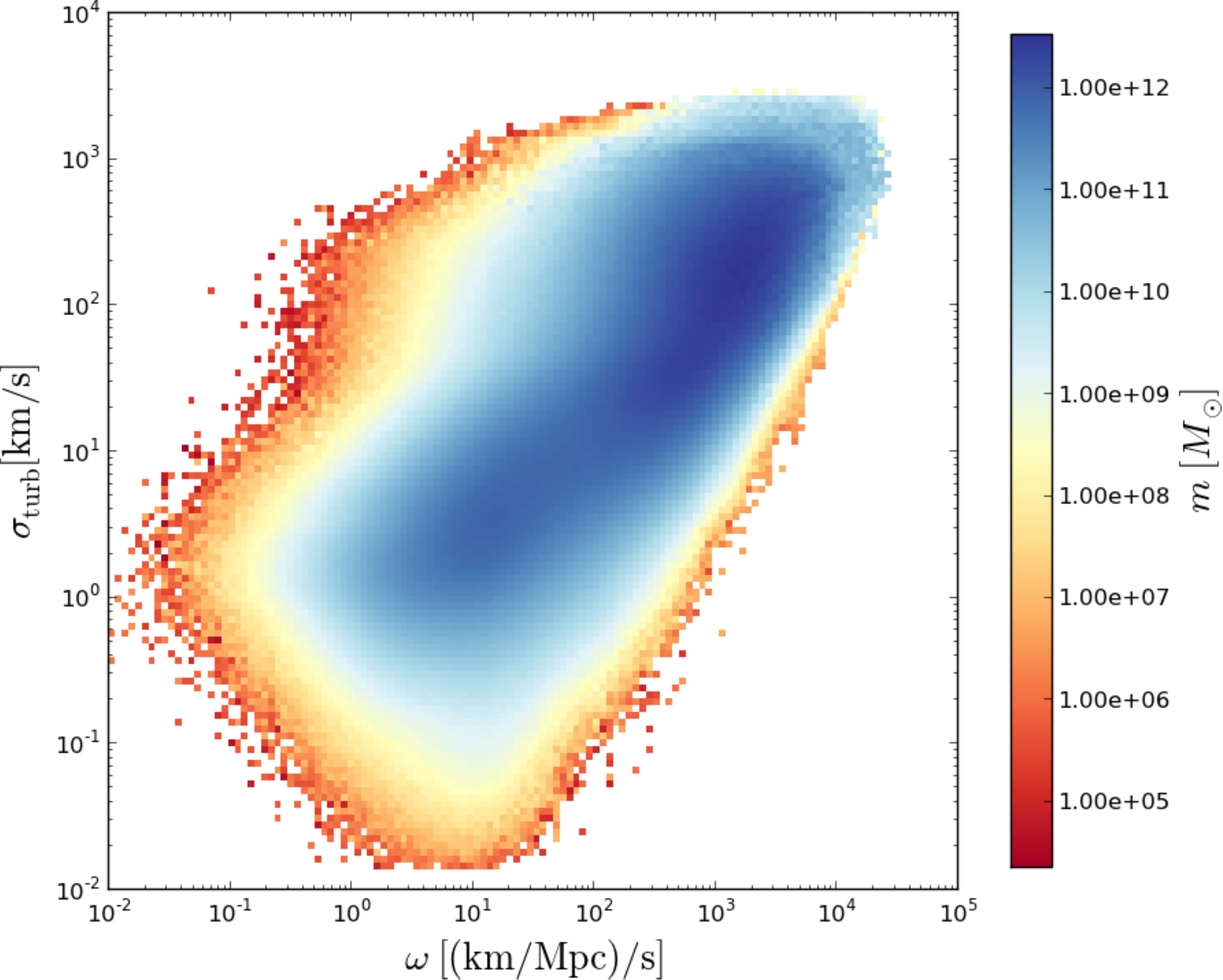}
\caption{Phase plot of the turbulent velocity dispersion $\sigma_{\rm turb}$ vs.\ the 
	vorticity modulus $\omega$ ($z=0.0$).}
\label{fig:sb_phase_vort}
\end{figure}

A central prediction of the shear-improved SGS model is the local turbulent velocity dispersion $\sigma_{\rm turb}$ defined by equation~(\ref{eq:sigma_turb}).
For the run with maximal resolution, a slice of $\sigma_{\rm turb}$ is shown in Fig.~\ref{fig:sb_slice_sigma_turb}. 
One can clearly discern the boundaries of strongly turbulent regions, which coincide with the
outer accretion shocks. The phase plot in Fig.~\ref{fig:sb_phase_vort} shows that large turbulent velocity dispersion is indeed associated
with strong vorticity. For the thick filaments, $10\;\mathrm{km/s}\lesssim\sigma_{\rm turb}\lesssim 100\;\mathrm{km/s}$ 
(in Fig.~\ref{fig:sb_slice_sigma_turb}, velocities around $50\;\mathrm{km/s}$ appear in white). In the voids, there is a 
residual turbulent velocity dispersion of a  few km/s, which is probably an artifact of the Kalman filter. In newly refined regions,
which appear as small quadratic or rectangular boxes with lower $\sigma_{\rm turb}$, the velocity fluctuations need a certain time to build up (see Apendix~\ref{sc:kalman}). But in the cluster gas, $\sigma_{\rm turb}$ is a valuable probe of turbulence.

\begin{figure*}
\centering
 \includegraphics[width=0.48\linewidth]{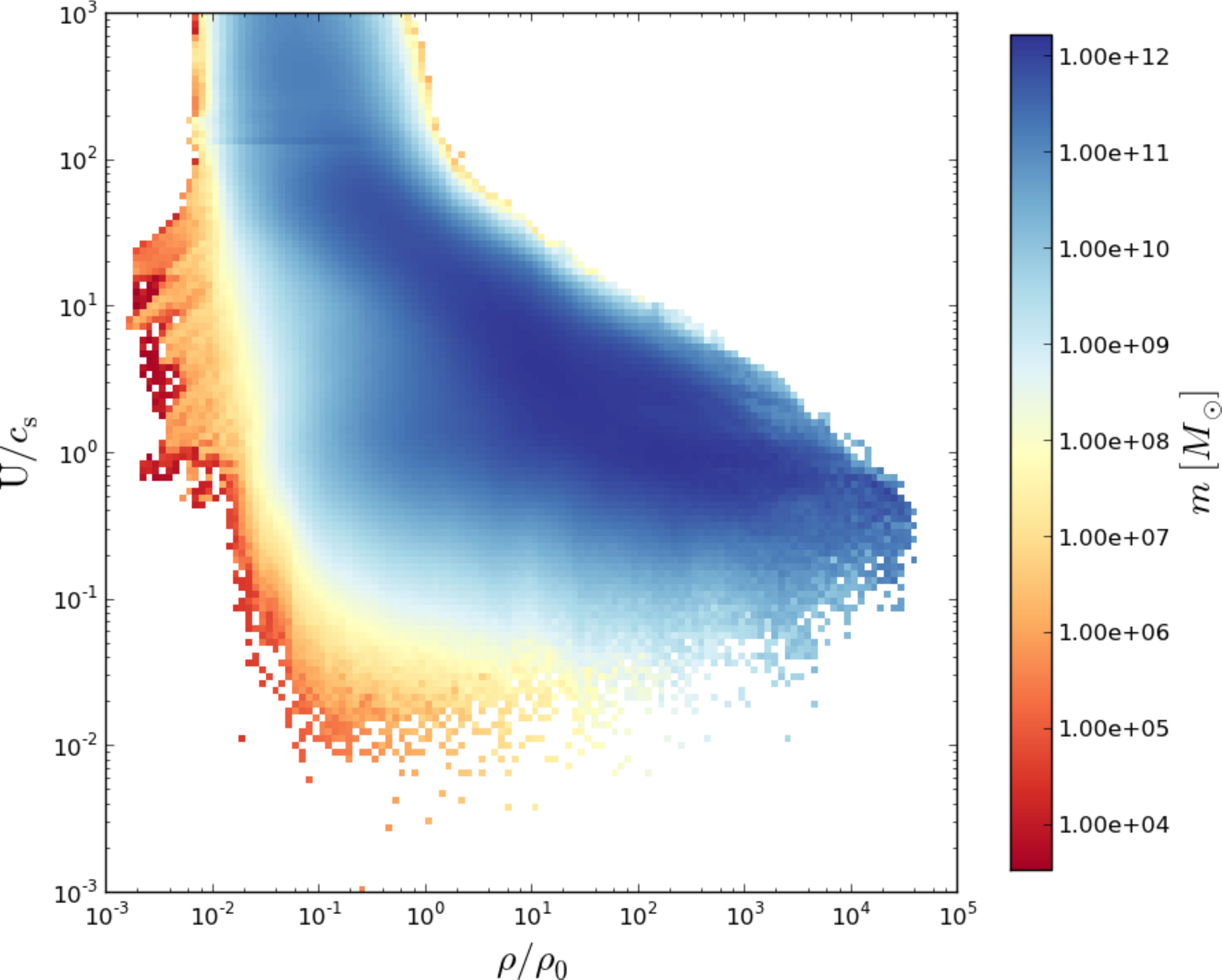}\quad
 \includegraphics[width=0.48\linewidth]{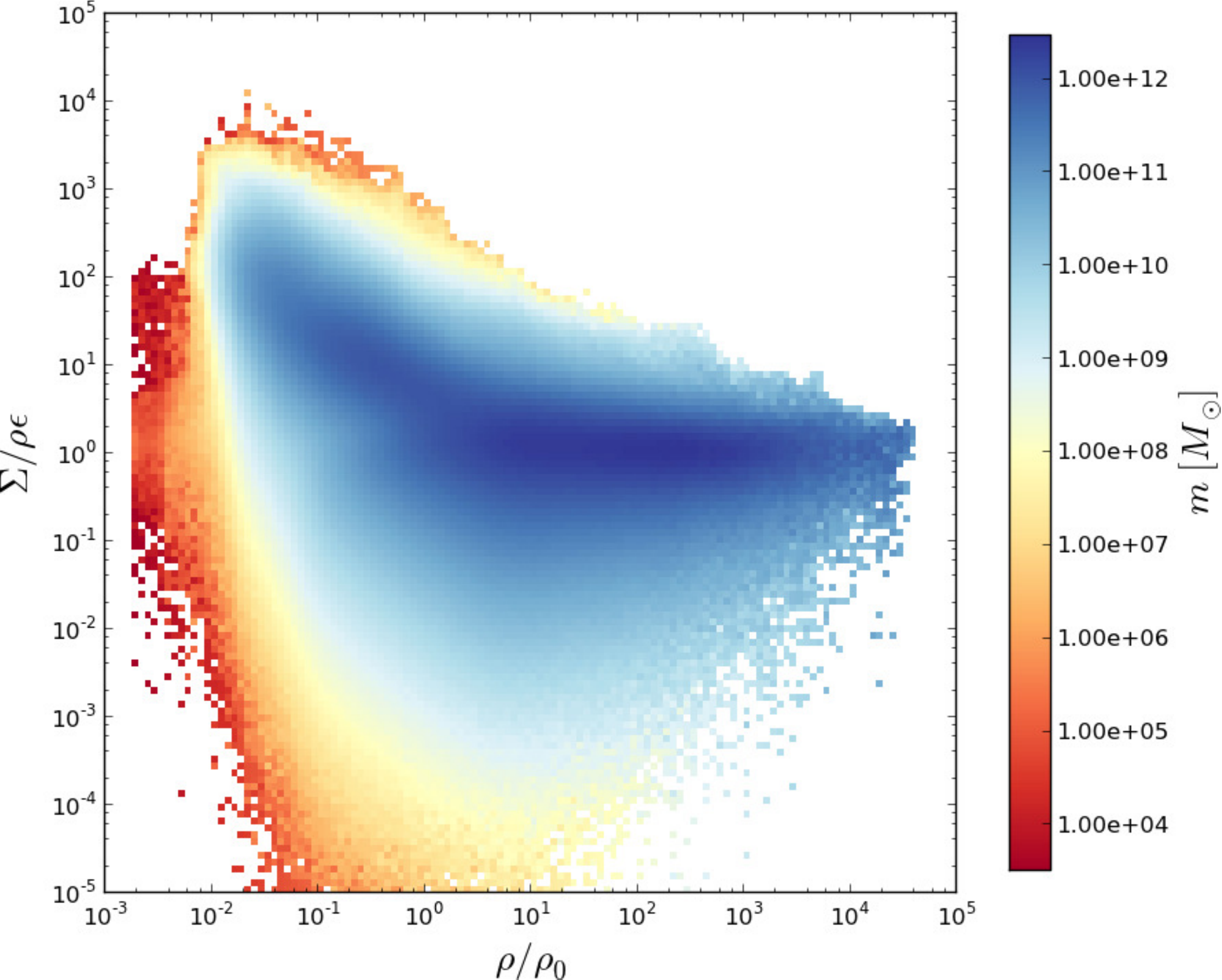}
\caption{Phase plots of the Mach number of the numerically resolved flow, $U/c_{\rm s}$, (left)
  and the SGS production-to-dissipation ratio (right) vs.\ the overdensity
  for the homogeneous SGS model  ($z=0.0$).}
\label{fig:sb_phase}
\end{figure*}

\begin{figure*}
\centering
 \includegraphics[width=0.48\linewidth]{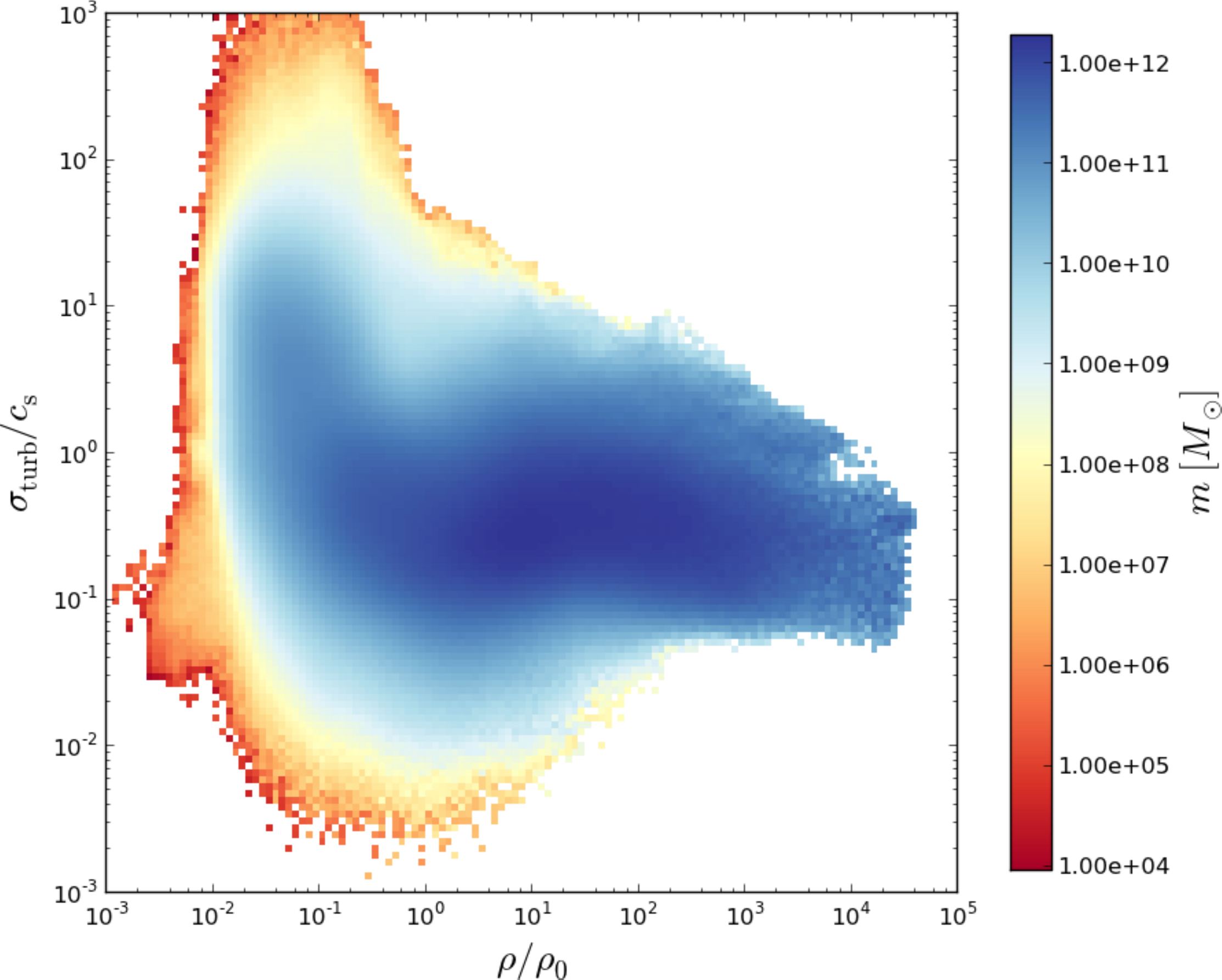}\quad
 \includegraphics[width=0.48\linewidth]{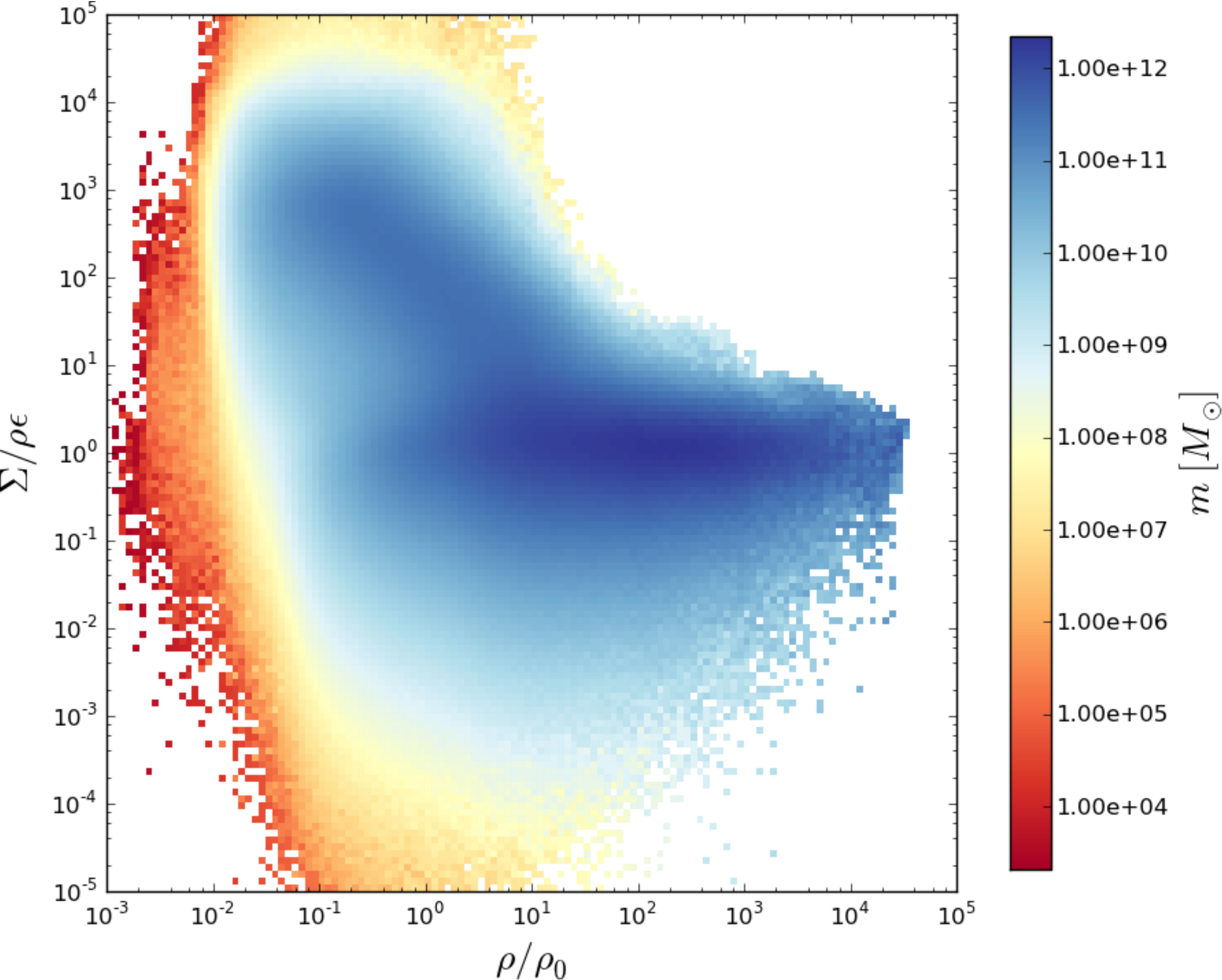}
\caption{Phase plots of the turbulent Mach number $\sigma_{\rm turb}/c_{\rm s}$  (left)
  and the SGS production-to-dissipation ratio (right) vs.\ the overdensity
  for the shear-improved SGS model  ($z=0.0$).}
\label{fig:sb_phase_si}
\end{figure*}

Let us first consider a phase plot of the Mach number of the resolved flow, $U/c_{\rm s}$, versus overdensity 
for the standard SGS model (see Fig.~\ref{fig:sb_phase}). 
Hypersonic Mach numbers $U/c_{\rm s}\gg 1$ are found for $\rho/\rho_0\lesssim 1$, while the flow is mainly transonic for large
overdensities. This is a consequence of shock heating of the gas that is pulled into the cluster and
passes the accretion shock fronts. However, since no heating and cooling is applied, the Mach numbers in the low-density
regions are largely overestimated. In the ICM, on the other hand, we would see higher Mach numbers if the effects of cooling
were incorporated into the simulation. For the dynamics of turbulence, it is particularly interesting to compare the rate
of turbulence energy production, $\Sigma$, to the dissipation rate $\rho\epsilon$. The plot of the production-to-dissipation ratio $\Sigma/\rho\epsilon$
against the overdensity in Fig.~\ref{fig:sb_phase} demonstrates that turbulence is close to equilibrium in the ICM, as
$\Sigma/\rho\epsilon\sim 1$ for most of the gas at high densities. There is also a significant fraction of gas with strong
dissipation. At lower densities ($\rho/\rho_0\lesssim 1$), 
$\Sigma$ tends to be greater than $\rho\epsilon$. This means that turbulence
is being produced, but it is not developed turbulence in the sense that an equilibrium between large-scale driving and small-scale
dissipation is mediated through the turbulent cascade. The corresponding plots for the shear-improved model are shown in
Fig.~\ref{fig:sb_phase_si}. In this case, we plot the turbulent Mach number, $\sigma_{\rm turb}/c_{\rm s}$. 
Compared to the phase plot of $U/c_{\rm s}$ in Fig.~\ref{fig:sb_phase}, the turbulent velocity fluctuations in overdense gas
are actually subsonic. Only a relatively small fraction exceeds the speed of sound. This result justifies the
application of the subsonic closure for the SGS stress tensor (equation~\ref{eq:tau_si}). 
The $\sigma$-shaped bluish region of the phase plot, which is occupied by most of the gas, 
can be interpreted as a signature of heating and turbulence production by the accretion shocks,
which convert low-density gas with little turbulence into hotter, compressed, and highly turbulent gas.

\begin{figure*}
\centering
 \includegraphics[width=0.48\linewidth]{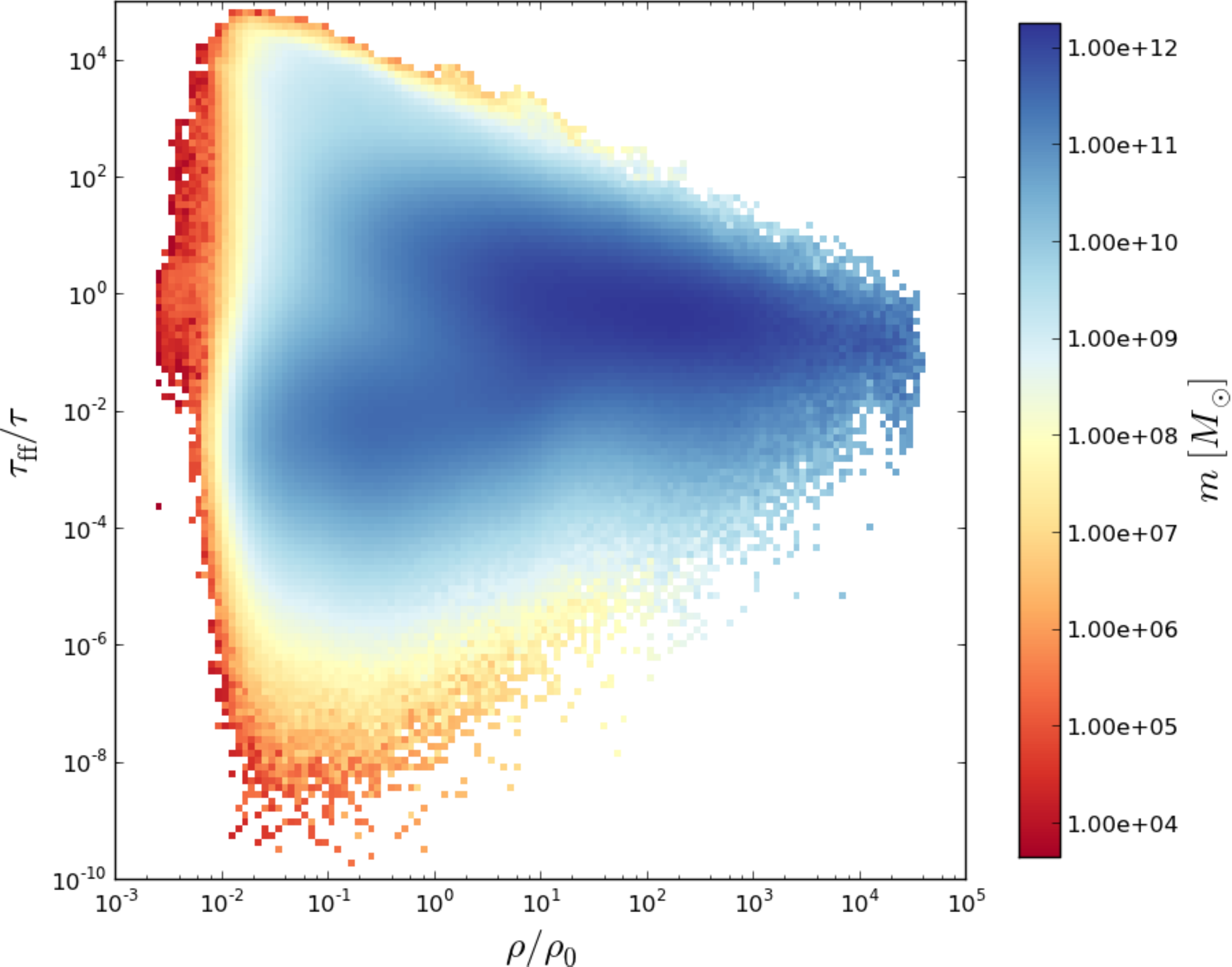}\quad
 \includegraphics[width=0.48\linewidth]{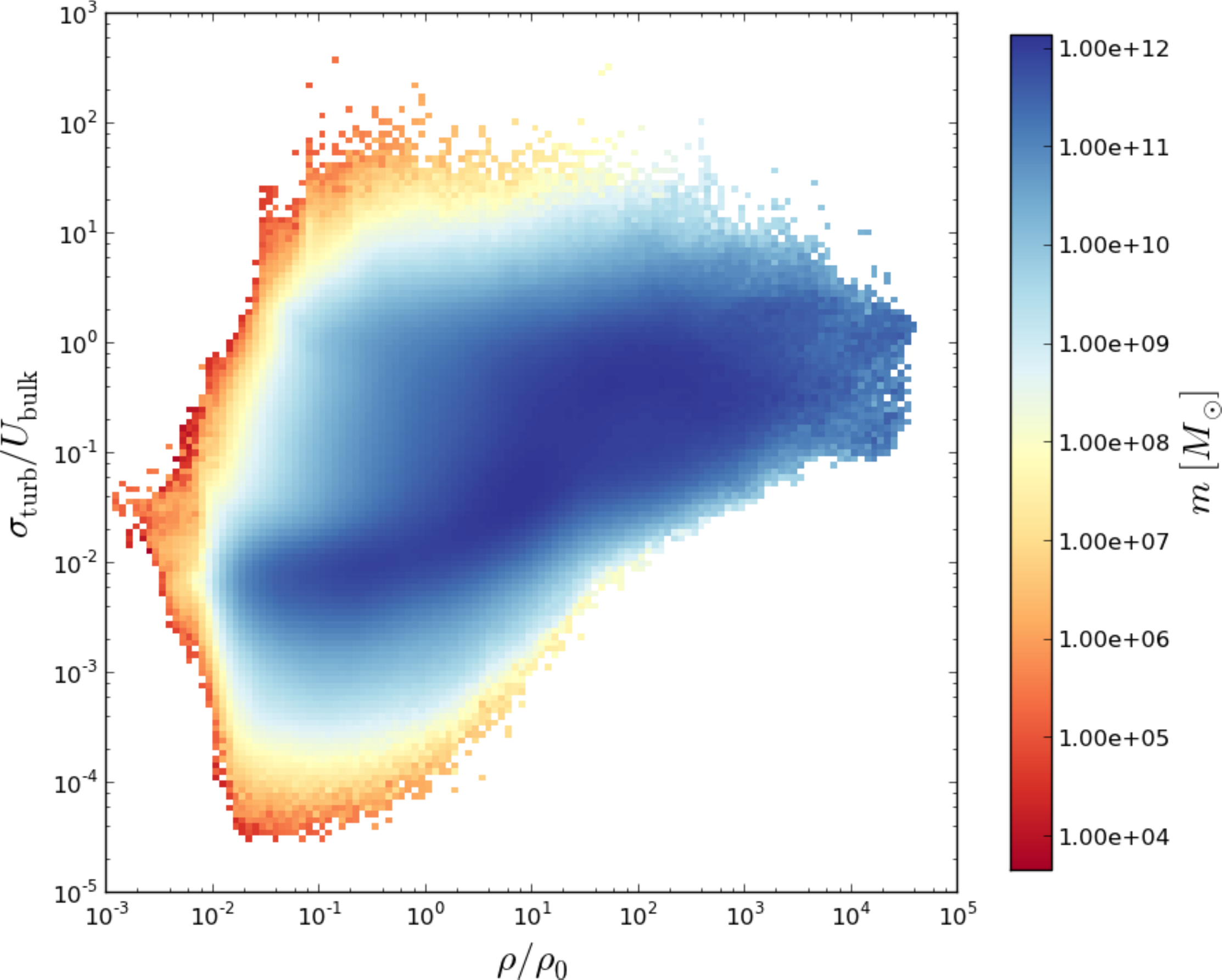}
\caption{Phase plots of the free-fall time scale to the dynamical time scale of the turbulent cascade (left)
  and the ratio of turbulent velocity dispersion to the velocity of the bulk flow (right) vs.\ the overdensity
  for the shear-improved SGS model  ($z=0.0$).}
\label{fig:sb_phase_si2}
\end{figure*}

The link between turbulence production 
and gas accretion by the cluster can also be seen in the phase plot of $\tau_{\rm ff}/\tau$ in Fig.~\ref{fig:sb_phase_si2},
where $\tau_{\rm ff}=\left[3/32\pi G(\rho+\rho_{\rm DM})\right]^{1/2}$ is the free-fall time scale and the dynamical time scale $\tau$ 
is defined by equation~(\ref{eq:tau_dyn}). While there is a very large spread of $\tau_{\rm ff}/\tau$ in underdense gas, 
the range of values becomes increasingly narrow toward high densities. There is a peak of the distribution around
$\rho/\rho_0\sim 0.1$, with $\tau_{\rm ff}$ being a few orders of magnitude smaller than $\tau$. This
region corresponds to the gravity-dominated accretion flows. For $\rho/\rho_0\sim 1$, there is an elongated
region appearing in dark blue, for which $\tau_{\rm ff}/\tau$ decreases from $\sim 10$ down to $0.1$ 
with increasing density. The shorter dynamical time scale indicates strong turbulence. If gravity were
balanced by turbulence, one would expect $\tau_{\rm ff}/\tau\sim 1$. This appears to be the case at moderate
overdensities, i.~e., for gas close to the accretion shocks. In the cluster core, however, virial equilibrium is
maintained mainly by the thermal pressure of the gas. The right plot in Fig.~\ref{fig:sb_phase_si2}
shows the ratio of the turbulent velocity dispersion $\sigma_{\rm turb}$ 
to the bulk-flow velocity magnitude $U_{\rm bulk}$. 
In underdense gas, the distribution of $U_{\rm bulk}$ peaks around $10^{-2}$, which indicates 
that coherent accretion flows are dominant. The
ratio increases around the mean density and is of order unity in the ICM, which corresponds well with
the results for $\tau_{\rm ff}/\tau\sim 1$.

\begin{figure*}
\centering
 \includegraphics[width=\linewidth]{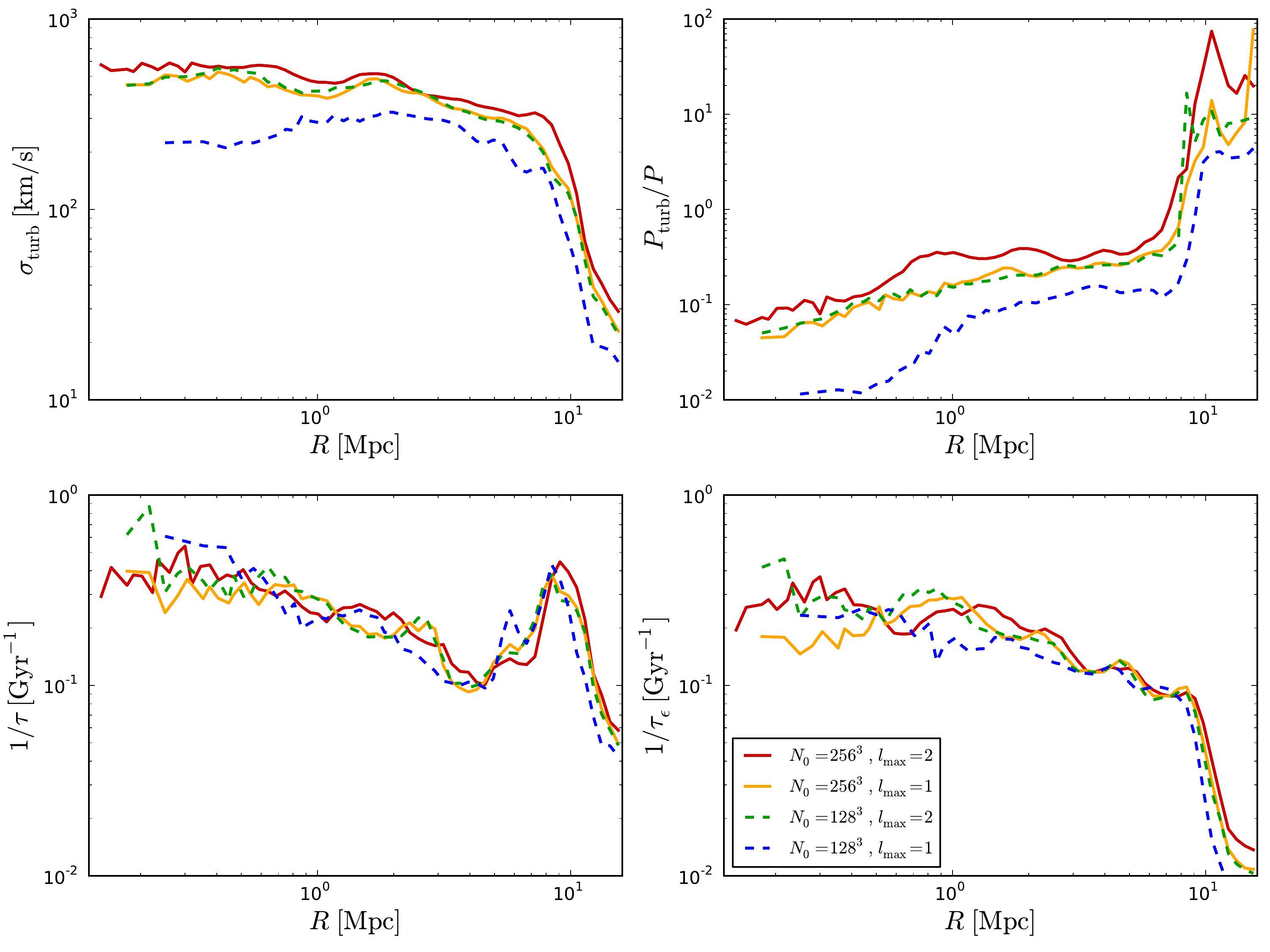}
\caption{Radial profiles of the turbulent velocity dispersion $\sigma_{\rm turb}$, the bulk velocity $U_{\rm bulk}$, 
	the inverse turbulence production time scale $1/\tau$ (see equation~\ref{eq:tau_dyn}), and the inverse dissipation 
	time scale $1/\tau_{\epsilon}$ (equation~\ref{eq:tau_eps})
	for simulations of the Santa Barbara cluster with the shear-improved model at different numerical resolutions ($z=0.0$) . The
	numerically unresolved turbulent velocity fluctuations, $\sigma_{\rm sgs}=\sqrt{2K}$ are shown as thin lines in the top left panel. }
\label{fig:profiles_sigma_turb}
\end{figure*}

\begin{figure}
\centering
  \includegraphics[width=\linewidth]{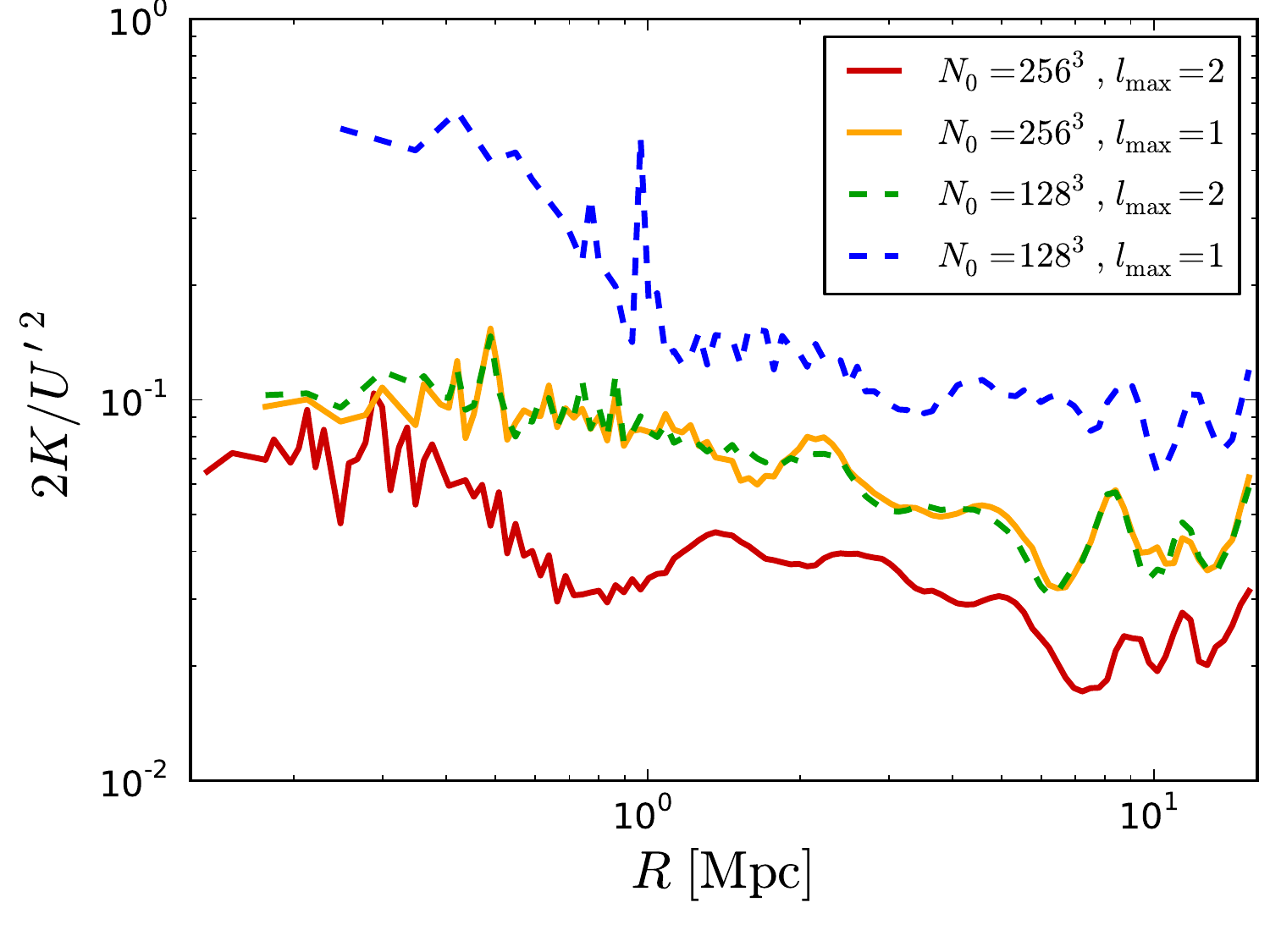}
\caption{Radial profiles of the ratio of the SGS turbulence energy $\rho K$ to the kinetic energy 
	$\frac{1}{2}\rho U^{\prime\,2}$ associated with the fluctuating component $\vecU^{\prime}$ of the numerically resolved flow.}
\label{fig:profiles_sgs_to_fluc}
\end{figure}

Although $\sigma_{\rm turb}$ is well correlated with the vorticity modulus $\omega$ (see Fig.~\ref{fig:sb_phase_vort}), 
it is fairly independent of numerical resolution, as shown by the radial profiles of $\sigma_{\rm turb}$, which are
plotted for different resolutions in Fig.~\ref{fig:profiles_sigma_turb}. This is
an important property of $\sigma_{\rm turb}$, which is in stark contrast to the scale-dependence of the vorticity modulus 
(see Fig.~\ref{fig:SB_profiles_res_si}).
One can also see that the averaged $\sigma_{\rm turb}$ varies little with radius inside the accretions shocks. 
The numerically unresolved fraction of the turbulence energy is of the order $0.1$,
as shown in Fig.~\ref{fig:profiles_sgs_to_fluc}. Moreover, this plot demonstrates that the subgrid scale turbulence energy
$\rho K$ decreases relative to $\frac{1}{2}\rho U^{\prime\,2}$ for higher resolution. This corresponds to
the scaling behavior shown in Fig.~\ref{fig:isoth_levels_evol} for the nested-grid simulations of forced turbulence. 
Since cluster turbulence is neither homogeneous nor stationary, however, there are more pronounced deviations 
from power-law scaling, particularly in the cluster core. Apart from that, a certain bias might be due to
the Kalman filter because $\Ub$ tends to be closer to the smoothed velocity $\Ub_{\rm bulk}$ as
the resolution decreases. As a result, $\Ub^\prime$ will be under-estimated if the resolution is too low.
This is also suggested by the relatively large deviation of $\sigma_{\rm turb}$ for a $128^3$ root grid and 
one level of refinement compared to the higher resolutions (see Fig.~\ref{fig:SB_profiles_res_si}).
Another important variable plotted in Fig.~\ref{fig:SB_profiles_res_si}
is the ratio of the turbulent pressure to the thermal pressure:
\[
  \frac{P_{\rm turb}}{P} = \frac{\sigma_{\rm turb}^2}{3c_{\rm s}^2},
\]
where $c_{\rm s}$ is the speed of sound. The top right plot in Fig.~\ref{fig:profiles_sigma_turb} shows that $P_{\rm turb}/P$ is roughly $0.1$ in the
central region of the cluster. At radial distances greater than $1\;$Mpc, $P_{\rm turb}/P\approx 0.4$ in the
maximally resolved run. Since the profile of $\sigma_{\rm turb}$ is rather flat, the larger $P_{\rm turb}/P$
is caused by lower temperatures, which could be identified with the warm-hot intergalactic medium (keeping in mind that the gas is adiabatic). Consequently, the turbulent pressure reaches a significant fraction of the thermal pressure. 

Reciprocals of the dynamical time scale defined by equation~(\ref{eq:tau_dyn}) and the dissipation time scale,
\begin{equation}
  \label{eq:tau_eps}
  \tau_{\varepsilon} = \frac{\sigma_{\rm turb}^2}{2\varepsilon}
  = \frac{\Delta}{C_{\varepsilon}K^{1/2}}\left[\left(\frac{U^{\prime\,2}}{2K}\right)+1\right]\,,
\end{equation}
are plotted in the bottom panels of Fig.~\ref{fig:profiles_sigma_turb}. These time scales should show no trend with numerical resolution
if the turbulence energy flux $\Sigma$ and the dissipation rate $\rho\epsilon$ are scale-invariant. Indeed, this can be seen in the plots.
Typical values in the cluster interior are a few Gyr, which is consistent with the
choice $T_{\rm c}=5\;\mathrm{Gyr}$ for the Kalman filter. Since, $\tau\sim\tau_{\epsilon}$, turbulence production and dissipation are nearly in 
equilibrium. The accretion shock is indicated by a pronounced peak of $1/\tau$, corresponding to
strong turbulence production, at radii around $10\;\mathrm{Mpc}$. Outside the accretion shock, $\tau_{\epsilon}$ is large compared to cosmological time scales,
and $\tau_{\epsilon}\gg\tau$ implies that no turbulent cascade has developed at $z=0$. This agrees with the phase plots for $\Sigma/\rho\epsilon$ 
shown in Fig.~\ref{fig:sb_phase_si}.

\begin{figure*}
\centering
 \includegraphics[width=\linewidth]{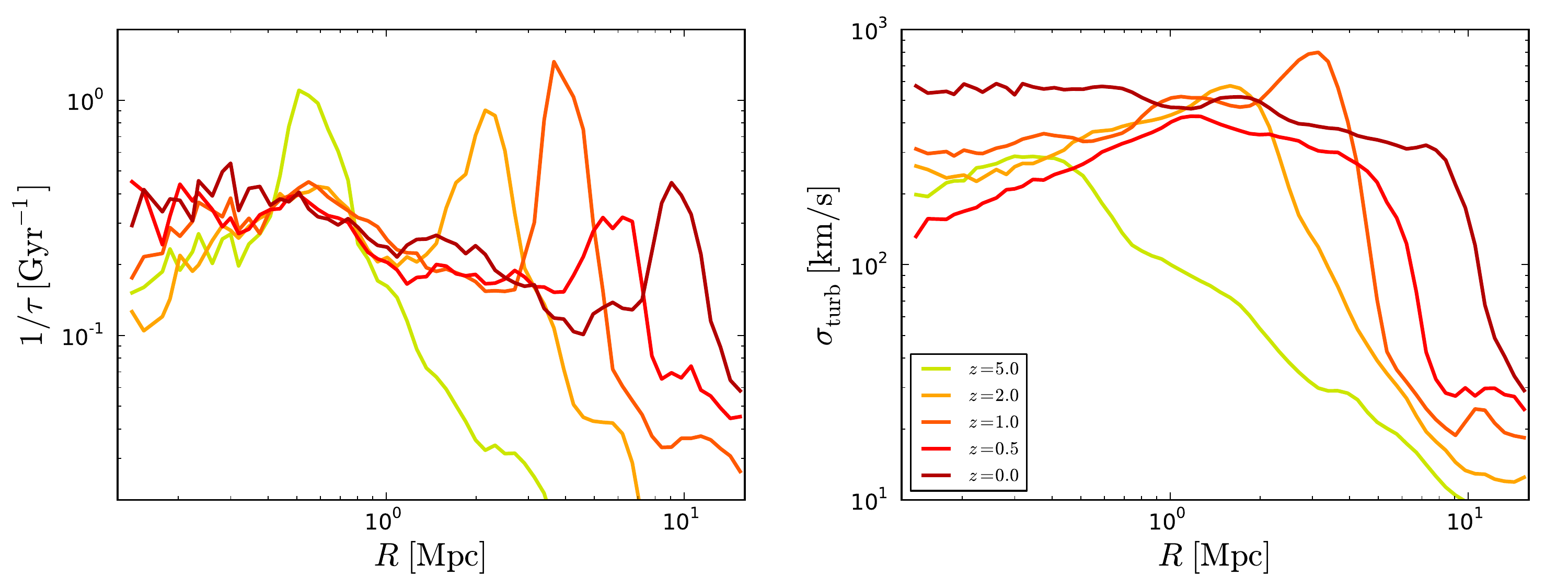}
\caption{Radial profiles of the inverse production time scale $1/\tau$ and 
	the turbulent velocity dispersion $\sigma_{\rm turb}$ at different redshits for the highest-resolution
	run with shear-improved SGS model.}
\label{fig:profiles_sigma_evol}
\end{figure*}

The evolution of $1/\tau$ and $\sigma_{\rm turb}$ with redshift is plotted for the highest-resolution case
in Fig.~\ref{fig:profiles_sigma_evol}. Basically, the mean values of these quantities do not vary much over most of the 
formation time of the cluster. In the left plot, one can see how the peak of $1/\tau$, which indicates the accretion
shock, moves from about $0.5\;\mathrm{Mpc}$ radial distance from the center at $z=5.0$ to about $10\;\mathrm{Mpc}$ as the cluster grows due to mergers and gas accretion. The turbulent velocity dispersion (right plot) reaches values of a few hundred km/s already at early stages in the cluster evolution, but the shape of the profiles gradually changes. These results imply that the Kalman filter parameters $T_{\rm c}$ and $U_{\rm c}$ are actually quite robust for the turbulent regions inside the outer accretion shocks. Only the average dynamical time scale at the shocks, which is less than one Gyr down to redshifts around one, is significantly lower than the smoothing time scale $T_{\rm c}$. As a result, the suppression of SGS turbulence energy production by the shear-improved model is somewhat too low at the shocks, but the discrepancy becomes small at low redshift. This is similar to the spurious production by the bow shock for the minor-merger scenario discussed in Section~\ref{sc:minor_merger}.

\begin{figure}
\centering
  \includegraphics[width=\linewidth]{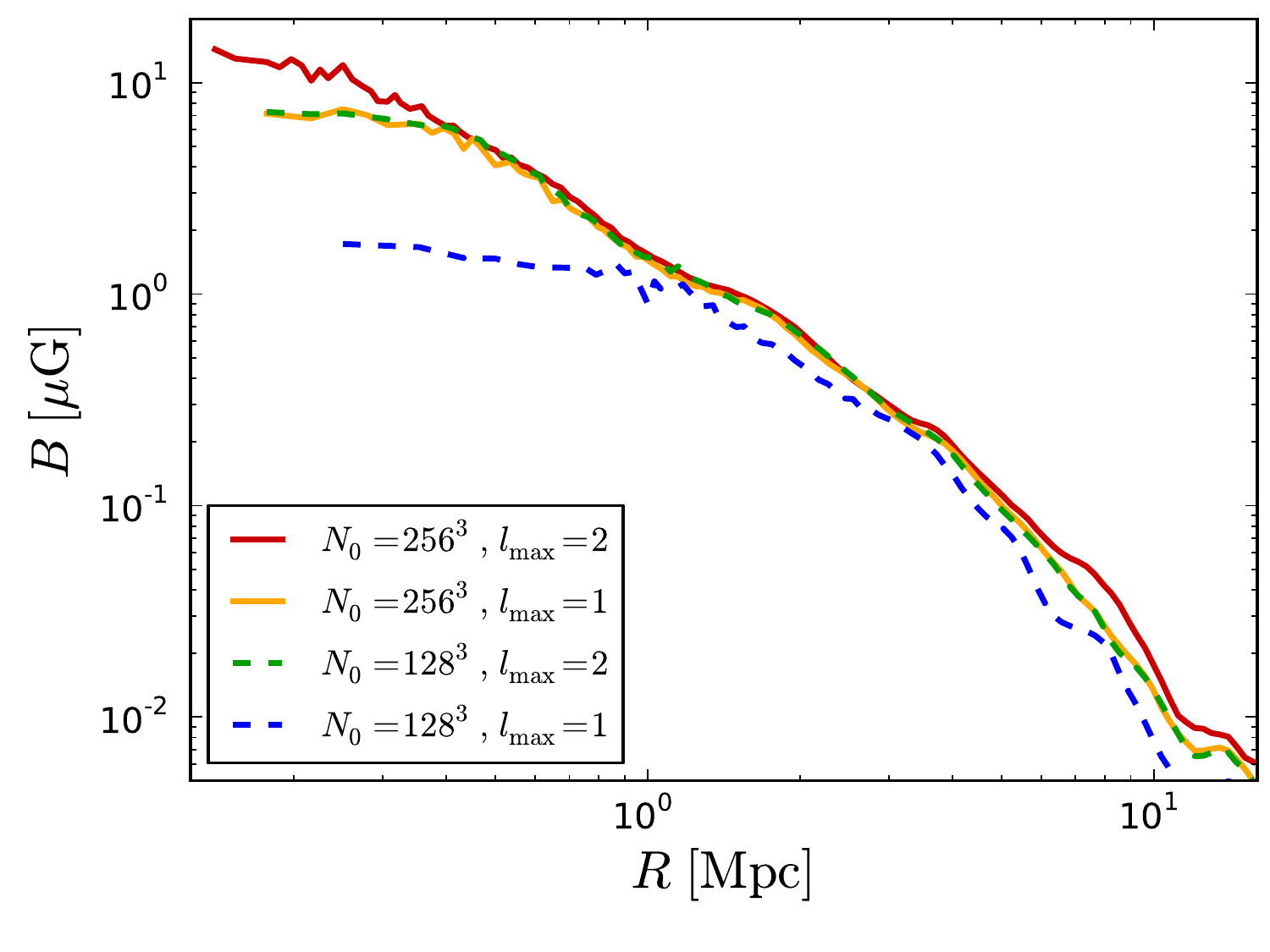}
\caption{Radial profiles of the magnetic field estimated from the turbulence energy
	(see equation~\ref{eq:magn_field}) for different numerical resolutions.}
\label{fig:profiles_magn}
\end{figure}

Furthermore, the turbulent velocity dispersion allows us to estimate the magnetic field produced by
turbulent dynamo action in the ICM by assuming that the magnetic energy is about 
$0.4$ times the turbulent energy in the saturated regime at subsonic Mach numbers \citep{RyuKang08,FederChab11}. 
This relation implies
\begin{equation}
	\label{eq:magn_field}
	B \simeq 2.24\rho^{1/2}\sigma_{\rm turb} = 2.24\sqrt{\rho\left(U^{\prime\,2} + 2K\right)}
\end{equation}
in Gaussian units. Radial profiles of $B$ calculated from the above equation are
plotted for $z=0$ in Fig.~\ref{fig:profiles_magn}. One can see that
the magnetic field assumes values of several $\mu$G in the cluster core,
which is consistent with observations, and rapidly falls off in the outer regions. Moreover,
the estimate of the field strength is nearly independent of resolution. However, the better-resolved density peak at maximal resolution leads to a stronger magnetics field in the cluster core. Outside the accretion shock fronts, 
the magnetic field is in the sub-nG range. Although our estimate is tentative, it suggests that a significant magnetization
of the intergalactic medium requires feedback processes that eject magnetic fields from the interiors of
galaxies. 

\section{Conclusions}
\label{sc:concl}

We propose a subgrid-scale model for numerically unresolved turbulence in cosmological AMR simulations that is based on
a partial differential equation for the energy density $\rho K$ of turbulent velocity fluctuations below the grid scale and
a separation of the numerically resolved flow into a fluctuating component $\Ub^\prime$ and a smoothed component $\Ub_{\rm bulk}$. 
The latter is interpreted as non-turbulent bulk flow. This separation enables us to approximate
the turbulent velocity dispersion $\sigma_{\rm turb}$ by equation~(\ref{eq:sigma_turb}). 
To calculate $\Ub_{\rm bulk}$, the velocity is smoothed by an adaptive temporal Kalman filter.
The smoothing is controlled by two parameters, which have to be calibrated for particular applications. Consistency with
the statistics obtained from the simulations fixes appropriate parameters even if no \emph{a priori} information is available. 
In contrast to previous attempts to combine LES and AMR \citep{MaierIap09},
our method is constructed such that momentum and the sum of the resolved and unresolved energy variables are globally
conserved on non-uniform grids. To avoid energy variables going negative, however, fallback options have to be included, which
cause small local violations of conservation. Apart from that, we use a physically well motivated 
compensation of energy differences due to changes in the grid scale, corresponding to refinement or de-refinement.

Although differences between ILES and LES can be discerned, for example, in our simulations of a subcluster in a wind,
the impact of the SGS model on statistical properties is generally small for subsonic adiabatic gas dynamics,
where the SGS turbulence energy is only a tiny fraction of the total energy.
As a proof of concept, we have run cosmological simulations of the Santa Barbara cluster with the \nyx\ code \citep{AlmBell12}.
In this case, quantities such as the turbulent velocity dispersion, the turbulent energy flux, and the rate of dissipation provide very powerful diagnostics of the turbulence in the cluster gas. The main results of this study are:
\begin{enumerate}
 \item Turbulence is confined inside the outer accretion shocks. The standard SGS model for homogeneous
  turbulence \citep{MaierIap09}, suffers from spurious production of small-scale turbulence in the 
  intergalactic medium  due to the shear associated with the accretion flow.
  The shear-improved model, on the other hand, predicts significant turbulence energy only in regions of high vorticity.
  It appears that the vorticity is mainly produced through accretion, with a peak at the shock fronts, with some contributions
  from previous mergers. However, the distribution of turbulence might change appreciably if feedback from galaxies is incorporated.
 \item The radial profiles of the turbulent velocity dispersion $\sigma_{\rm turb}$ predicted by the Kalman filter and the 
  shear-improved SGS model is nearly flat in the cluster and declines sharply around a radius of $10\;\mathrm{Mpc}$.
  Inside this radius, the turbulent velocity dispersion increases from about $300\;\mathrm{km/s}$ close to the outer accretion shocks
  to more than $500\;  \mathrm{km/s}$ in the cluster core. The numerically unresolved fraction is about $100\;\mathrm{km/s}$ in the
  central region, and $\sigma_{\rm turb}/U_{\rm bulk}\sim 1$. In the voids, this ratio tends toward zero. 
 \item The ratio of the turbulent and thermal pressures, $P_{\rm turb}/P\propto(\sigma_{\rm turb}/c_s)^2$, 
  is of the order $0.1$ in the cluster core, and reaches
  $0.4$ at radii of several Mpc. For the moderate overdensities at these radii, which correspond to the warm-hot intergalactic
  medium, the turbulent pressure will be overestimated if it is calculated from the total velocity. 
  As a consequence, 
  turbulence in the ICM is subsonic and the contribution to virial equilibrium is subdominant, albeit non-negligible.
  At radii greater than about 10 Mpc, very large values of $P_{\rm turb}/P$ are obtained, which is due to the 
  unphysical temperatures of the adiabatic gas. If heating sources were added, the ratio of the pressures
  would decrease to more realistic values. In the light of the studies by \citet{SchmColl13} and \citet{LatSchl13c}, however, 
  it is unclear to what degree turbulence enhances the support of the gas against gravity, particularly in the
  supersonic regime. 
 \item The rate of turbulence energy production $\Sigma$ is comparable to the dissipation rate $\rho\epsilon$ in the centre of the
  cluster, which indicates that turbulence is nearly in a steady state. In the outer regions, on the other hand, the dissipation rate
  tends to be small compared to the production rate. Consequently, the flow is far from equilibrium.
  Correspondingly, the time scales $\tau$ (equation~\ref{eq:tau_dyn}) and $\tau_{\epsilon}$ (equation~\ref{eq:tau_eps}) 
  associated with $\Sigma$ and $\rho\epsilon$, respectively, range from about $2.5$ to $10\;\mathrm{Gyr}$ 
  inside the accretion shocks, but are much greater outside. The smallest dynamical time scale is encountered 
  in the vicinity of the outer accretion shocks, where $\tau\sim 2\;\mathrm{Gyr}$, averaged over all directions. 
 \item The turbulence energy spectra are roughly consistent with a Kolmogorov spectrum, with deviations due to
  the very narrow range between turbulence driving and numerical dissipation. The spectral distribution of
  $\Ub^{\prime}$ suggests a driving scale of a few Mpc. 
 \item From equation~(\ref{eq:magn_field}), we estimate an average magnetic field between $1$ and $10\;\mu$G in the ICM.
  Towards larger radii, the magnetic field steadily decreases and falls below $1\;$nG beyond the accretions shocks. 
  The central part of the radial profile is roughly comparable to the results from MHD simulations of a cluster in the $\Lambda$CDM
  cosmology by \citet{XuLi09}. 

\end{enumerate}

Apart from the implications summarized above, a stronger impact of the SGS model on the flow properties can be expected
in applications where the gas cools significantly. This has already been demonstrated for LES of the gravitational collapse in 
atomic cooling halos \citep{LatSchl13,LatSchl13c}. In these studies the SGS model of \citet{MaierIap09} was applied. Remarkably,
the turbulent viscosity of unresolved velocity fluctuations gives rise to disk-like structures around collapsing
objects, which tend to be much less pronounced or are even absent in ILES. Moreover, by calculating the
rate of compression \citep[see][]{ZhuFeng10,IapSchm11,SchmColl13}, it was demonstrated that the turbulent pressure predicted by the SGS model contributes significantly to the support of the gas against gravity. The same effects are likely to play a role
for the formation of disk galaxies, particularly the condensation of the cold phase in the ISM. To a lesser degree, there
might be a direct influence on turbulence in the warm-hot intergalactic medium in galaxy clusters with realistic cooling \citep{IapiViel13}. This issue will be addressed in a follow-up study.
Moreover, it is important to note that turbulent regions are almost completely refined up to the highest level in the simulations of the Santa Barbara cluster. For this reason, turbulence is insignificant at the 
fine-coarse boundaries. In general, however, it is too costly to completely refine turbulence in cosmological simulations.
Refinement is often restricted to some objects of interest, for example, by means of the zoom-in technique. 
In this case, the method of turbulence energy compensation described in Appendix~\ref{sc:refine} is advantageous
because turbulent regions cross the boundaries between different refinement levels. 

Furthermore, an SGS model for turbulence can be utilized for stellar feedback models. Cosmological simulations
cannot resolve the length scales of turbulence energy injection by supernovae. In addition to heating, supernovae
also effectively produce turbulent pressure on the grid scale, which is given by $P_{\rm sgs}=\frac{3}{2}\rho K$.
\citet{JoungLow09} propose a model similar to ours, with a source term that accounts for driving by supernovae, but
no turbulence production by shear.  A much more detailed model of feedback and turbulence production on
sub-galactic scales was recently formulated by \citet{BraunSchm12}. For simulations of cosmological structure formation, some
variant in between these two approaches will be suitable. Another exciting prospect is the
numerical modelling of magnetic field amplification in clusters if the methodology could be generalized to magnetohydrodynamics.
Conventional MHD simulations with purely numerical resistivity, cannot account for dynamo effects, which
are expected to be strongest on length scales close to the physical dissipation length scale \citep{SchlSchob13,LatSchl13b}, 
provided that this scale is small compared to the grid resolution. 
As pointed out in the introduction, there are still many open question regarding the very nature of turbulence
in the ICM and the physical dissipation mechanism. There is a wide field of fundamental research that remains to be done.
Eventually, an SGS model might turn out to be capable of approximating turbulent dynamo action on small scales
to better understand the generation of the strong cosmic magnetic fields that are ubiquitously observed.

\section*{Acknowledgments}

We thank Peter Nugent and his team for supporting the development of \nyx\ at the Computational Cosmology Center at LBNL.
In particular, we thank Zarija Luki\'{c} for the setup of the Santa Barbara cluster.
We are indebted to Emmanuel L\'ev\^eque, who proposed the application of the shear-improved model to treat inhomogenous
turbulence. During a stay at \'Ecole Normale Sup\'erieure de Lyon, W. Schmidt had the opportunity to discuss this approach in detail. We thank the referee and Andrey Kravtsov for many useful comments and suggestions, which helped to improve
this article.
Furthermore, we are grateful for discussions about turbulent magnetic field amplification with Dominik Schleicher.
W.~Schmidt, H.~Braun, J.~F.~Engels, and J.~C.~Niemeyer acknowledge financial support by the German Research Council and the
Faculty of Physics in G\"{o}ttingen for visits at LBNL. R.~R.~Mekuria is grateful for an AstroMundus scholarship. 
The work at LBNL was supported by the SciDAC Program and the Applied Mathematics Program of the U.S. Department of Energy under 
Contract No. DE-AC02-05CH11231.
The simulations presented in this article were performed with supercomputing resources of HLRN (project nip00020) and 
GWDG in Germany. We also acknowledge the yt toolkit by \citet{TurkSmith11} that was used for our analysis of numerical data. 
We thank Muhammad Latif for providing scripts for plotting profiles and David Collins for assistance with the
computation of Fourier spectra. 

\bibliography{CFEARLESS}

\begin{thebibliography}{}

\bibitem[\protect\citeauthoryear{{Almgren}, {Beckner}, {Bell}, {Day}, {Howell},
  {Joggerst}, {Lijewski}, {Nonaka}, {Singer} \& {Zingale}}{{Almgren}
  et~al.}{2010}]{AlmBeck10}
{Almgren} A.~S.,  {Beckner} V.~E.,  {Bell} J.~B.,  {Day} M.~S.,  {Howell}
  L.~H.,  {Joggerst} C.~C.,  {Lijewski} M.~J.,  {Nonaka} A.,  {Singer} M.,
  {Zingale} M.,  2010, \apj, 715, 1221

\bibitem[\protect\citeauthoryear{{Almgren}, {Bell}, {Lijewski}, {Luki{\'c}} \&
  {Van Andel}}{{Almgren} et~al.}{2013}]{AlmBell12}
{Almgren} A.~S.,  {Bell} J.~B.,  {Lijewski} M.~J.,  {Luki{\'c}} Z.,    {Van
  Andel} E.,  2013, \apj, 765, 39

\bibitem[\protect\citeauthoryear{{Aspden}, {Nikiforakis}, {Dalziel} \&
  {Bell}}{{Aspden} et~al.}{2008}]{AspNiki08}
{Aspden} A.~J.,  {Nikiforakis} N.,  {Dalziel} S.~B.,    {Bell} J.,  2008,
  {Communications in Applied Mathematics and Computational Science}, 3, 103

\bibitem[\protect\citeauthoryear{{Biffi}, {Dolag} \& {B{\"o}hringer}}{{Biffi}
  et~al.}{2013}]{BiffiDol12}
{Biffi} V.,  {Dolag} K.,    {B{\"o}hringer} H.,  2013, \mnras, 428, 1395

\bibitem[\protect\citeauthoryear{{Bonafede}, {Feretti}, {Murgia}, {Govoni},
  {Giovannini}, {Dallacasa}, {Dolag} \& {Taylor}}{{Bonafede}
  et~al.}{2010}]{BonaFer10}
{Bonafede} A.,  {Feretti} L.,  {Murgia} M.,  {Govoni} F.,  {Giovannini} G.,
  {Dallacasa} D.,  {Dolag} K.,    {Taylor} G.~B.,  2010, \aap, 513, A30

\bibitem[\protect\citeauthoryear{{Bonafede}, {Govoni}, {Feretti}, {Murgia},
  {Giovannini} \& {Br{\"u}ggen}}{{Bonafede} et~al.}{2011}]{BonaGov11}
{Bonafede} A.,  {Govoni} F.,  {Feretti} L.,  {Murgia} M.,  {Giovannini} G.,
  {Br{\"u}ggen} M.,  2011, \aap, 530, A24

\bibitem[\protect\citeauthoryear{{Borgani} \& {Kravtsov}}{{Borgani} \&
  {Kravtsov}}{2011}]{BorKravt11}
{Borgani} S.,  {Kravtsov} A.,  2011, Advanced Science Letters, 4, 204

\bibitem[\protect\citeauthoryear{{Braun} \& {Schmidt}}{{Braun} \&
  {Schmidt}}{2012}]{BraunSchm12}
{Braun} H.,  {Schmidt} W.,  2012, \mnras, 421, 1838

\bibitem[\protect\citeauthoryear{{Br{\"u}ggen} \& {Scannapieco}}{{Br{\"u}ggen}
  \& {Scannapieco}}{2009}]{BrueggScann09}
{Br{\"u}ggen} M.,  {Scannapieco} E.,  2009, \mnras, 398, 548

\bibitem[\protect\citeauthoryear{{Cahuzac}, {Boudet}, {Borgnat} \&
  {L{\'e}v{\^e}que}}{{Cahuzac} et~al.}{2010}]{CahuBou10}
{Cahuzac} A.,  {Boudet} J.,  {Borgnat} P.,    {L{\'e}v{\^e}que} E.,  2010,
  Physics of Fluids, 22, 125104

\bibitem[\protect\citeauthoryear{{Cahuzac}, {Boudet}, {Borgnat} \&
  {L{\'e}v{\^e}que}}{{Cahuzac} et~al.}{2011}]{CahuBou11}
{Cahuzac} A.,  {Boudet} J.,  {Borgnat} P.,    {L{\'e}v{\^e}que} E.,  2011,
  Journal of Physics Conference Series, 318, 042047

\bibitem[\protect\citeauthoryear{{Colella}}{{Colella}}{1990}]{colella1990}
{Colella} P.,  1990, Journal of Computational Physics, 87, 171

\bibitem[\protect\citeauthoryear{{Colella} \& {Glaz}}{{Colella} \&
  {Glaz}}{1985}]{ColGlaz85}
{Colella} P.,  {Glaz} H.~M.,  1985, Journal of Computational Physics, 59, 264

\bibitem[\protect\citeauthoryear{{Federrath}, {Chabrier}, {Schober},
  {Banerjee}, {Klessen} \& {Schleicher}}{{Federrath}
  et~al.}{2011}]{FederChab11}
{Federrath} C.,  {Chabrier} G.,  {Schober} J.,  {Banerjee} R.,  {Klessen}
  R.~S.,    {Schleicher} D.~R.~G.,  2011, Physical Review Letters, 107, 114504

\bibitem[\protect\citeauthoryear{{Ferrari}, {Govoni}, {Schindler}, {Bykov} \&
  {Rephaeli}}{{Ferrari} et~al.}{2008}]{FerrGov08}
{Ferrari} C.,  {Govoni} F.,  {Schindler} S.,  {Bykov} A.~M.,    {Rephaeli} Y.,
  2008, \ssr, 134, 93

\bibitem[\protect\citeauthoryear{Frisch}{Frisch}{1995}]{Frisch}
Frisch U.,  1995, {Turbulence. The legacy of A.N. Kolmogorov}.
Cambridge: Cambridge University Press

\bibitem[\protect\citeauthoryear{{Gaspari} \& {Churazov}}{{Gaspari} \&
  {Churazov}}{2013}]{GasChur13}
{Gaspari} M.,  {Churazov} E.,  2013, \aap, 559, A78

\bibitem[\protect\citeauthoryear{Germano}{Germano}{1992}]{Germano92}
Germano M.,  1992, J. Fluid Mech., 238, 325

\bibitem[\protect\citeauthoryear{{Heitmann}, {Ricker}, {Warren} \&
  {Habib}}{{Heitmann} et~al.}{2005}]{HeitRick05}
{Heitmann} K.,  {Ricker} P.~M.,  {Warren} M.~S.,    {Habib} S.,  2005, \apjs,
  160, 28

\bibitem[\protect\citeauthoryear{{Iapichino}, {Adamek}, {Schmidt} \&
  {Niemeyer}}{{Iapichino} et~al.}{2008}]{IapiAda08}
{Iapichino} L.,  {Adamek} J.,  {Schmidt} W.,    {Niemeyer} J.~C.,  2008,
  \mnras, 388, 1079

\bibitem[\protect\citeauthoryear{{Iapichino} \& {Niemeyer}}{{Iapichino} \&
  {Niemeyer}}{2008}]{IapNie08}
{Iapichino} L.,  {Niemeyer} J.~C.,  2008, \mnras, 388, 1089

\bibitem[\protect\citeauthoryear{{Iapichino}, {Schmidt}, {Niemeyer} \&
  {Merklein}}{{Iapichino} et~al.}{2011}]{IapSchm11}
{Iapichino} L.,  {Schmidt} W.,  {Niemeyer} J.~C.,    {Merklein} J.,  2011,
  \mnras, 414, 2297

\bibitem[\protect\citeauthoryear{{Iapichino}, {Viel} \& {Borgani}}{{Iapichino}
  et~al.}{2013}]{IapiViel13}
{Iapichino} L.,  {Viel} M.,    {Borgani} S.,  2013, \mnras

\bibitem[\protect\citeauthoryear{{Joung}, {Mac Low} \& {Bryan}}{{Joung}
  et~al.}{2009}]{JoungLow09}
{Joung} M.~R.,  {Mac Low} M.-M.,    {Bryan} G.~L.,  2009, \apj, 704, 137

\bibitem[\protect\citeauthoryear{{Kritsuk}, {Norman}, {Padoan} \&
  {Wagner}}{{Kritsuk} et~al.}{2007}]{KritNor07}
{Kritsuk} A.~G.,  {Norman} M.~L.,  {Padoan} P.,    {Wagner} R.,  2007, {\apj},
  665, 416

\bibitem[\protect\citeauthoryear{{Kritsuk}, {Ustyugov}, {Norman} \&
  {Padoan}}{{Kritsuk} et~al.}{2010}]{KritUsty10}
{Kritsuk} A.~G.,  {Ustyugov} S.~D.,  {Norman} M.~L.,    {Padoan} P.,  2010, in
  {Pogorelov} N.~V.,  {Audit} E.,   {Zank} G.~P.,  eds, Numerical Modeling of
  Space Plasma Flows, Astronum-2009 Vol.~429 of Astronomical Society of the
  Pacific Conference Series, {Self-organization in Turbulent Molecular Clouds:
  Compressional Versus Solenoidal Modes}.
p.~15

\bibitem[\protect\citeauthoryear{{Latif}, {Schleicher}, {Schmidt} \&
  {Niemeyer}}{{Latif} et~al.}{2013a}]{LatSchl13c}
{Latif} M.~A.,  {Schleicher} D.~R.~G.,  {Schmidt} W.,    {Niemeyer} J.,  2013a,
  \mnras, 433, 1607

\bibitem[\protect\citeauthoryear{{Latif}, {Schleicher}, {Schmidt} \&
  {Niemeyer}}{{Latif} et~al.}{2013b}]{LatSchl13}
{Latif} M.~A.,  {Schleicher} D.~R.~G.,  {Schmidt} W.,    {Niemeyer} J.,  2013b,
  \mnras, 430, 588

\bibitem[\protect\citeauthoryear{{Latif}, {Schleicher}, {Schmidt} \&
  {Niemeyer}}{{Latif} et~al.}{2013c}]{LatSchl13b}
{Latif} M.~A.,  {Schleicher} D.~R.~G.,  {Schmidt} W.,    {Niemeyer} J.,  2013c,
  \mnras, 432, 668

\bibitem[\protect\citeauthoryear{{Lazarian}}{{Lazarian}}{2006}]{Lazar06}
{Lazarian} A.,  2006, \apjl, 645, L25

\bibitem[\protect\citeauthoryear{{L{\'e}v{\^e}que}, {Toschi}, {Shao} \&
  {Bertoglio}}{{L{\'e}v{\^e}que} et~al.}{2007}]{LevTosch07}
{L{\'e}v{\^e}que} E.,  {Toschi} F.,  {Shao} L.,    {Bertoglio} J.-P.,  2007,
  Journal of Fluid Mechanics, 570, 491

\bibitem[\protect\citeauthoryear{{Maier}, {Iapichino}, {Schmidt} \&
  {Niemeyer}}{{Maier} et~al.}{2009}]{MaierIap09}
{Maier} A.,  {Iapichino} L.,  {Schmidt} W.,    {Niemeyer} J.~C.,  2009, \apj,
  707, 40

\bibitem[\protect\citeauthoryear{{Miller} \& {Colella}}{{Miller} \&
  {Colella}}{2002}]{miller-colella}
{Miller} G.~H.,  {Colella} P.,  2002, Journal of Computational Physics, 183, 26

\bibitem[\protect\citeauthoryear{{Narayan} \& {Medvedev}}{{Narayan} \&
  {Medvedev}}{2001}]{NaraMed01}
{Narayan} R.,  {Medvedev} M.~V.,  2001, ApJ, 562, L129

\bibitem[\protect\citeauthoryear{{Oppenheimer} \& {Dav{\'e}}}{{Oppenheimer} \&
  {Dav{\'e}}}{2009}]{OppDave09}
{Oppenheimer} B.~D.,  {Dav{\'e}} R.,  2009, \mnras, 395, 1875

\bibitem[\protect\citeauthoryear{{Parrish}, {McCourt}, {Quataert} \&
  {Sharma}}{{Parrish} et~al.}{2012}]{ParrCourt12}
{Parrish} I.~J.,  {McCourt} M.,  {Quataert} E.,    {Sharma} P.,  2012, \mnras,
  422, 704

\bibitem[\protect\citeauthoryear{{Paul}, {Iapichino}, {Miniati}, {Bagchi} \&
  {Mannheim}}{{Paul} et~al.}{2011}]{PaulIap11}
{Paul} S.,  {Iapichino} L.,  {Miniati} F.,  {Bagchi} J.,    {Mannheim} K.,
  2011, \apj, 726, 17

\bibitem[\protect\citeauthoryear{{Rauch}, {Becker}, {Viel}, {Sargent},
  {Smette}, {Simcoe}, {Barlow} \& {Haehnelt}}{{Rauch}
  et~al.}{2005}]{RauchBeck05}
{Rauch} M.,  {Becker} G.~D.,  {Viel} M.,  {Sargent} W.~L.~W.,  {Smette} A.,
  {Simcoe} R.~A.,  {Barlow} T.~A.,    {Haehnelt} M.~G.,  2005, \apj, 632, 58

\bibitem[\protect\citeauthoryear{{Rauch}, {Sargent}, {Barlow} \&
  {Carswell}}{{Rauch} et~al.}{2001}]{RauchSar01}
{Rauch} M.,  {Sargent} W.~L.~W.,  {Barlow} T.~A.,    {Carswell} R.~F.,  2001,
  \apj, 562, 76

\bibitem[\protect\citeauthoryear{{Rebusco}, {Churazov}, {Sunyaev} \&
  {B{\"o}hringer} H. an d~{Forman}}{{Rebusco} et~al.}{2008}]{RebusChur08}
{Rebusco} P.,  {Churazov} E.,  {Sunyaev} R.,    {B{\"o}hringer} H. an
  d~{Forman} W.,  2008, MNRAS, 384, 1511

\bibitem[\protect\citeauthoryear{{Reynolds}, {McKernan}, {Fabian}, {Stone} \&
  {Vernaleo}}{{Reynolds} et~al.}{2005}]{ReyKer05}
{Reynolds} C.~S.,  {McKernan} B.,  {Fabian} A.~C.,  {Stone} J.~M.,
  {Vernaleo} J.~C.,  2005, \mnras, 357, 242

\bibitem[\protect\citeauthoryear{{Robertson} \& {Goldreich}}{{Robertson} \&
  {Goldreich}}{2012}]{RobGold12}
{Robertson} B.,  {Goldreich} P.,  2012, \apjl, 750, L31

\bibitem[\protect\citeauthoryear{{Roediger}, {Kraft}, {Forman}, {Nulsen} \&
  {Churazov}}{{Roediger} et~al.}{2013}]{RoedKraft13}
{Roediger} E.,  {Kraft} R.~P.,  {Forman} W.~R.,  {Nulsen} P.~E.~J.,
  {Churazov} E.,  2013, \apj, 764, 60

\bibitem[\protect\citeauthoryear{{Ryu}, {Kang}, {Cho} \& {Das}}{{Ryu}
  et~al.}{2008}]{RyuKang08}
{Ryu} D.,  {Kang} H.,  {Cho} J.,    {Das} S.,  2008, Science, 320, 909

\bibitem[\protect\citeauthoryear{Sagaut}{Sagaut}{2006}]{Sagaut}
Sagaut P.,  2006, {Large eddy simulation for incompressible flows: An
  introduction}.
Berlin: Springer-Verlag

\bibitem[\protect\citeauthoryear{{Saltzman}}{{Saltzman}}{1994}]{saltzman}
{Saltzman} J.,  1994, Journal of Computational Physics, 115, 153

\bibitem[\protect\citeauthoryear{{Sanders}, {Fabian} \& {Smith}}{{Sanders}
  et~al.}{2011}]{SandFab11}
{Sanders} J.~S.,  {Fabian} A.~C.,    {Smith} R.~K.,  2011, \mnras, 410, 1797

\bibitem[\protect\citeauthoryear{{Santos-Lima}, {de Gouveia Dal Pino}, {Kowal},
  {Falceta-Gon{\c c}alves}, {Lazarian} \& {Nakwacki}}{{Santos-Lima}
  et~al.}{2014}]{SantosGouv13}
{Santos-Lima} R.,  {de Gouveia Dal Pino} E.~M.,  {Kowal} G.,  {Falceta-Gon{\c
  c}alves} D.,  {Lazarian} A.,    {Nakwacki} M.~S.,  2014, \apj, 781, 84

\bibitem[\protect\citeauthoryear{{Schleicher}, {Schober}, {Federrath}, {Bovino}
  \& {Schmidt}}{{Schleicher} et~al.}{2013}]{SchlSchob13}
{Schleicher} D.~R.~G.,  {Schober} J.,  {Federrath} C.,  {Bovino} S.,
  {Schmidt} W.,  2013, New Journal of Physics, 15, 023017

\bibitem[\protect\citeauthoryear{{Schmidt}}{{Schmidt}}{2014}]{Schmidt}
{Schmidt} W.,  2014, Numerical Modelling of Astrophysical Turbulence.
SpringerBriefs in Astronomy, Springer

\bibitem[\protect\citeauthoryear{{Schmidt}, {Collins} \& {Kritsuk}}{{Schmidt}
  et~al.}{2013}]{SchmColl13}
{Schmidt} W.,  {Collins} D.~C.,    {Kritsuk} A.~G.,  2013, \mnras, 431, 3196

\bibitem[\protect\citeauthoryear{{Schmidt} \& {Federrath}}{{Schmidt} \&
  {Federrath}}{2011}]{SchmFeder11}
{Schmidt} W.,  {Federrath} C.,  2011, \aap, 528, A106

\bibitem[\protect\citeauthoryear{{Schmidt}, {Federrath}, {Hupp}, {Kern} \&
  {Niemeyer}}{{Schmidt} et~al.}{2009a}]{SchmFeder09a}
{Schmidt} W.,  {Federrath} C.,  {Hupp} M.,  {Kern} S.,    {Niemeyer} J.~C.,
  2009a, \aap, 494, 127

\bibitem[\protect\citeauthoryear{{Schmidt}, {Federrath}, {Hupp}, {Kern} \&
  {Niemeyer}}{{Schmidt} et~al.}{2009b}]{SchmFeder09}
{Schmidt} W.,  {Federrath} C.,  {Hupp} M.,  {Kern} S.,    {Niemeyer} J.~C.,
  2009b, \aap, 494, 127

\bibitem[\protect\citeauthoryear{Schmidt, Hillebrandt \& Niemeyer}{Schmidt
  et~al.}{2006}]{SchmHille06}
Schmidt W.,  Hillebrandt W.,    Niemeyer J.~C.,  2006, {Comp. Fluids.}, 35, 353

\bibitem[\protect\citeauthoryear{{Schmidt}, {Niemeyer} \&
  {Hillebrandt}}{{Schmidt} et~al.}{2006}]{SchmNie06}
{Schmidt} W.,  {Niemeyer} J.~C.,    {Hillebrandt} W.,  2006, \aap, 450, 265

\bibitem[\protect\citeauthoryear{Sunyaev, Norman \& Bryan}{Sunyaev
  et~al.}{2003}]{SunyNor03}
Sunyaev R.~A.,  Norman M.~L.,    Bryan G.~L.,  2003, Astronomy Letters, 29, 783

\bibitem[\protect\citeauthoryear{{Turk}, {Smith}, {Oishi}, {Skory}, {Skillman},
  {Abel} \& {Norman}}{{Turk} et~al.}{2011}]{TurkSmith11}
{Turk} M.~J.,  {Smith} B.~D.,  {Oishi} J.~S.,  {Skory} S.,  {Skillman} S.~W.,
  {Abel} T.,    {Norman} M.~L.,  2011, \apjs, 192, 9

\bibitem[\protect\citeauthoryear{{Vazza}, {Br{\"u}ggen} \& {Gheller}}{{Vazza}
  et~al.}{2013}]{VazzaBruegg13}
{Vazza} F.,  {Br{\"u}ggen} M.,    {Gheller} C.,  2013, \mnras, 428, 2366

\bibitem[\protect\citeauthoryear{{Vazza}, {Brunetti}, {Gheller}, {Brunino} \&
  {Br{\"u}ggen}}{{Vazza} et~al.}{2011}]{VazzaBrun11}
{Vazza} F.,  {Brunetti} G.,  {Gheller} C.,  {Brunino} R.,    {Br{\"u}ggen} M.,
  2011, \aap, 529, A17

\bibitem[\protect\citeauthoryear{{Vogt} \& {En{\ss}lin}}{{Vogt} \&
  {En{\ss}lin}}{2005}]{VogtEns05}
{Vogt} C.,  {En{\ss}lin} T.~A.,  2005, \aap, 434, 67

\bibitem[\protect\citeauthoryear{{Xu}, {Li}, {Collins}, {Li} \& {Norman}}{{Xu}
  et~al.}{2009}]{XuLi09}
{Xu} H.,  {Li} H.,  {Collins} D.~C.,  {Li} S.,    {Norman} M.~L.,  2009, \apjl,
  698, L14

\bibitem[\protect\citeauthoryear{{Zhu}, {Feng} \& {Fang}}{{Zhu}
  et~al.}{2010}]{ZhuFeng10}
{Zhu} W.,  {Feng} L.,    {Fang} L.,  2010, \apj, 712, 1

\end{thebibliography}

% Details of numerics that are NOT covered by the Nyx code paper
%------------------------------------------------------------------------

\appendix

\section{Numerical algorithms}
\label{sc:algo}

\subsection{Time integration}
\label{sc:time_integr}

For adiabatic gas dynamics with $\gamma = 5/3$, the system of PDEs~(\ref{eq:momt_les})--(\ref{eq:k_les}) 
for the state $\Qb = (\rho, a \rho \Ub, a^2 \rho E, a^2 \rho e, a^2 \rho K)$ can be written
as\footnote{In contrast to \citet{AlmBell12}, we find it convenient to use the symbol
	$\Ub$ for the velocity vector, while the state vector is denoted by $\Qb$ in this paper.}
\begin{equation}
	\frac{\partial\Qb}{\partial t} = -\nabla\cdot\Fb + S_e + S_g + S_{\rm sgs}\,.
\end{equation}
Here, 
\[
	\Fb = \left(\rho \Ub/a,\; \rho \Ub\otimes\Ub,\; a (\rho \Ub E + p \Ub),\; a \rho \Ub e,\; a\rho \Ub K\right)
\]
are the advective fluxes due to the numerically resolved flow and
\begin{align*}
	S_e = (& 0,\; 0,\; 0,\; -a p \nabla \cdot \Ub,\;0, \\
	S_g = (& 0,\; \rho \gb,\; a \rho \Ub \cdot \gb,\; 0,\; 0), \\
	S_{\rm sgs} = (& 0,\; \nabla \cdot \boldsymbol{\tau},\; 
		a [\nabla \cdot (\Ub\cdot\boldsymbol{\tau}) - (\Sigma - \rho \varepsilon)],\;
		a\rho \varepsilon,\; \\
		&a [\nabla \cdot \left(\rho \kappa_{\rm sgs}\nabla K\right) + (\Sigma - \rho \varepsilon)]),
\end{align*}
represent, respectively, adiabatic heating ($d=\nabla \cdot \Ub < 0$)
or cooling ($d > 0$), the gravitational sources and the SGS sources.
For the homogenous part of the equations, the unsplit Godunov solver of \nyx\ with piecewise
linear \citep{colella1990,saltzman} or piecewise parabolic \citep{miller-colella} reconstruction is applied.
For brevity, we refer to these methods as PLM and PPM, where it is understood that in this article PPM differs form the
standard PPM method with directional splitting. Both reconstruction methods include reference states,
which is important for stability. 
Moreover, the Riemann solver features a two-shock approximation
to robustly treatment hypersonic flows \citep{ColGlaz85}, 
which occur in the voids in cosmological simulations.
The source terms are added with the predictor-corrector framework of \nyx\ 
(see Section~3.6 of \citealt{AlmBell12}), where
flux-like sources such as $\nabla \cdot \boldsymbol{\tau}$ are discretized as pseudo fluxes 
with face-centered finite differences. 

If $K$ becomes negative in the intermediate state $\Qb^{n+1,\ast}$ that results from the
predictor step or in the corrected state $\Qb^{n+1}$, the following
negative-energy correction is executed. We impose a lower bound $K_{\rm min}$ and 
define
\begin{align}
	\Delta (\rho K) &= \max\left[\rho (K_{\rm min}  - K),\, 0\right], \\
	\Delta (\rho K)_{\rm kin} &= \min\left[\Delta (\rho K),\, \rho(E-e)\right].
\end{align}
The corrected energy variables are given by
\begin{align}
	(\rho K)' &= \rho K + \Delta (\rho K) \\
	(\rho E)' &= \rho E - \Delta (\rho K) \\
	(\rho e)' &= \max\left[\rho e - \Delta (\rho K) + \Delta (\rho K)_{\rm kin},\, \rho e_{\min}\right]
\end{align}
If enough kinetic energy is available, i.~e., $\Delta (\rho K) < \rho(E-e)$, 
we have $\Delta (\rho K)_{\rm kin} = \Delta (\rho K)$. 
In this case, the internal energy remains unaffected. Otherwise, the undershoot of $\rho K$ is
partially compensated by internal energy. In very rare cases, a non-conservative
cutoff at $e_{\min}$ has to be applied, where $e_{\min}$ corresponds to the minimum temperature $T_{\min}$ 
in \nyx\ (for adiabatic cluster simulations without radiative heating, we use $T_{\min}=10\,\mathrm{K}$). 
To restore the consistency of the kinetic energy, the velocities are adjusted by
\begin{equation}
	\Ub^\prime = \beta\Ub\quad\mbox{where}\ \beta = \sqrt{1 - \frac{2\Delta (\rho K)_{\rm kin}}{\rho U^2}}.
\end{equation}
Since 
\[
	(\rho e)' +\frac{1}{2}\rho U^{\prime\,2} + (\rho K)' = \rho e+\frac{1}{2}\rho U^2 + \rho K,
\]
the sum of resolved and unresolved energies is preserved by the
correction except if the internal energy floor is reached. 
We also adjust the production rate $\Sigma$ for data output 
or statistics:
\[
	\Sigma^{\,\prime} = \Sigma +  \Delta (\rho K) / \Delta t.
\]

\subsection{Interpolation and restriction}
\label{sc:refine}

Creation of new fine grid cells from coarse data and the construction of boundary data 
for fine grids require data to be interpolated from coarse resolution (lower refinement level)
to fine resolution (higher refinement level). This entails differences in the kinetic energy between 
the refinement levels, which have to be compensated. The inverse problem arises when the fine-grid
data are projected to the coarser grids via conservative
averaging. This operation is called restriction. 
In this section, we describe two possible approaches for mapping between levels: 
the fully turbulent energy compensation (FTEC), and the constrained turbulent energy
compensation (CTEC).

\subsubsection{Fully turbulent energy compensation}
\label{sc:ftec}

We denote the values of the state variables in the coarse grid cells by $\rho_{\rm crs}$, $(\rho\Ub)_{\rm crs}$, 
$(\rho K)_{\rm crs}$, etc. For grid refinement, the values obtained by interpolation from the coarse grid to the
fine grid are denoted by $\rho_{n}^{\ast}$, $(\rho\Ub)_{n}^{\ast}$, $(\rho K_{n})^{\ast}$, etc., 
where $n$ runs from 1 to 8 for a refinement factor $r=2$ or from 1 to 64 for $r=4$. The 
interpolated values of the SGS turbulence energy satisfy
\begin{equation}
	\label{eq:rho_K_crs}
	 \frac{1}{N}\sum_{n}(\rho K)_n^{\ast} = (\rho K)_{\rm crs}\,.
\end{equation}
The difference between the kinetic energy computed from the interpolated momenta and
the kinetic energy on the coarse grid is 
\begin{equation}
	\label{eq:delta_rhoK}
	\Delta(\overline{\rho K}) =
	\frac{1}{N}\sum_{n}\frac{1}{2}\frac{|(\rho\Ub)_{n}^{\ast}|^2}{\rho_{n}^{\ast}} -
	\frac{1}{2}\frac{(\rho U)_{\rm crs}^2}{\rho_{\rm crs}}\,.
\end{equation}
To maintain energy conservation, we set
\begin{equation}
	\label{eq:energy_sgs_refined}
	(\rho K)_n = (\rho K)_n^{\ast} - \Delta(\rho K)_n\,,
\end{equation}
%\end{equation}
where
\[
	\Delta(\rho K)_n = \frac{(\rho K)_n^{\ast}}{(\rho K)_{\rm crs}}\,\Delta(\overline{\rho K})\,.
\]
From the above expression and equation~(\ref{eq:rho_K_crs}), it immediately follows that 
\begin{equation}
	\label{eq:correct_tot}
	\frac{1}{N}\sum_{n}\Delta(\rho K)_n = \Delta(\overline{\rho K})
\end{equation}
and, hence, summation of equation~(\ref{eq:energy_sgs_refined}) over all $i$ yields equation~(\ref{eq:rho_K_balance}).

However, to preserve the positivity of the SGS turbulence energy, we require
$(\rho K)_n>\rho K_{\rm min}$ for all $i$. From this constraint
follows the maximal energy difference
\begin{equation}
	\label{eq:delta_energy_max_def}
	\Delta(\overline{\rho K})_{\mathrm{max}} = 
	\min_n\left(\frac{(\rho K)_{\rm crs}}{(\rho K)_n^{\ast}}\left[(\rho K)_n^{\ast} - \rho K_{\rm min}\right]\right).
\end{equation}
If $\Delta(\overline{\rho K})$ given by equation~(\ref{eq:delta_rhoK}) exceeds the above maximum,
the interpolated momenta have to be corrected so that the energy difference between the
finer and coarser levels becomes sufficiently small. Otherwise, we set $\Ub_{n}=\Ub_{n}^{\ast}$.
For the correction, the generalized expression 
\begin{equation}
	\label{eq:vel_refined_corr}
	(\rho\Ub)_n = (\rho\Ub)_{\rm crs} + \alpha\Delta(\rho\Ub)_n\,,
\end{equation}
where
\[
  \quad \Delta(\rho\Ub)_n=(\rho\Ub)_n^{\ast} - (\rho\Ub)_{\rm crs}
\]
is used, with $\alpha\in[0,1]$. 
The case without correction of the interpolated momenta, $\Ub_{n}=\Ub_{n}^{\ast}$, 
corresponds to $\alpha = 1$. 
Since 
\begin{equation}
	\frac{1}{N}\sum_{n}\Delta(\rho\Ub)_n = 0
\end{equation}
we have $\overline{\rho\Ub}=(\rho\Ub)_{\rm crs}$, i.~e., momentum is conserved. The values of the total
resolved energy in the fine-grid cells then have to be set to
\begin{equation}
	(\rho E)_n = (\rho e)_n + \frac{1}{2}\rho_n U_n^2. 
\end{equation}
 
To calculate the corrected momenta $(\rho\Ub)_n$, 
we need to determine $\alpha$ from equation~(\ref{eq:delta_energy}) with
$\Delta(\overline{\rho K})$ replaced by $\Delta(\overline{\rho K})_{\mathrm{max}}$,
where $\Delta(\overline{\rho K})_{\mathrm{max}}$ is given by equation~(\ref{eq:delta_energy_max_def}).
Consequently, $\alpha$ must be chosen such that
\begin{equation}
\begin{split}
	\label{eq:delta_energy_max}
	\Delta(\overline{\rho K})_{\mathrm{max}} = &
	\frac{1}{2}\left(\frac{1}{N}\sum_{n}\rho_{n}U_{n}^2 -
	\frac{(\rho U)_{\rm crs}^2}{\rho_{\rm crs}}\right) \\
	=&\frac{1}{2}\left(\frac{1}{N}\sum_{n}
	\frac{|(\rho\Ub)_{\rm crs} + \alpha\Delta(\rho\Ub)_n|^2}{\rho_{n}} -
	\frac{(\rho U)_{\rm crs}^2}{\rho_{\rm crs}}\right).
\end{split}
\end{equation}
This is a quadratic equation in $\alpha$,
\begin{equation}
\label{eq:alpha}
A \alpha^{2} + B \alpha + C = 0,
\end{equation}
where
\begin{align*}
A &= \frac{1}{2N}\sum_{n}\frac{|\Delta(\rho\Ub)_n|^2}{\rho_{n}}, \\
B &= \frac{1}{N}\sum_{n}\frac{(\rho\Ub)_{\rm crs}\cdot\Delta(\rho\Ub)_n}{\rho_{n}}, \\
C &= \frac{1}{2}\left(\frac{1}{N}\sum_{n}\frac{1}{\rho_{n}} -
	\frac{1}{\rho_{\rm crs}}\right)(\rho U)_{\rm crs}^2 -
	\Delta(\overline{\rho K})_{\mathrm{max}}\,.
\end{align*}
A real solution exists if the discriminant $D = B^{2} - 4AC\ge 0$. For a constant mass density, we have $B=0$, 
because summing up $\Delta(\rho\Ub)_n$ for all $i$ yields zero. In this case, $D=4A\Delta(\overline{\rho K})_{\mathrm{max}}$ 
is always non-negative, because $A\ge 0$ and $\Delta(\overline{\rho K})_{\mathrm{max}}\ge 0$. For compressible flows with 
strong small-scale fluctuations of the mass density, however, the discriminant $D$ can become negative. 

If a real solution in the range $0\le\alpha\le 1$ does not exist (typically, this is the case for a few
percent of the corrections), the interpolated momenta are simply reduced by a factor $\beta$:
\begin{equation}
	(\rho\Ub)_n = \beta(\rho\Ub)_{n}^{\ast},
\end{equation}
where
\[
	\beta^2 = \left(\frac{1}{2}\frac{(\rho U)_{\rm crs}^2}{\rho_{\rm crs}} + \Delta (\overline{\rho K})_{\mathrm{max}}\right)
	\left(\frac{1}{N}\sum_{n}\frac{1}{2}\frac{|(\rho\Ub)_{n}^{\ast}|^2}{\rho_{n}^{\ast}}\right)^{-1}.
\]
With this definition and $\overline{\rho K}=(\rho K)_{\rm crs} - \Delta(\overline{\rho K})_{\mathrm{max}}$, the energy balance equation~(\ref{eq:energy_balance}) is fulfilled. 

To complete the AMR algorithm, we have to consider restrictions from a fine grid to the covering coarse grid 
(coarse-grid values are overwritten by the data from higher refinement levels). The
values obtained by conservative averaging of the fine-grid values are $\overline{\rho}$, $\overline{\rho\Ub}$, $\overline(\rho K)$,
etc. The energy compensation is now simply given by equation~(\ref{eq:rho_K_balance}), which is the inverse operation
to equation~(\ref{eq:energy_sgs_refined}).

\subsubsection{Constrained turbulent energy compensation}
\label{sc:ctec}

As motivated in Section~\ref{sc:consrv}, a difficulty arises if there is a large non-turbulent component of the flow.
In this case, the energy difference given by equation~(\ref{eq:delta_rhoK}) does not correspond to the difference in the
turbulent energies when the cutoff scale is shifted from $\Delta_l$ to $\Delta_l+1=\Delta_l/r$. To estimate the fraction 
of energy that is due to the turbulent cascade, we use power-law scaling as a proxy for equation~(\ref{eq:delta_rhoK}):
\begin{equation}
	\label{eq:delta_rhoK_ctec}
	\Delta(\overline{\rho K}) = \left(1-r^{-2\eta}\right)(\rho K)_{\rm crs}\,.
\end{equation}
For Kolmogorov scaling, $\eta=1/3$. This definition constrains the amount of energy that can be shifted from the
resolved to the SGS budget. 

Now, energy conservation can only be fulfilled if $\Delta(\overline{\rho K})$
is complemented by a correction of the internal energy. Thus, we set 
\begin{align}
	\label{eq:energy_sgs_refined_ctec}
	(\rho K)_n &= (\rho K)_n^{\ast} - \frac{(\rho K)_n^{\ast}}{(\rho K)_{\rm crs}}\,\Delta(\overline{\rho K})
	=r^{-2\eta}(\rho K)_n^{\ast}\,,\\
	\label{eq:energy_int_refined_ctec}
	(\rho e)_n &= (\rho e)_n^{\ast} - \frac{(\rho e)_n^{\ast}}{(\rho e)_{\rm crs}}\,\Delta(\overline{\rho e})\,,
\end{align}
where
\begin{equation}
	\label{eq:delta_rhoe_ctec}
	\Delta(\overline{\rho e}) = 
	\frac{1}{N}\sum_{n}\frac{1}{2}\frac{|(\rho\Ub)_{n}^{\ast}|^2}{\rho_{n}^{\ast}} -
	\frac{1}{2}\frac{(\rho U)_{\rm crs}^2}{\rho_{\rm crs}}
	-\Delta(\overline{\rho K})\,.
\end{equation}
One can immediately see that equation~(\ref{eq:delta_rhoK}) corresponds to $\Delta(\overline{\rho e})=0$ (FTEC). 
Equation~(\ref{eq:energy_sgs_refined_ctec})
is formally the same as the correction rule for the SGS turbulence energy in \citet{MaierIap09}. Our method, however, is fully consistent
\emph{without} further changing the fine-cell momenta $(\rho\Ub)_n^{\ast}$ because we interpret $\Delta(\overline{\rho K})$ as the
turbulent fraction of the inter-level energy difference given by the first two terms on the right-hand side of equation~(\ref{eq:delta_rhoe_ctec}). 
The non-turbulent fraction then gives rise to the residual $\Delta(\overline{\rho e})$.
In contrast to \citet{MaierIap09}, the sum of resolved and SGS energies as well as the momentum are conserved with our approach. 

To avoid negative values of $(\rho K)_n$ or $(\rho e)_n$, we supplement the maximum SGS energy
difference (equation~\ref{eq:delta_energy_max_def}) by an upper bound on the internal energy correction,
\begin{equation}
	\Delta(\overline{\rho e})_{\mathrm{max}} = 
	\min_n\left(\frac{(\rho e)_{\rm crs}}{(\rho K)_n^{\ast}}\left[(\rho e)_n^{\ast} - \rho e_{\rm min}\right]\right)\,
\end{equation}
and apply the negative-energy correction algorithms explained in Sect~\ref{sc:ftec} with
\[
	\min\left[\Delta(\overline{\rho K})_{\mathrm{max}},\Delta(\overline{\rho K})\right] +
	\min\left[\Delta(\overline{\rho e})_{\mathrm{max}},\Delta(\overline{\rho e})\right]
\]
as the total energy difference on the left-hand side of equation~(\ref{eq:delta_energy_max}).

For projections from a fine grid to the covering coarse grid, the SGS turbulence energy difference
is given by
\begin{equation}
\begin{split}
	\label{eq:delta_rhoK_ctec_proj}
	\Delta&(\overline{\rho K}) =\\ 
	&\min\left[\left(r^{2\eta}-1\right)\overline{\rho K},\,
	\frac{1}{N}\sum_{n}\frac{1}{2}\frac{|(\rho\vecU_{n})|^2}{\rho_{n}} -
	\frac{1}{2}\frac{(\rho U)_{\rm crs}^2}{\rho_{\rm crs}}\right].
\end{split}	
\end{equation}
This definition ensures $\Delta(\overline{\rho e})>0$ for projections. Without the cutoff
at the full kinetic energy difference, overshoots of the power-law estimate can
can result in unphysical reductions of the internal energy in some cases.
To obtain consistent energy variables in the coarse-grid cells, we set
\begin{align}
	\label{eq:energy_int_crse_ctec}
	(\rho K)_{\rm crs} &= \overline{\rho K} + \Delta(\overline{\rho K})\, ,\\
	(\rho e)_{\rm crs} &= \overline{\rho e} + \Delta(\overline{\rho e})\, ,\\
	(\rho E)_{\rm crs} &= \overline{\rho E} - \Delta(\overline{\rho K})\, ,
\end{align}
where $\Delta(\overline{\rho e})$ is given by equation~(\ref{eq:delta_rhoe_ctec}). By substituting
equation~(\ref{eq:delta_rhoK_ctec_proj}) into the first equation, we obtain
$(\rho K)_{\rm crs}=r^{2\eta}\,\overline{\rho K}$, which is the inverse of 
equation~(\ref{eq:energy_sgs_refined_ctec}) summed over all $i$.

\begin{figure}
\centering
  \includegraphics[width=\linewidth]{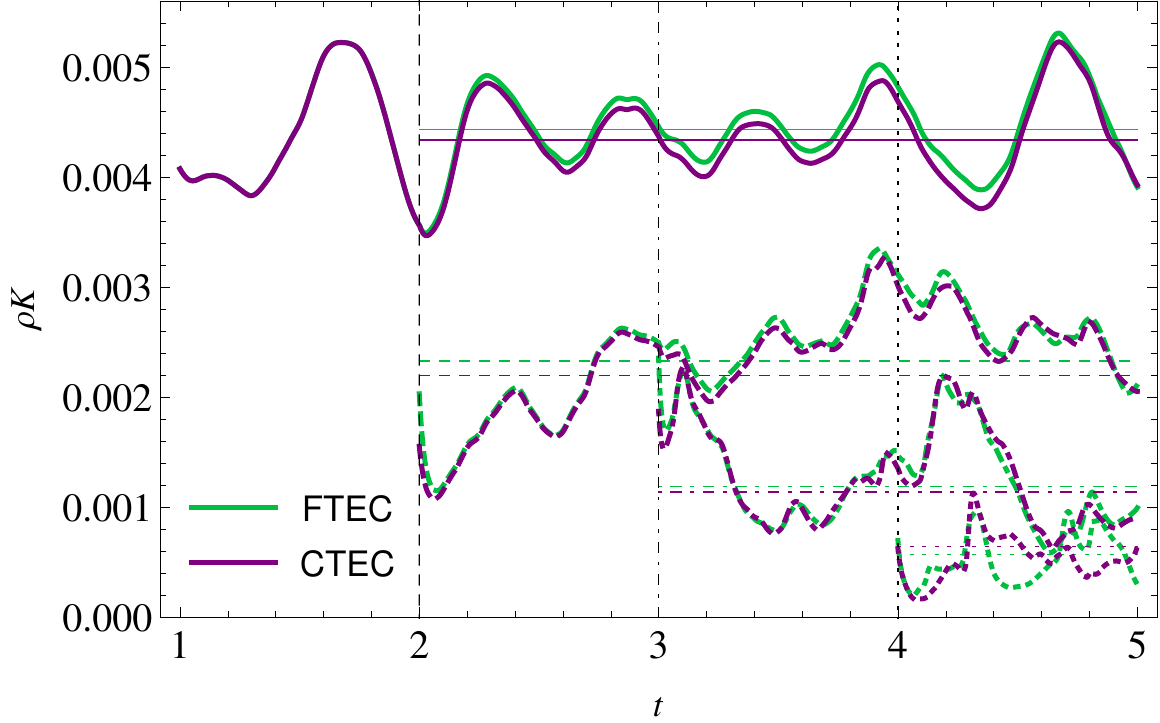}
\caption{Level-wise mean values of the SGS turbulence energy $\rho K$ as functions of time for 
	LES of forced isothermal turbulence ($\mathcal{M}_{\rm rms}\approx 2.0$)
    	with two different refinement methods. The thin horizontal lines indicated the time averages
	in the statistically stationary regime. The vertical lines indicate the insertion of nested refined grids with $1/2$,
  	$1/2$ and $1/8$ of the root-grid cell size at $t=2.0$, $3.0$, and $4.0$, respectively.}
\label{fig:rhoK_FvsC}
\end{figure}

\subsubsection{Comparison of FTEC and CTEC}

\begin{figure*}
  \begin{center}
    \mbox{\subfigure[FTEC, all levels]{\includegraphics[width=0.49\linewidth]{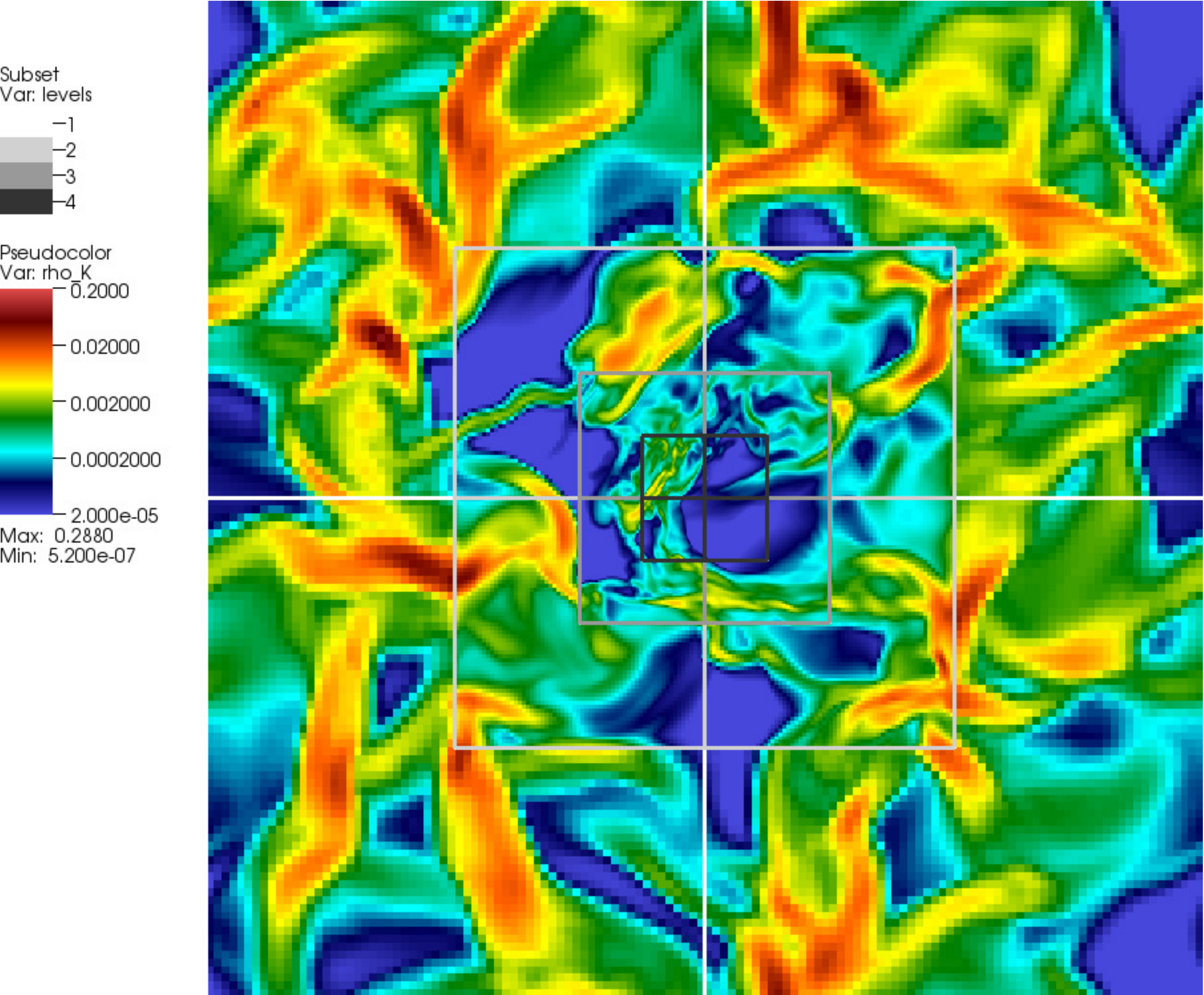}}\qquad
          \subfigure[FTEC, root grid]{\includegraphics[width=0.49\linewidth]{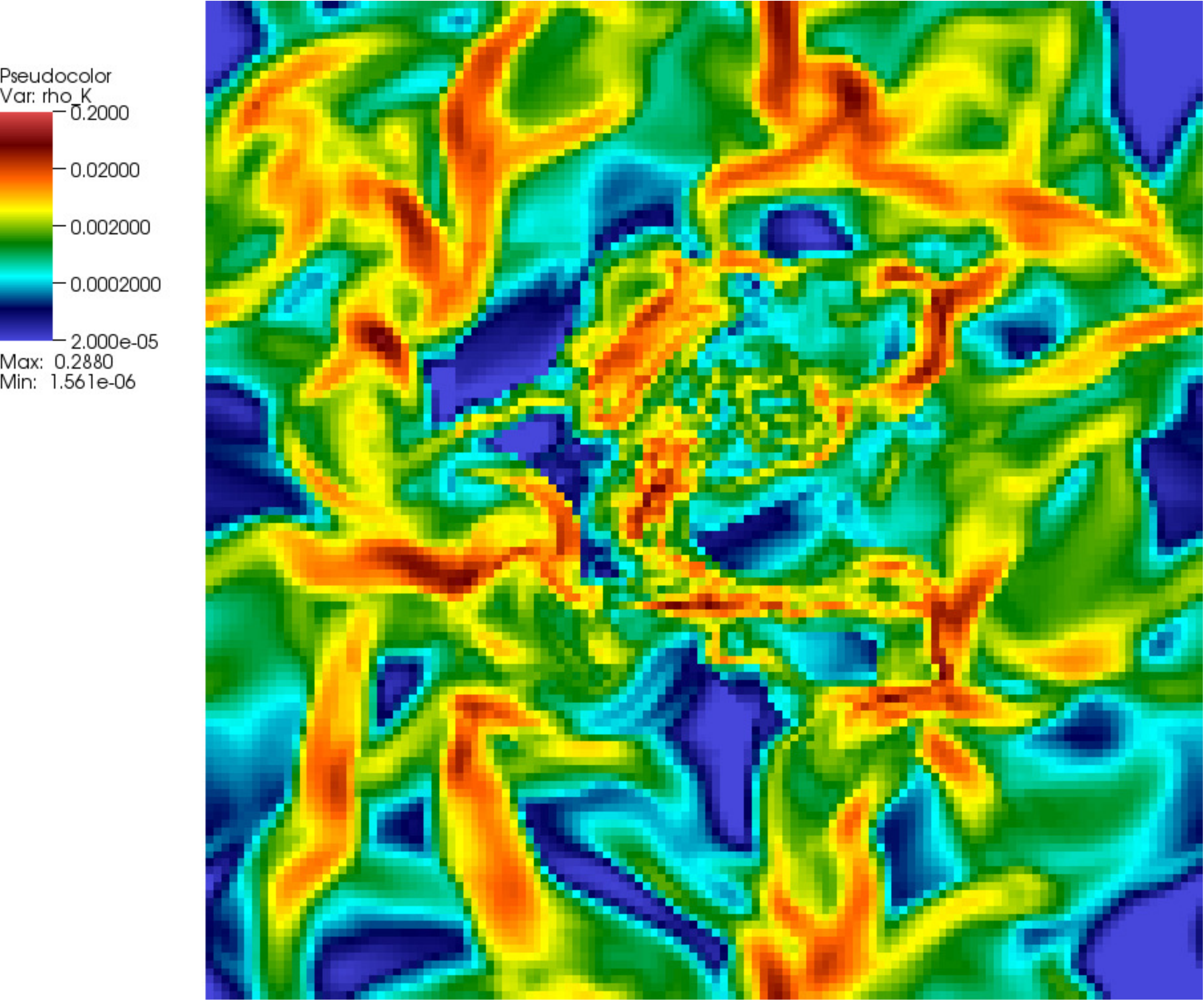}}}
    \mbox{\subfigure[CTEC, all levels]{\includegraphics[width=0.49\linewidth]{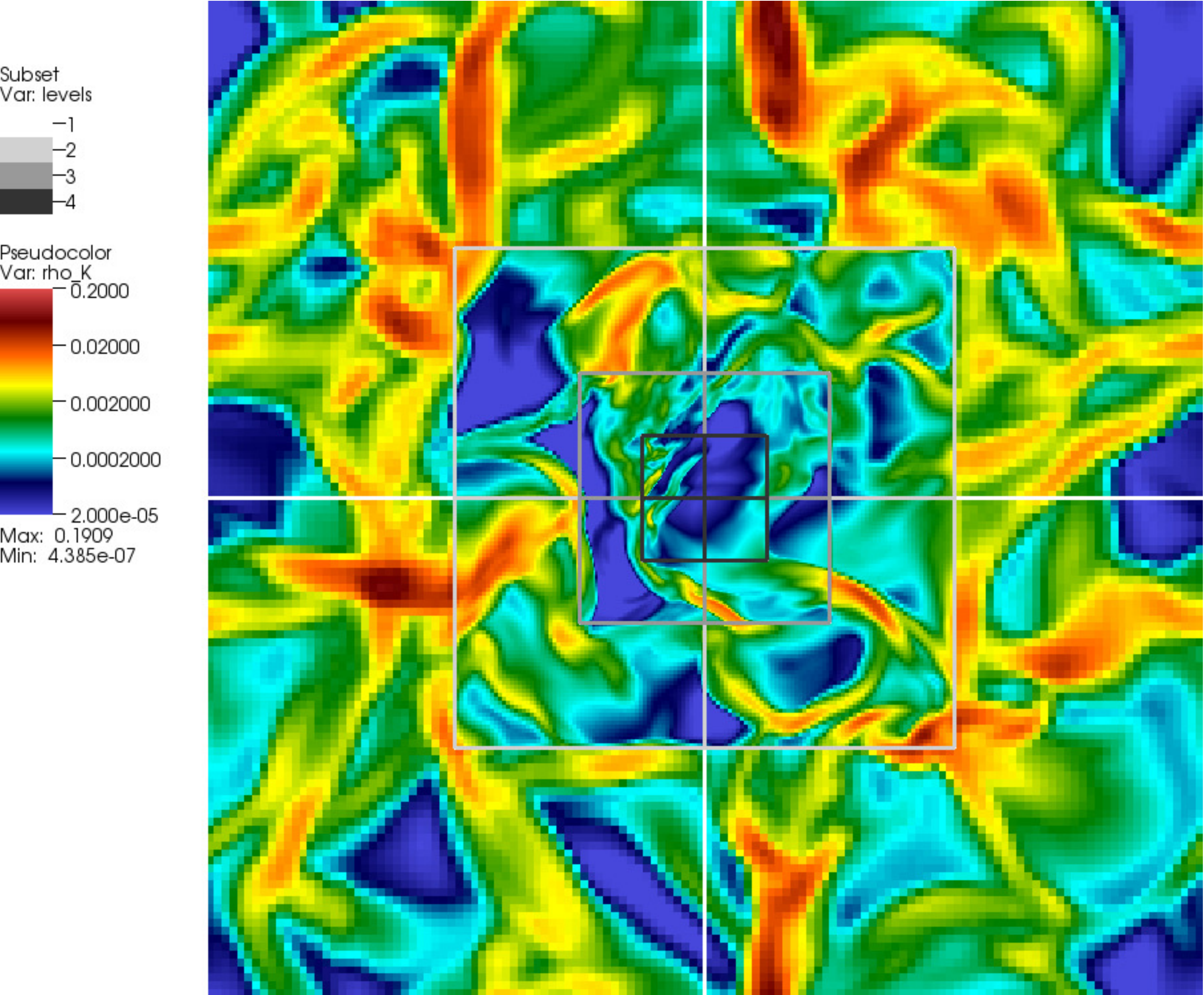}}\qquad
          \subfigure[CTEC, root grid]{\includegraphics[width=0.49\linewidth]{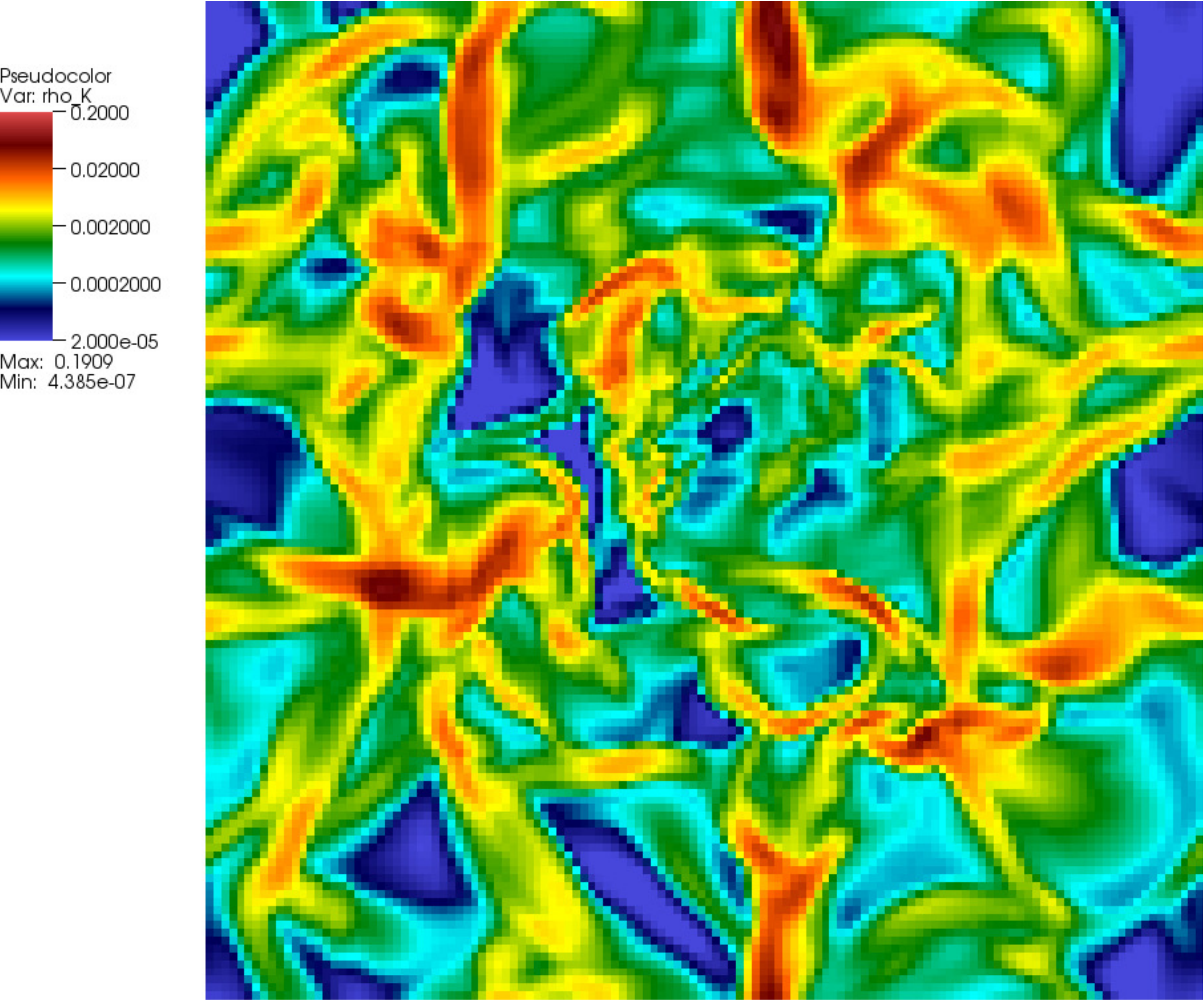}}}
    \caption{Slices of $\rho K$ in LES of forced isothermal turbulence ($\mathcal{M}_{\rm rms}\approx 2.3$)
    		with three nested-grid levels at time $t/T=5.0$, where the SGS turbulence energy
    		differences between levels are deterimened by equations~(\ref{eq:delta_rhoK}) 
    		and~(\ref{eq:delta_rhoK_ctec}) for the top and bottom plots,
		respectively. The grey-shaded lines in the plots on the left show
		the boundaries of refined grid boxes. The plots on the right show the root-grid representation, 
		where data from higher refinement levels are successively projected to coarser levels. }
    \label{fig:isoth_slices_FvsC}
  \end{center}
\end{figure*}

We compare the alternative interpolation strategies for a turbulent box simulation. If the underlying assumption of CTEC
is correct, we should see about the same statistical behavior of the SGS turbulence energy as in the FTEC case.
Fig.~\ref{fig:rhoK_FvsC} shows that this is indeed the case for the particular example of driven isothermal turbulence with 
$\mathcal{M}_{\rm rms}\approx 2.3$ (see Section~\ref{sc:turb_box}). The averages of $\rho K$ at the different refinement levels are 
quite close and also the initial values when new nested grids are inserted are comparable.\footnote{For the 
creation of the second refinement level at time $t=3.0$, there is only a small energy difference 
(for FTEC, the initial energy is even higher than the average at the first level).
However, this is merely coincidence that can be attributed to the intermittency of turbulence. 
At this instance, turbulent fluctuations in the region that is refined to the second level 
just happen to be strong in comparison to typical magnitude.}
The slices of $\rho K$ plotted in Fig.~\ref{fig:isoth_slices_FvsC} demonstrate that coarsening from the finer grids to the
root grid yields fields that are nearly smooth across the boundaries between the refinement levels in both cases. 
However, one can see enhanced small-scale structure even in the root-grid representations because of the reduced
numerical dissipation in the refined regions. Although the large-scale structure, which is imprinted by the forcing, 
is roughly comparable, it is also clear that the flow realizations differ. This is a natural consequence of the 
non-linearity of the system, which amplifies even small deviations such as those introduced by the different refinement techniques. 
In conclusion, the results suggest that consistent statistics are obtained for statistically homogeneous and stationary turbulence
both with FTEC and CTEC.

\begin{figure*}
  \begin{center}
    \mbox{\subfigure[FTEC, all levels]{\includegraphics[width=0.49\linewidth]{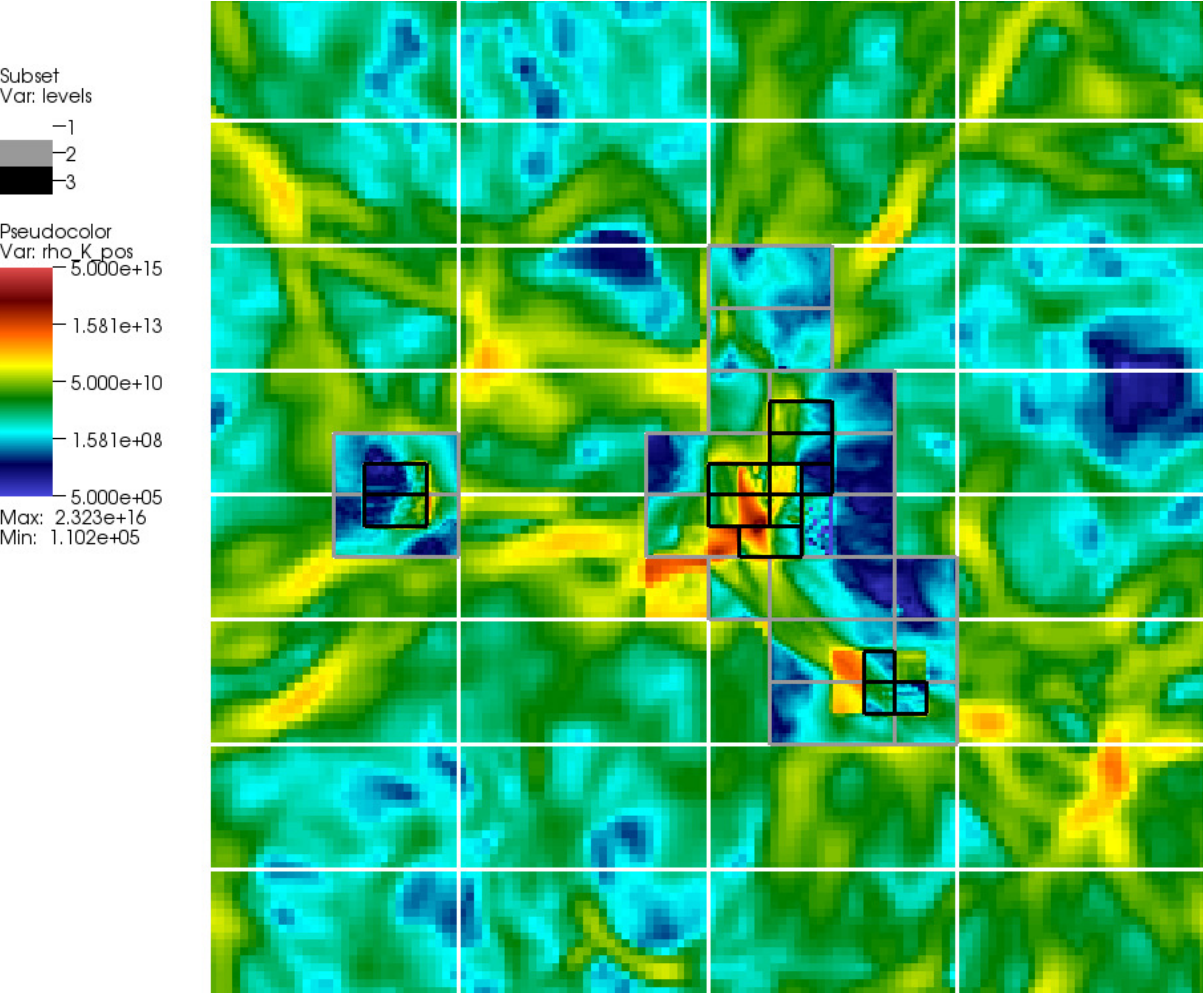}}\qquad
          \subfigure[FTEC, root grid]{\includegraphics[width=0.49\linewidth]{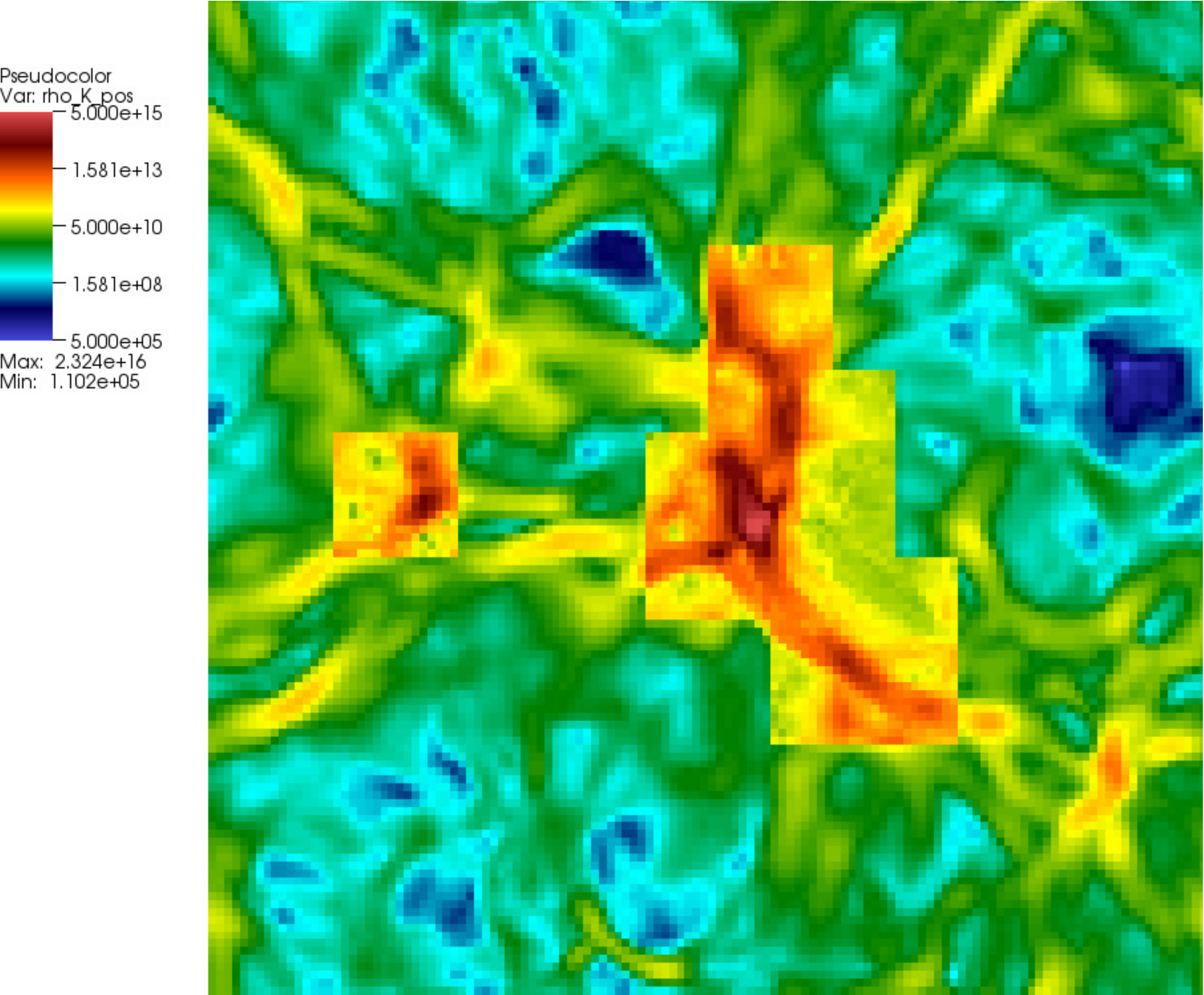}}}
    \mbox{\subfigure[CTEC, all levels]{\includegraphics[width=0.49\linewidth]{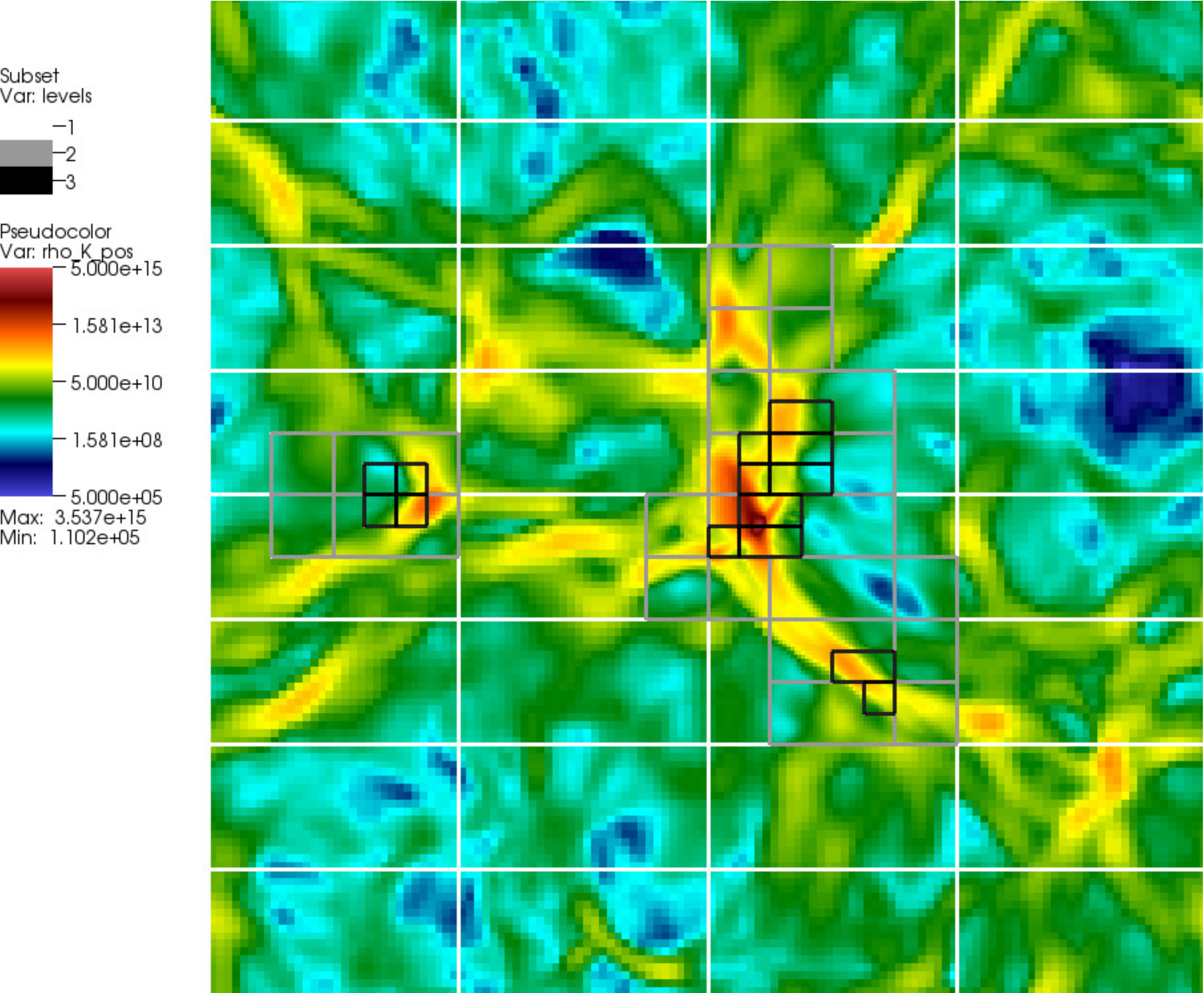}}\qquad
          \subfigure[CTEC, root grid]{\includegraphics[width=0.49\linewidth]{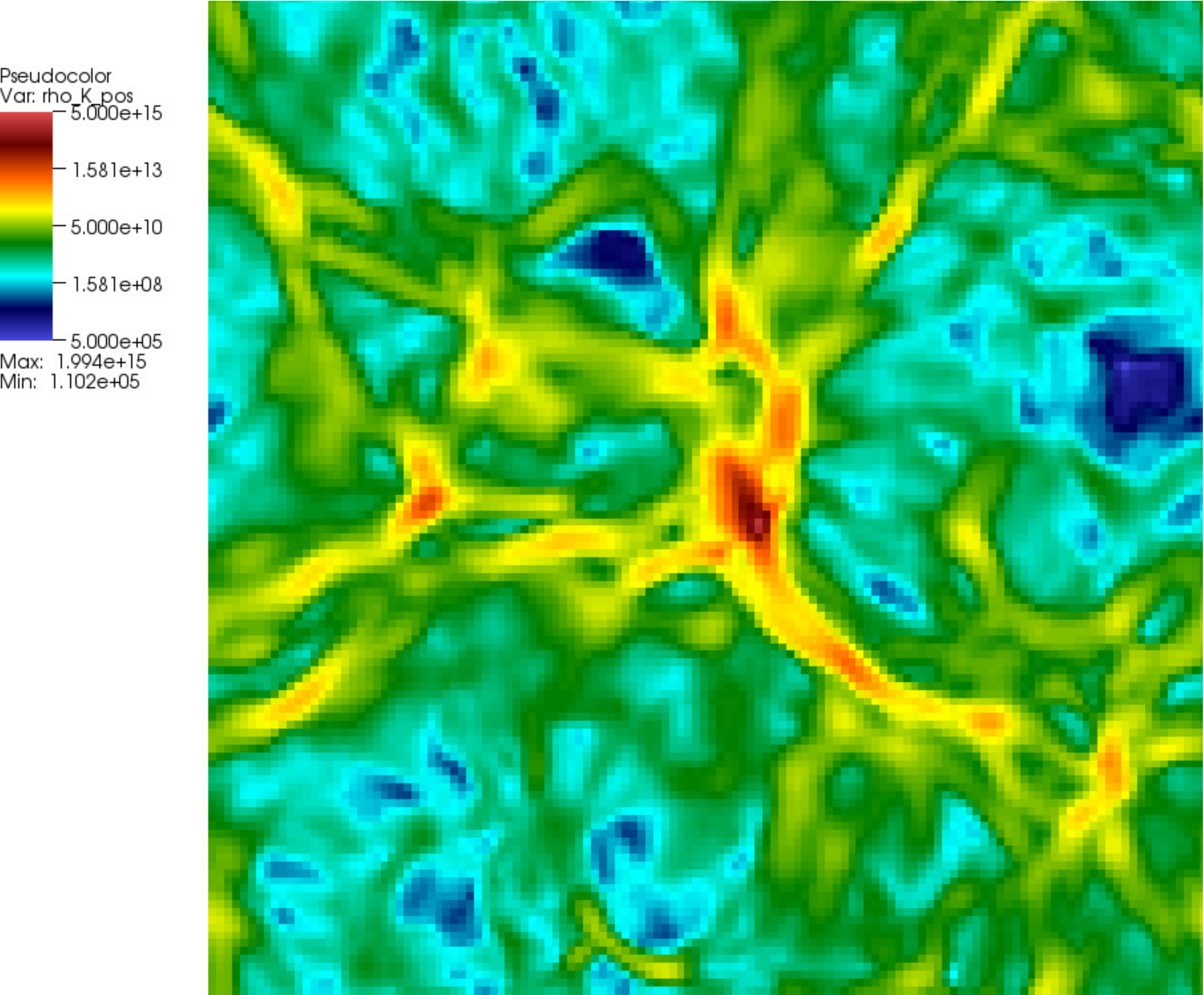}}}
    \caption{Slices of $\rho K$ in LES of the Santa Barbara cluster with two 
	  levels of refinement at $z=5.0$, as in Fig.~(\ref{fig:isoth_slices_FvsC}).}
    \label{fig:SB_slices_FvsC}
  \end{center}
\end{figure*}

\begin{figure}
\centering
  \includegraphics[width=\linewidth]{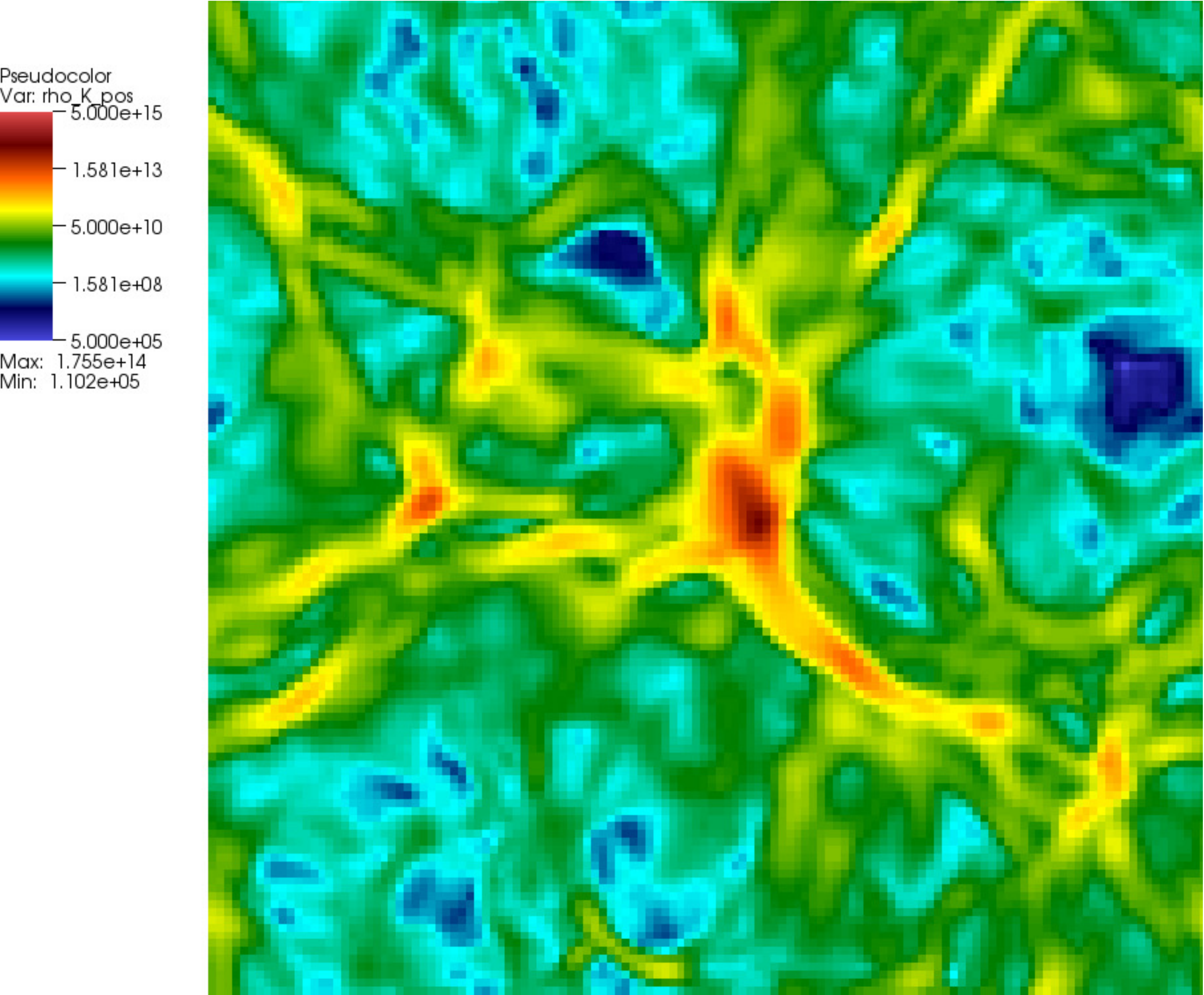}
\caption{Slices of $\rho K$ in an LES of the Santa Barbara cluster on a uniform grid ($z=5.0$).}
\label{fig:rhoK_SB_slice_uniform}
\end{figure}

For LES of the Santa Barbara cluster (see Section~\ref{sc:sb_sgs}), however, severe problems become apparent if FTEC is used. 
In Fig.~\ref{fig:SB_slices_FvsC}, slices of the SGS turbulence energy are shown for the two methods in simulations with
a $128^3$ root grid and two levels of refinement. Particularly at the boundaries of the first refinement level, an extreme drop 
of the SGS turbulence energy over several orders of magnitude can be seen. Apart from that, spurious fluctuations can be seen in the
refined regions, which have no relation to the resolved flow at all. The root-grid representation, on the other hand,
reveals that the projected SGS turbulence energy values in refined regions are systematically too high compared to the surroundings.
This is immediately clear by comparing to Fig.~\ref{fig:rhoK_SB_slice_uniform}, which shows the result for a uniform-grid simulation 
with the same resolution as the root grid in the AMR simulation. What is wrong with FTEC in this case? The problem is 
that the dominant component of the velocity field is non-turbulent gravity-driven flow, especially at high redshift. 
Nevertheless, there are significant gradients associated with this flow, which give rise to large inter-level energy differences, 
while the SGS turbulence energy is still very low. Consequently, the changes in the SGS turbulence energy resulting from
interpolations or projections are far too large if equation~(\ref{eq:delta_rhoK}) is applied. As demonstrated by the
bottom panels in Fig.~\ref{fig:SB_slices_FvsC}, the discrepancy between the unresolved and resolved energy
variables is resolved by CTEC, where it is assumed that the SGS turbulence energy follows the scaling law of the turbulent 
cascade (equation~(\ref{eq:delta_rhoe_ctec}). Of course, this is a rather crude assumption in regions where turbulence
is building up and strong anisotropies are present, but the results obtained with CTEC are much more consistent:
the root-grid representation (bottom right panel in Fig.~\ref{fig:SB_slices_FvsC}) shows no systematic deviation of the
projected SGS turbulence energy in refined regions in comparison to the uniform-grid field (Fig.~\ref{fig:rhoK_SB_slice_uniform}).

\subsection{Shear-improved model}
\label{sc:kalman}

The basic idea of a shear-improved SGS model for inhomogeneous flow is that the turbulent stresses should be given by 
the shear of the velocity fluctuations relative to the mean flow. In the case of statistically homogenous turbulence,
the mean flow vanishes. For the eddy-viscosity closure applied in this work, the shear-improved SGS turbulence stress tensor 
is defined by equation~(\ref{eq:tau_si}). The problem, then, is to compute the mean flow. In LES, the only feasible option
is to approximate the mean flow in each cell by applying a temporal low-pass filter. \citet{LevTosch07} use a simple exponential
smoothing algorithm with a given time scale that separates slow changes associated with the mean flow from fast oscillations.
They show that LES of plane-channel flows with the shear-improved Smagorinsky model and exponential smoothing reproduces 
the results from direct numerical simulations very well. A more sophisticated smoothing algorithm that has the
capability to adapt to an unsteady mean flow is applied by \citet{CahuBou10}. This so-called Kalman filtering technique performs
well in LES of turbulence produced by the flow past a cylinder \citep{CahuBou11}, which is a standard test case.

The Kalman filter can be regarded as an adaptive exponential filter, where estimates of the mean-flow 
$[\Ub]$ at two consecutive time steps are related by:
\begin{equation}
	\label{eq:exp_smooth}
	[U_i]^{(n+1)} = \left(1-\alpha_i^{(n+1)}\right)[U_i]^{(n)} + \alpha_i^{(n+1)}U_i^{(n+1)}.
\end{equation}
For a simple exponential filter, 
\[
	\alpha_i^{(n)} = \frac{2\pi\Delta t^{(n)}}{\sqrt{3}\,T_{\rm c}}\,.
\]
Changes occurring on time scales smaller than the smoothing scale $T_{\rm c}$ are
suppressed in $[\Ub]$ (equivalently, Fourier modes of $\Ub$ above the cut-off 
frequency $f_{\rm c}=1/T_{\rm c}$ are suppressed; see \citealt{CahuBou10}). A fixed smoothing time scale is appropriate
for statistically inhomogeneous, but steady flows. The adaptive Kalman filter, on the other
hand, adjusts the weighting factor $\alpha_i^{(n)}$ to the evolution of the mean flow. This
is achieved by setting $\alpha_i^{(n)}$ equal to the Kalman gain $K_i^{(n)}$, which is defined by the
ratio between the error variance of the smoothed component and the total error variance, 
including the fluctuating component. Since the error variances have to be evaluated
at time $t^{(n+1)}$, a predictor-corrector scheme is used:
\begin{enumerate}
\item Given the error variance $P_i^{(n)}$ at time $t^{(n)}$, the prediction for $t^{(n+1)}$ is
	\begin{equation}
		P_i^{(n+1)*} = P_i^{(n)} + \sigma^{2\,(n)}_{\delta[U_i]}\,,
	\end{equation}
	where
	\[
		\sigma_{\delta[U_i]}^{(n)} = \frac{2\pi\Delta t^{(n)}}{\sqrt{3}\,T_{\rm c}}\,U_{\rm c}\,.
	\]
	Here it is assumed that the typical correction of the mean flow, $\delta[U_i]^{(n)}=[U_i]^{(n)}-[U_i]^{(n-1)}$,
	is of the order $2\pi\Delta t^{(n)}U_{\rm c}/(\sqrt{3}\,T_{\rm c})$.
\item The Kalman gain is then given by
	\begin{equation}
		\label{eq:kalman_gain}
		\alpha_i^{(n+1)} = K_i^{(n+1)} = \frac{P_i^{(n+1)*}}{P_i^{(n+1)*} + \sigma^{2\,(n)}_{\delta U_i}}\,,
	\end{equation}	
	where
	\[
		\label{eq:sigma_fluc}
		\sigma^{2\,(n)}_{\delta U_i}= 
		\max\left(\left|\delta U_i^{(n)}\right|,0.1U_{\rm c}\right)U_{\rm c}
	\]	
	is the contribution of the fluctuating component $\delta U_i^{(n)}\equiv U_i^{\prime\,(n)}=U_i^{(n)}-[U_i]^{(n)}$ 
	to the error variance.
	The lower bound on $\sigma^{2\,(n)}_{\delta U_i}$ is necessary to obtain non-vanishing
	fluctuations from an initially smooth flow with $[U_i]=U_i$. 
\item The corrected error variance for the next step is given by
	\begin{equation}
		P_i^{(n+1)} = \left(1 - K_i^{(n+1)}\right)P_i^{(n+1)*}\,.
	\end{equation}
\end{enumerate}
In a statistically stationary state, the velocity fluctuations should be of the order 
$\sigma^{(n)}_{\delta U_i}\simeq U_{\rm c}$. In this case the Kalman filter corresponds to simple exponential smoothing. 

Whenever new grid cells are created, the Kalman filtering is initialized with
\[
	[U_i]^{(0)}=U_i^{(0)}\quad\mbox{and}\quad
	P^{(0)}=\sigma^{2\,(0)}_{\delta[U_i]}\,.
\]
This means that the initial velocity fluctuations in newly refined regions are always zero. As a consequence,
the initial turbulent energy production is also zero (equation~\ref{eq:tau_si}). This entails an initial
relaxation phase, in which fluctuations relative to the mean flow and the associated turbulent stresses
gradually grow and the down-scaled SGS turbulence energy adjusts to the new grid scale. This also means that
the resolved flow experiences only a negligible turbulent viscosity in the initial phase after grid refinement, 
which supports the growth of small-scale velocity fluctuations that were unresolved on the coarser grid
scale prior to refinement. 

\begin{figure*}
  \begin{center}
    \mbox{\subfigure[{$[U]$ for $T_{\rm c}=2\;\mathrm{Gyr}$}]{\includegraphics[width=0.49\linewidth]{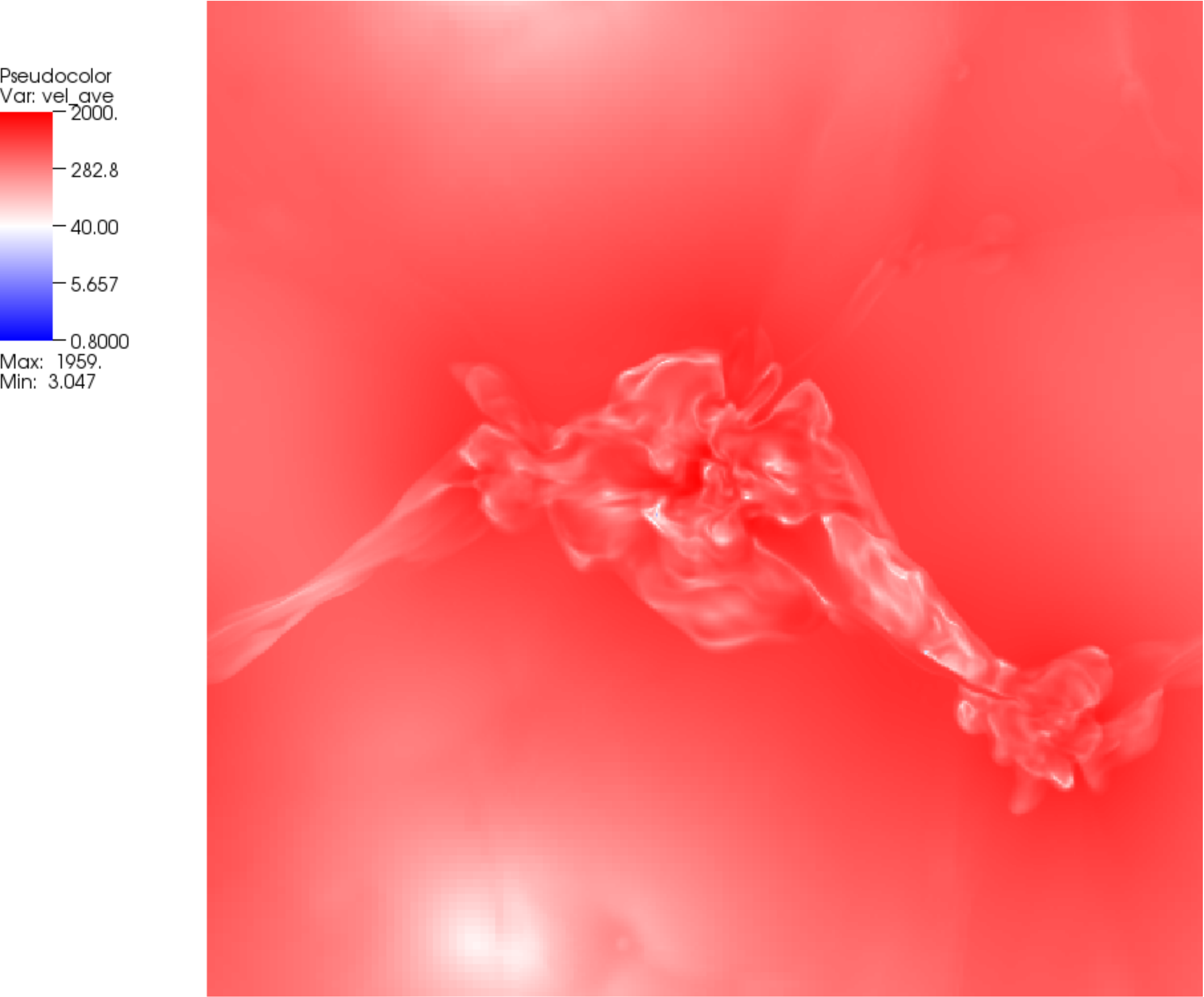}}\qquad
          \subfigure[{$U^\prime$ for $T_{\rm c}=2\;\mathrm{Gyr}$}]{\includegraphics[width=0.49\linewidth]{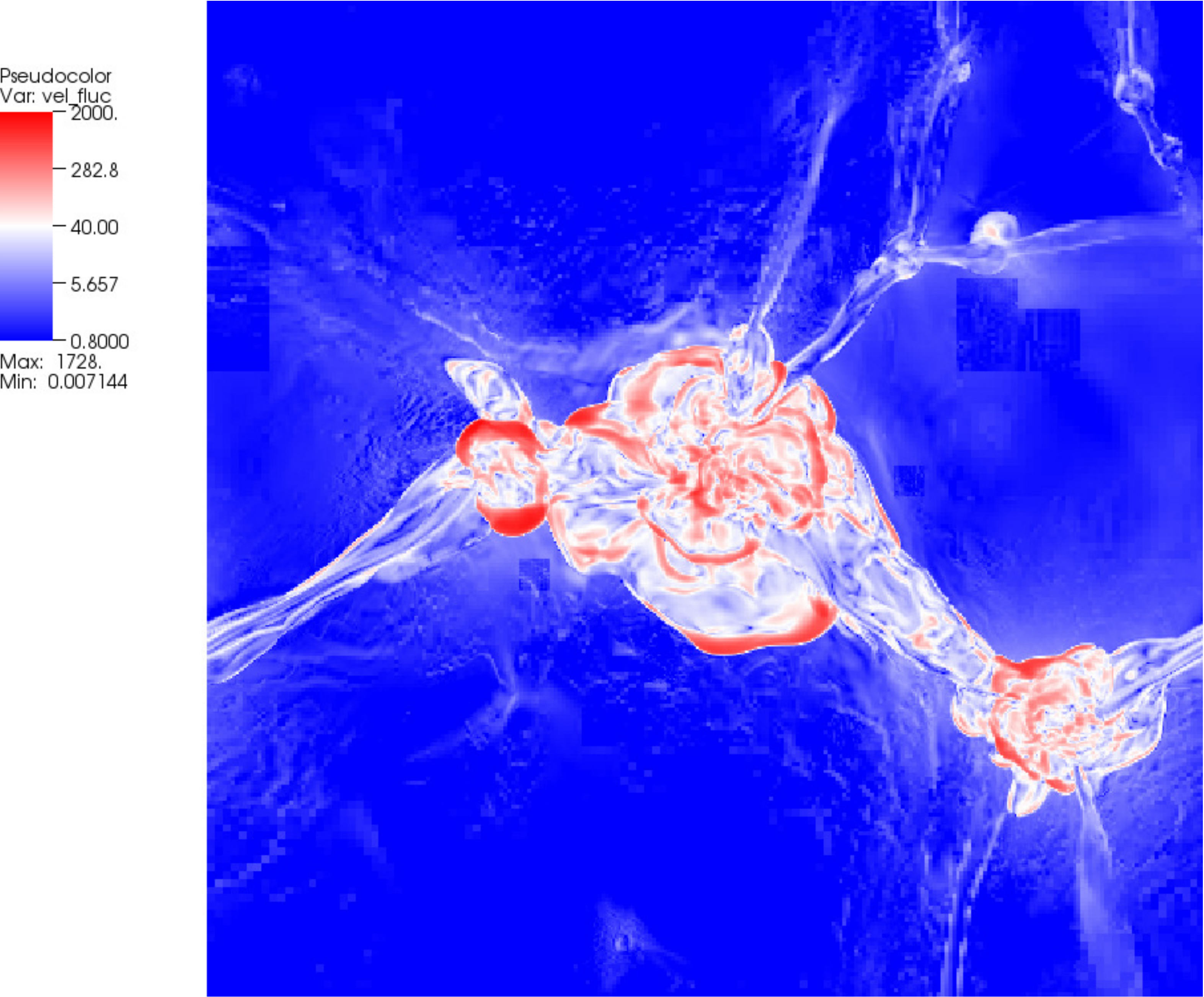}}}
    \mbox{\subfigure[{$[U]$ for $T_{\rm c}=10\;\mathrm{Gyr}$}]{\includegraphics[width=0.49\linewidth]{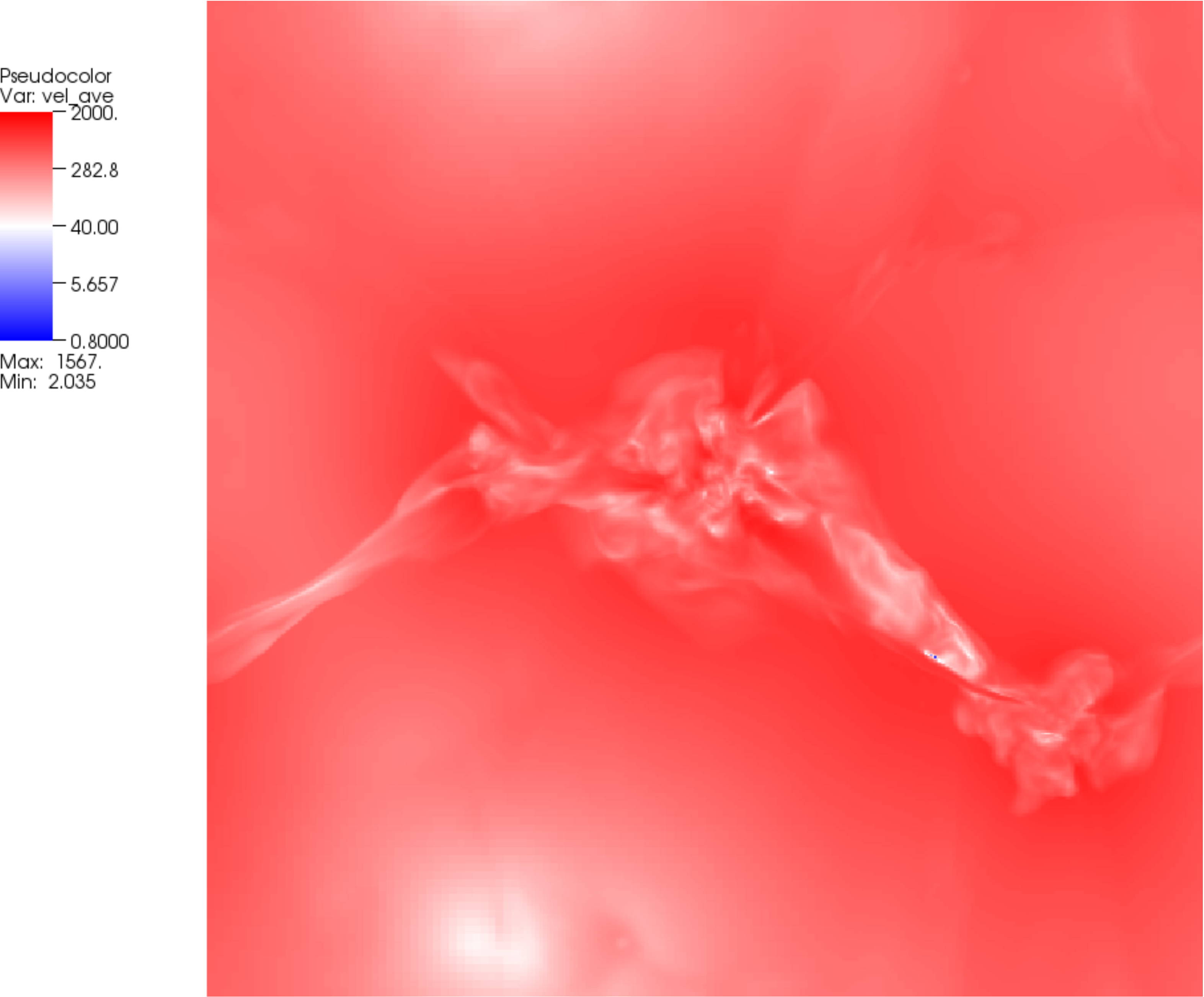}}\qquad
          \subfigure[{$U^\prime$ for $T_{\rm c}=10\;\mathrm{Gyr}$}]{\includegraphics[width=0.49\linewidth]{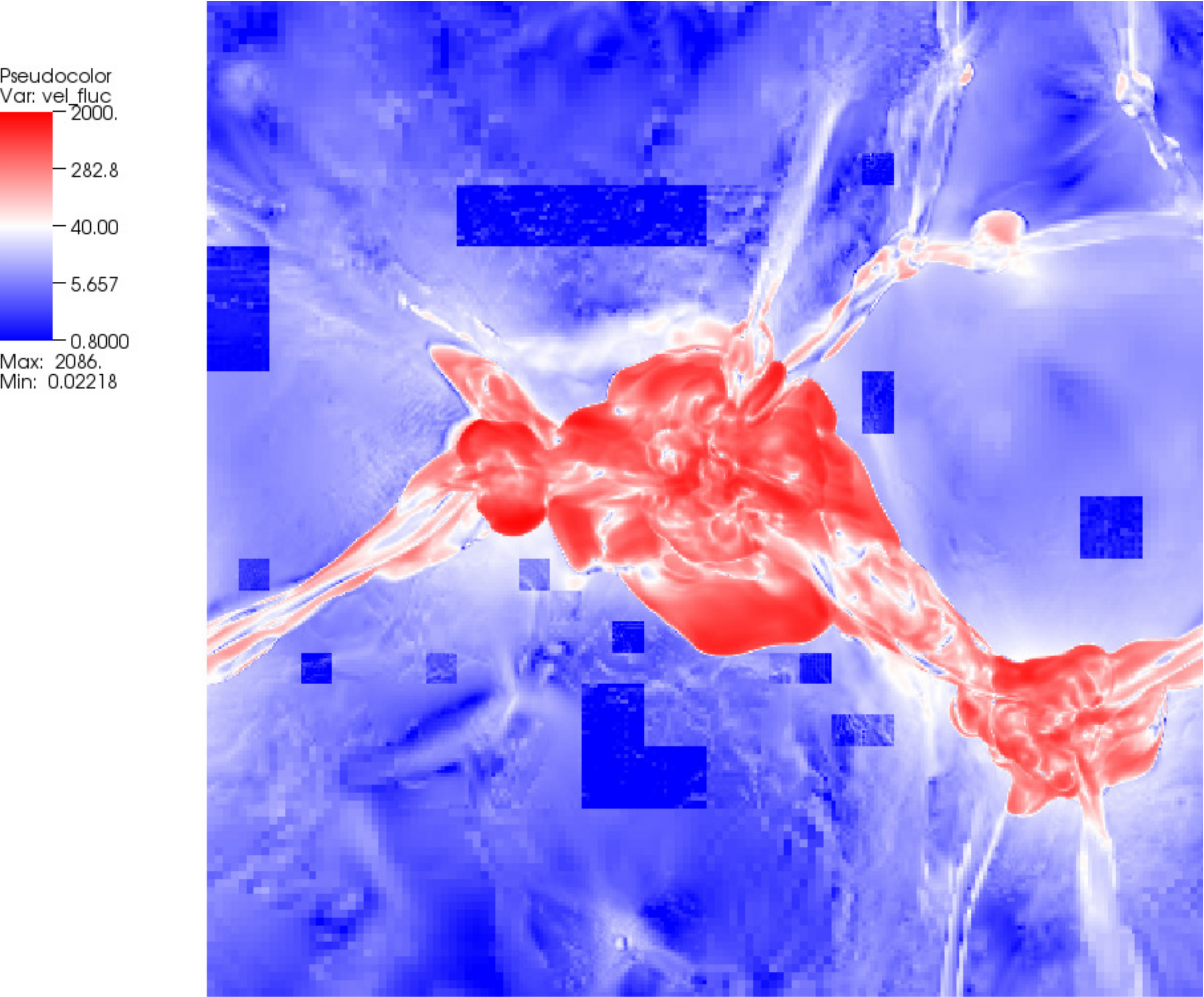}}}
    \caption{Slices of the filtered (left) and fluctuating (right) velocity components in shear-improved LES 
    		of the Santa Barbara cluster at $z=0.0$ with $U_{\rm c}=400\;\mathrm{km/s}$ and two different smoothing
		time scales. Velocities around the lower bound $0.1_{\rm c}=40\;\mathrm{km/s}$ in
		equation~(\ref{eq:sigma_fluc}) appear in white, smaller fluctuations in blue, and stronger fluctuations in red
		(online version).}
    \label{fig:SB_slices_vel}
  \end{center}
\end{figure*}

The two free parameters, $T_{\rm c}$ and $U_{\rm c}$, have to be chosen such that
$U_{\rm c}$ is roughly the integral velocity of turbulence if the flow enters a steady state and
$T_{\rm c}$ is a characteristic time scale over which the flow evolves. 
In Fig.~\ref{fig:SB_slices_vel}, we show the filtered velocities and the fluctuating component for test runs
of the Santa Barbara cluster with the fiducial turbulent velocity $U_{\rm c}=400\;\mathrm{km/s}$ (see Section~\ref{sc:sb_sgs}). 
For $T_{\rm c}=2\;\mathrm{Gyr}$ (top plots), which is below the typical time scale inferred from the turbulence production
rate, one can see that the mean flow $[\Ub]$ picks up steep fluctuations in the ICM, while $\Ub'$ is reduced to 
a magnitude of $\sim 0.1 U_{\rm c}\;\mathrm{km/s}$, with the exception of accretion shocks and the center of the cluster. 
This is clearly inconsistent. For $T_{\rm c}=10\;\mathrm{Gyr}$ (bottom plots), on the other hand, $\Ub'$ has roughly 
the expected magnitude and the Kalman filter mostly extracts the coherent accretion flows, although the fluctuations 
are slightly too high in the voids. One can also discern the reduced velocity fluctuation 
in regions that were recently refined. As shown in Section~\ref{sc:sb_calibr}, the optimal choice is in between these two cases. 

\subsection{Refinement criteria}
\label{sc:refinement}

\begin{figure*}
  \begin{center}\rm dm
    \mbox{\subfigure[{$z=1$, refinement by $(\rho+\rho_{\rm dm})\Delta^3$}]{\includegraphics[width=0.49\linewidth]{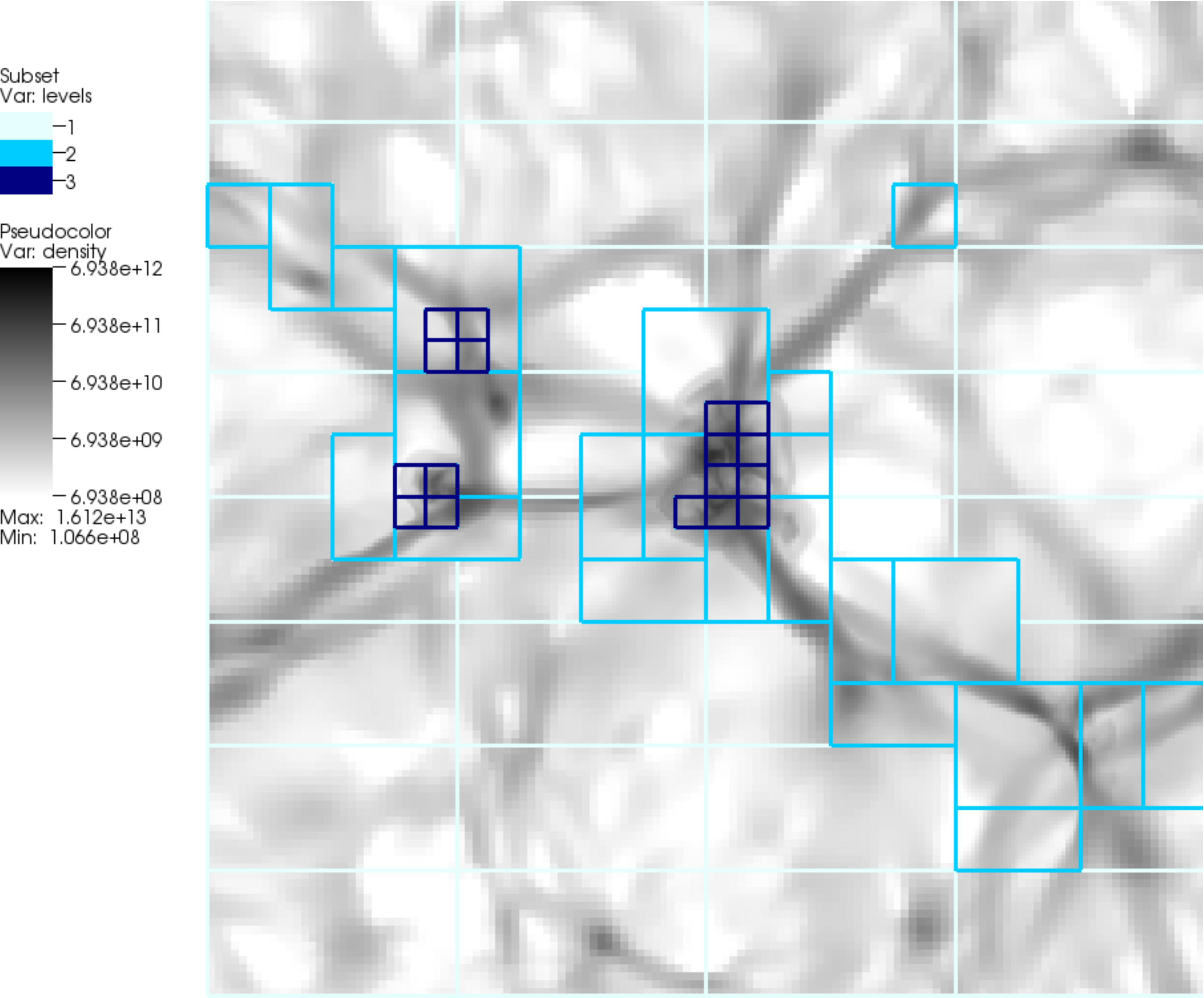}}\qquad
          \subfigure[{$z=1$, refinement by $(\rho+\rho_{\rm dm})\Delta^3$ and $\omega$}]{\includegraphics[width=0.49\linewidth]{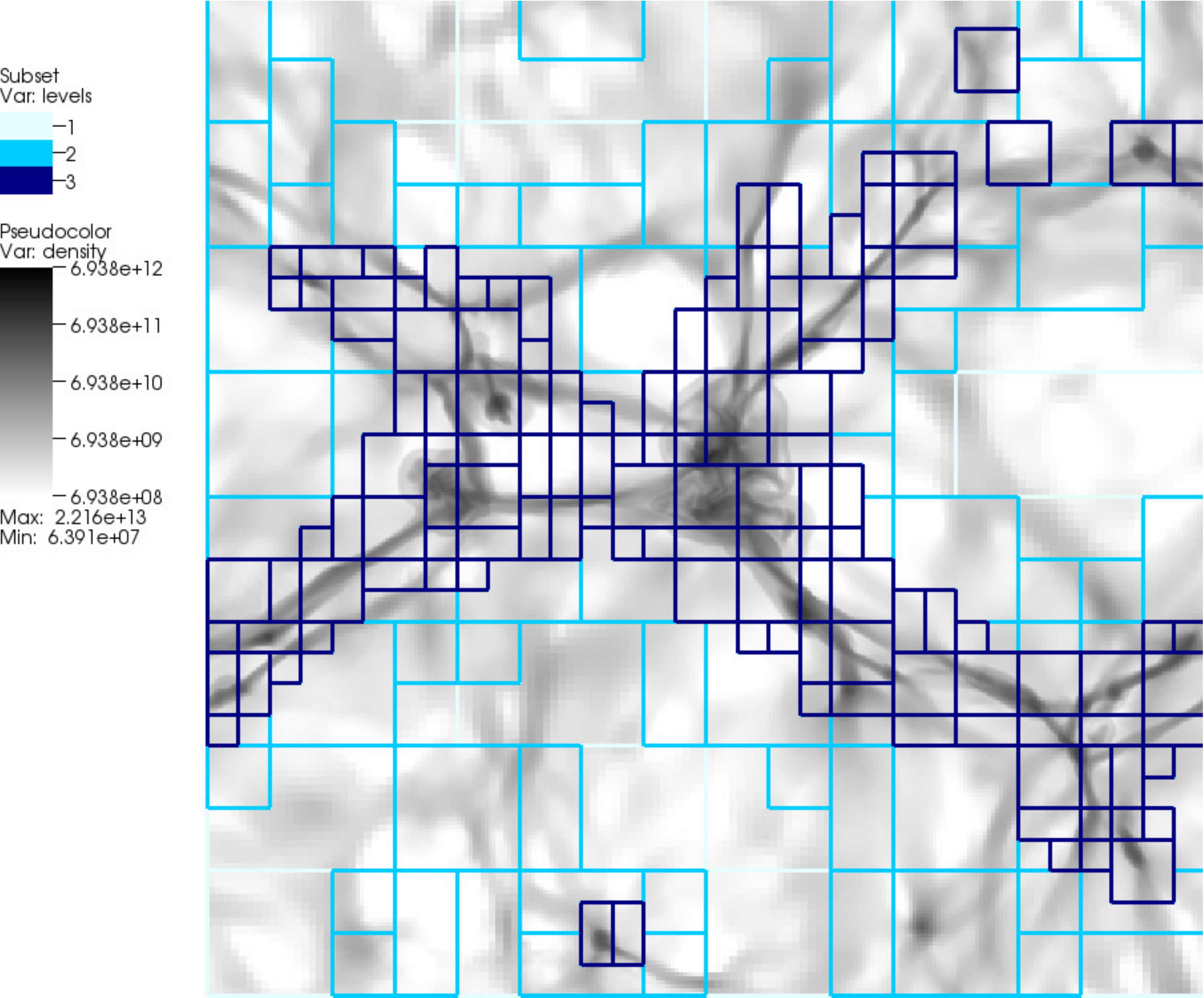}}}
    \mbox{\subfigure[{$z=0$, refinement by $(\rho+\rho_{\rm dm})\Delta^3$}]{\includegraphics[width=0.49\linewidth]{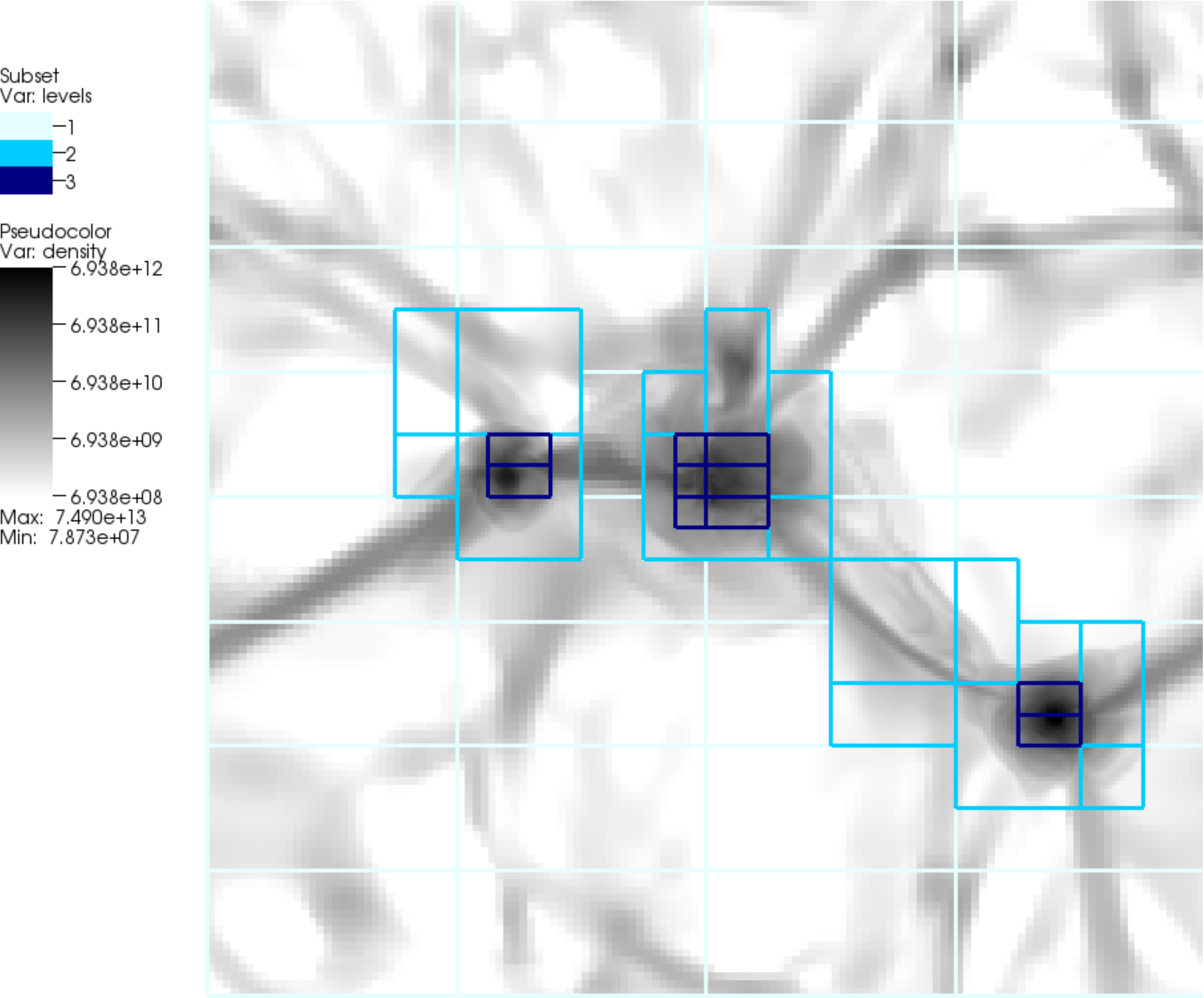}}\qquad
          \subfigure[{$z=0$, refinement by $(\rho+\rho_{\rm dm})\Delta^3$ and $\omega$}]{\includegraphics[width=0.49\linewidth]{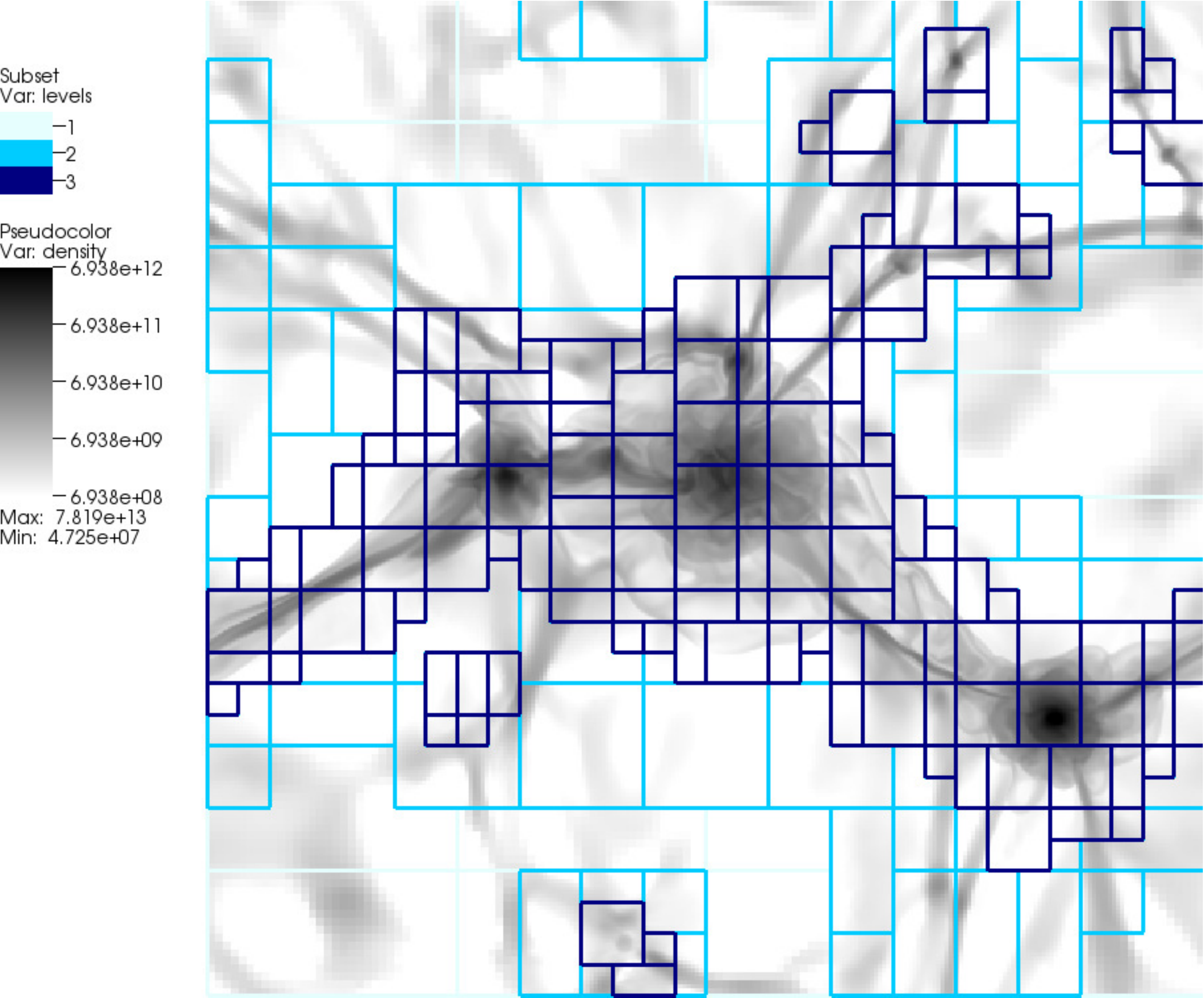}}}
    \caption{Slices of the mass density for refinement by cell mass only (left) and additional refinement by the vorticity
	modulus (right) at redshifts $z=0$ and $1$. The rectangular grids boxes are colored corresponding to the refinement level.
	In both simulations, the shear-improved SGS model with $T_{\rm c}=5\;\mathrm{Gyr}$ and
	$U_{\rm c}=400\;\mathrm{km/s}$ is applied.}
    \label{fig:SB_slices_dens}
  \end{center}
\end{figure*}

As in \cite{AlmBell12}, we set a threshold for the total mass in a cell, $(\rho+\rho_{\rm dm})\Delta^3$, above which cells
are tagged for refinement in simulations of the Santa-Barbara cluster. For root grid with $256^3$ cells,
the mass threshold is $3.5\times 10^9\;M_{\odot}$. 
For the lower-resolution test runs, the minimal mass is raised by a factor of $8$, corresponding to the eight times larger volume
of the root-grid cells. In both cases, cells at level $l$ are tagged for refinement if 
\[
	\delta + \delta_{\rm dm}\ge 3.228\times 8^{l},
\]
where $\delta+\delta_{\rm dm}=(\rho+\rho_{\rm dm})/\rho_0$ is the overdensity of gas and dark matter. 
The gas density and the grid structure at two different redshifts obtained with
this refinement criterion are shown in the left plots of Fig.~\ref{fig:SB_slices_dens}. 

To refine turbulent regions in AMR simulations, \citet{SchmFeder09} propose refinement by the vorticity modulus and
the compression rate, which is defined by the negative substantial time derivative of the divergence. The thresholds
are determined by computing averages and standard deviations of these control variables at each refinement level.
As shown by \citet{IapiAda08}, this does an excellent job of reproducing the turbulent wake produced in
idealized simulations of minor merger events, where a subhalo plunging into the ICM of a big cluster is modeled
by a gravitationally bound gas cloud in a wind. In this work, we apply refinement by the vorticity modulus
in addition to the refinement by overdensity, as defined above. For the vorticity
modulus we set absolute thresholds of $200$ and $500\;(\mathrm{km/Mpc)/s}$ for root-grid resolutions of $128^3$ and $256^3$,
respectively. Only if these thresholds are exceeded, we use the statistical threshold
\begin{equation}
	\label{eq:omega_thresh}
	\omega \ge \langle\omega\rangle_l + \max\left(\langle\omega\rangle_l,\mathrm{std}_l\,\omega\right),
\end{equation} 
where the $\langle\omega\rangle_l$ is the average and $\mathrm{std}_l\,\omega$ the standard deviation of
$\omega$ over all cells at level $l$. The absolute thresholds avoid refinement of large volume fractions of the
domain at early stages of structure formation. Turbulence in the cluster gas and the main filaments, however, is
covered by refined grids, as demonstrated by the right plots in Fig.~\ref{fig:SB_slices_dens}. 

\label{lastpage}

\end{document}